\documentclass[3p]{elsarticle}

\usepackage{lineno,hyperref}
\usepackage{epsfig}
\usepackage{natbib}
\modulolinenumbers[5]

\journal{Journal of \LaTeX\ Templates}

\begin{document}

\title{Evolutionary potential games on lattices}

\author[mfa]{Gy{\"o}rgy Szab{\'o}}
\ead{szabo@mfa.kfki.hu}

\author[mfa]{Istv\'an Borsos}
\ead{borsos@mfa.kfki.hu}

\address[mfa]{Institute for Technical Physics and Materials
Science, Centre for Energy Research, Hungarian Academy of Sciences, \\ P.O. Box 49, H-1525 Budapest, Hungary}

\begin{abstract}
Game theory provides a general mathematical background to study the effect of pair interactions and evolutionary rules on the macroscopic behavior of multi-player games where players with a finite number of strategies may represent a wide scale of biological objects, human individuals, or even their associations. In these systems the interactions are characterized by matrices that can be decomposed into elementary matrices (games) and classified into four types. The concept of decomposition helps the identification of potential games and also the evaluation of the potential that plays a crucial role in the determination of the preferred Nash equilibrium, and defines the Boltzmann distribution towards which these systems evolve for suitable types of dynamical rules. This survey draws parallel between the potential games and the kinetic Ising type models which are investigated for a wide scale of connectivity structures. We discuss briefly the applicability of the tools and concepts of statistical physics and thermodynamics. Additionally the general features of ordering phenomena, phase transitions and slow relaxations are outlined and applied to evolutionary games. The discussion extends to games with three or more strategies. Finally we discuss what happens when the system is weakly driven out of the "equilibrium state" by adding non-potential components representing games of cyclic dominance.
\end{abstract}

\date{\today}


\maketitle

\tableofcontents

\section{INTRODUCTION}
\label{sec:intro}

Most games can be considered as simplified real-life situations in which the number of players as well as their options is limited. The simplest case is represented by two-player non-cooperative games where the intelligent and selfish players wish to maximize their own payoff that is quantified by a real number and depending on their simultaneous choices while their utilities are quantified by two payoff matrices \citep{neumann_44}.

The original concept and application of payoff matrices have been extended basically within the framework of evolutionary game theory \cite{zeeman_80, maynard_82, hofbauer_88, hofbauer_98, gintis_00, cressman_03, nowak_06, sigmund_10, sandholm_10}. In biological systems the payoff matrices characterize the resultant fitness (more precisely, the capability to create offspring) for the interacting pair of species (representing the strategies for this terminology) and serve as a fundamental quantity in the population dynamics developed in the spirit of the Darwinian selection \cite{maynard_n73, taylor_p_mb78, hofbauer_jtb79, schuster_jtb83}. In human systems, however, the imitation of the more successful player can control the time-dependent frequency of players following a given strategy, in close analogy to biological systems \cite{axelrod_s81, axelrod_84, helbing_pa92, schlag_jet98, hummert_mbs14}.

The first systematic investigations of evolutionary games were performed in well-mixed population of players for a wide scale of games and strategies. A progressively expanding research area was initiated by \citet{nowak_n92b} who introduced a model where the players are distributed on a square lattice and their incomes come from one-shot games with their neighbors. In this cellular automaton type model the players changed their strategy simultaneously (in discrete time steps) by imitating the strategy of the most successful neighbor. The numerical investigations of these deterministic models \cite{nowak_ijbc93} have demonstrated the advantage of spatial structures for short-range interactions in the maintenance of cooperation among selfish players for the prisoner's dilemma. Since that time numerous followers have clarified relevant effects and phenomena supporting the maintenance of cooperative behaviors in real-life situations described by different versions of social dilemmas as surveyed by \citet{nowak_06, sigmund_10, szabo_pr07, allen_ems14}.

The systematic investigations involve systems where the authors studied the effects of different sets of strategies, the connectivity structures described by lattices or graphs, and a wide scale of possible evolutionary rules including co-evolutionary games where all the relevant ingredients of the model are allowed to evolve together with the strategy distribution \cite{perc_bs10}. Up to now most of the relevant two-person games are studied within the framework of evolutionary games. New features can be investigated via versions of group interactions that cannot be decomposed into the sum of pair interactions \cite{pacheco_prsb09, perc_jrsi13}.

Within this wide scale of systems less efforts are focused on the potential games \cite{monderer_geb96}, which are related intimately to the statistical physics \cite{blume_l_geb93}. For a more detailed exposition of potential games see the relevant chapter in the book by \citet{sandholm_10}. The decomposition of the matrix games into the linear combination of elementary games was discussed previously by \citet{candogan_mor11, hwang_sh_arx11}. The latter approach allows us to identify different types of pair interactions and also to determine the existence of potential as well as to evaluate it if it exists \cite{szabo_pre14b}. Very recently \citet{cheng_dz_a14} has suggested another method for the evaluation of the potential. The applications and perspectives  of potential games within the economy are reviewed by \citet{mallozzi_cejor13}.

In this review we survey briefly the fundamental concepts and background material necessary for understanding the recent and future developments in the research of evolutionary potential games. Due to the wide scale of research fields involved and the limited length of this article we cannot give a comprehensive picture. Instead of mathematical rigorousness we use a concise style with an effort to give a wide picture of the set of phenomena and relationships that are already studied in the fields of statistical physics.

The structure of this review is as follows. The next four Sections summarize the basic concepts and methods of traditional and evolutionary game theory including the general features of potential games and the way of the evaluation of potential. Subsequently we discuss the intimate relationship between multi-agent evolutionary potential games and Ising type models. Afterwards we briefly survey the phenomena ({e.g.}, order-disorder phase transitions) explored by the application of the Ising type models. In Sec. \ref{sec:ordproc} we describe briefly the ordering processes characterizing the ways how the systems tend towards the final stationary ordered structure at low noise levels. Section \ref{sec:dfepg} is addressed to discuss what happens when the system behavior deviates from the thermodynamical equilibrium due to the presence of non-potential components representing cyclic dominance in the payoff matrices. Finally we outline some challenging questions for the continuation of research in the near future.

\section{BRIEF SURVEY OF GAMES}
\label{sec:bsg}

First we give a concise and tutorial description of the basic definitions and concepts of normal games we use throughout this work. The reader can find further details of the traditional game theory in standard textbooks like \citet{fudenberg_91, gibbons_92, hofbauer_98, weibull_95, samuelson_97, gintis_00, cressman_03, sandholm_10, sigmund_10} that provide a more general outline of a wider scale of games and illustrate their application in biology, economics, and behavioral sciences.

\subsection{Players, strategies, payoffs, and potential}
\label{sec:psp}

The normal game is an abstract formulation of a decision situation where each of $N$ players must choose simultaneously (and independent of each other) one of their own possible options to receive payoffs dependent on the choices of all of them. Each intelligent and selfish (rational) player wishes to maximize her own payoff with the assumption that the co-players are also intelligent. In other words, the players are capable of deducing the best possible way of playing the game and they can handle the problem of the common knowledge implying the fact that all players know that the others wish to maximize their own utility, and all players know that all the others know that all of them wish to maximize their own payoff, etc.

In multi-player games the players will be denoted by $x=1, \ldots , N$. Each player $x$ can choose a pure strategy $s_x$ from her set of strategies indexed by an integer, $s_x \in \{1, 2, \ldots , n_x\}$ where $n_x$ denotes the number of strategies of player $x$. The choices of all the players are defined by the strategy profile ${\bf s}=(s_1, \ldots , s_N)$ determining their payoff. In this notation each player $x$ receives a payoff $u_x({\bf s})=u_x(s_1, \ldots , s_N)$ quantifying her utility. The above strategies are pure strategies. In many games (e.g., poker, matching pennies, or rock-scissors-paper game) the players can also play mixed strategies where they select one of their pure strategies with a given probability in each decision instance. Henceforth our discussion is limited to the cases where the players are constrained to use pure strategies.

In the above games each of the $N$ players wishes to optimize her own payoff regardless of the others, which is complicated by the fact that, in general, none of the players can achieve her maximum payoff because of the counter-interest of the others. Despite it, the players can be satisfied if they choose a strategy profile $s^*=(s_1^*,\dots,s_N^*)$, called Nash equilibrium, that satisfies the following conditions:
\begin{equation}\label{eq:NE}
\forall\,x, \forall\, s_x^{}\ne s_x^* : \quad
u_x^{}(s_x^*,s_{-x}^*) \ge u_x^{}(s_x^{},s_{-x}^*).
\end{equation}
where $s_{-x}=(s_1, \ldots ,s_{x-1},s_{s+1},\ldots , s_N)$ denotes the strategy profile of the co-players. If the inequality is strict then $s^*$ is called a strict Nash equilibrium. For the Nash equilibrium the players do not have unilateral incentive to choose another strategy. In other words, each player is satisfied with her own choice, because there is no way to receive a higher payoff for the given choices of the others. Nash's theorem \cite{nash_pnas50, nash_am51} states that in normal games there exists at least one Nash equilibrium, possibly involving mixed strategies.

For most of the cases the game has more than one Nash equilibrium. In these situations we need additional criteria for the strategy selection. For example we can choose those Nash equilibria that provides the higher sum of the individual utilities. Another way of selection, based on the consideration of risk dominance, was introduced by \citet{harsanyi_88}. The latter method suggests choosing the strategy that yields higher expected payoff against an opponent playing a mixed strategy when all feasible strategies are chosen with equal probability. These additional criteria may give contradictory advices for some games.

Some of the games have attracted huge attention due to their curiosity and the high frequency we face them day by day. The best known example is the prisoner's dilemma representing a real-life situation when the two players have only two options, called cooperation and defection, and the game has only one Nash equilibrium, the mutual defection, while the mutual cooperation would result in a higher payoff for both players (hence the dilemma). The analysis of similar situations attracted progressive activity in different fields of science, including biology, economics, social sciences, and physics \cite{nowak_06, sigmund_10, szabo_pr07}.

In normal games each player has her own utility function $u_x({\bf s})$ to consider. For the potential games we can introduce a single function $V({\bf s})$, called potential, that involves the strategic incentives of all players in the following way suggested by \citet{monderer_geb96}. Namely, the difference of the potential $V({\bf s})=V(s_1,s_2,\ldots,s_N)$, when player $x$ modifies her strategy (from $s_x$ to $s_x^{\prime}$), is equal to the utility function difference for the given player, that is,
\begin{equation}\label{eq:potential}
u_x(s_x^\prime;s_{-x})-u_x(s_x;s_{-x}) =
V(s_x^\prime;s_{-x})-V(s_x;s_{-x})
\end{equation}
for $\forall\,x, \forall\, s_x, s_x^{\prime}$, and $\forall\, s_{-x}$. This means that the potential is constructed from the payoff variations of those players changing their strategies. Evidently, the existence of a potential for a normal game is strongly limited by the above conditions. In other words, Eq.\ (\ref{eq:potential}) gives strict constraints for the possible values of payoffs. The difficulties can be well illustrated by introducing a dynamical graph \cite{schnakenberg_rmp76} where each node represents a strategy profile ${\bf s}$ and the edges denote transitions between two strategy profiles when only a single player modifies her strategy. The sum of the given individual payoff differences is zero along all loops of this dynamical graph for potential games.

Notice that if only one player modifies her strategy along a loop, {\it e.g.}, $s_{x}(1) \to s_{x}(2) \to \ldots \to s_{x}(k) \to s_{x}(1)$) (while $s_{-x}$ is quenched) then the sum of the payoff variation:
\begin{equation}\label{eq:loopV0}
\sum_{i=1}^{k} \left[u_x(s_{x}(i+1);s_{-x})-u_x(s_{x}(i);s_{-x})\right] = 0
\end{equation}
because in this sum all payoff components appear twice with opposite signs as $s_{x}(k+1)$ equals $s_{x}(1)$. \citet{monderer_geb96} have proved that a normal game admits a potential if and only if over any four-edge loops of the dynamical graph the sum of the changes in the deviator's payoffs equals zero. Analogously to the potential energy in physical systems the potential is unique also here up to addition of a constant. In the literature of game theory one can find a wide range of examples satisfying the conditions of potential games from genetic competition \cite{fisher_30, hofbauer_88} to congestion game on network \cite{beckmann_56, rosenthal_ijgt73, facchini_td97, sandholm_wh_jet01} or oligopoly games \cite{slade_jie94}.

The existence of the potential simplifies the analysis. For example, the strategy profile where $V({\bf s})$ achieves its maximum is a preferred (pure) Nash equilibrium that plays a distinguished role in evolutionary game theory like the ground state in physical systems.

If the edges of the above mentioned dynamical graph are directed (called flow graph henceforth) by pointing towards the higher potential value, then the nodes (strategy profiles) with only incoming edges are Nash equilibria. Similar flow graphs can be derived for the so-called ordinal potential games where a weaker condition should be satisfied. Namely, if
\begin{equation}\label{eq:ordpot}
u_x(s_x^\prime;s_{-x})-u_x(s_x;s_{-x})> 0 \nonumber
\end{equation}
then
\begin{equation}\label{eq:ordpot2}
V(s_x^\prime;s_{-x})-V(s_x;s_{-x})>0 \,.
\end{equation}
Evidently, the existence of a potential prohibits the presence of directed loops here. The advantage of the latter feature will be demonstrated later. In the literature of game theory one can find other classes of potential games, that we will not discuss here. For example, besides the exact and ordinal potential games mentioned above,  \citet{monderer_geb96} have introduced the weighted and generalized ordinal potential games; \citet{voorneveld_el00} has studied the best-response potential games; and \citet{morris_jet05} investigated the robust sets of equilibria for the generalized potentials.

The most relevant application of potential games was discovered by \citet{blume_l_geb93, blume_l_geb95}, who proved that for a certain set of evolutionary rules the system evolves into a Boltzmann-Gibbs ensemble and the stationary state can be well investigated by the tools of statistical physics. This feature justifies the importance of the preferred Nash equilibrium.

Henceforth our analysis will be focused on those multi-player games where the players are allowed to use only pure strategies and the interaction is composed of two-player games surveyed in the following section.

\subsection{Two-player games}
\label{sec:2pg}

For a two-player game with players $x$ and $y$ the payoffs are defined by payoff tables and we can apply the terminology of matrices. In many cases the strategy labels $i$ and $j$ ($i=1, 2, \ldots, n_x$ and $j=1, 2, \dots, n_y$) are sufficient to identify the pure strategies selected by players $x$ and $y$. For the expression of payoffs, however, it is convenient to denote the strategies of player $x$ by ($n_x$-dimensional) unit vectors, as
\begin{equation}
\label{eq:purestrats}
{\bf s}_x={\bf s}_{x1}=\left( \matrix{1 \cr 0 \cr \vdots \cr 0 \cr}\right)\,,
    {\bf s}_{x2}=\left( \matrix{0 \cr 1 \cr \vdots \cr 0 \cr}\right)\,,
    \ldots \,,
    {\bf s}_{xn_x}=\left( \matrix{0 \cr  \vdots \cr 0 \cr 1 \cr}\right)\,,
\end{equation}
and similar expressions define the strategies of player $y$. For this notation the payoffs of player $x$ and $y$ can be expressed by the following matrix products as
\begin{eqnarray}\label{eq:po}
u_x(s_x,s_y)&=&{\bf s}_x \cdot {\bf A} {\bf s}_y \,, \\
u_y(s_x,s_y)&=&{\bf s}_y \cdot {\bf B} {\bf s}_x = {\bf s}_x \cdot
{\bf B}^{T} {\bf s}_y \,, \nonumber
\end{eqnarray}
where the matrix elements $A_{ij}$ and $B_{ij}$ define the payoffs of players $x$ and $y$ (we used the relation $B_{ij}^T=B_{ji}$). Here we have to mention that the above expressions define the average payoffs for mixed strategies, {\it e.g.}, ${\bf s}_x=\sum_i \rho_i {\bf s}_{xi}$, when player $x$ uses her $i$th strategy with a probability $\rho_i$ ($\sum_i \rho_i =1$). The latter notation is useful for games having mixed Nash equilibrium. If the players are different (like in male-female, young-old, buyer-seller, sender-receiver interactions) then the so-called asymmetric two-player normal games are specified by two payoff matrices (${\bf A}$ and ${\bf B}$) and it is customary to use its bi-matrix form, ${\bf G}=({\bf A},{\bf B}^T)$, where the payoffs are given as
\begin{equation}\label{eq:bi-matrix}
{\bf G}=\left(\matrix{ (A_{11},B_{11}^T) & \cdots &
(A_{1n_y},B_{1n_y}^T) \cr
                 \vdots & \ddots & \vdots \cr
                 (A_{n_x1},B_{n_x1}^T) & \cdots &
(A_{n_xn_y},B_{n_xn_y}^T) \cr}\right).
\end{equation}
For identical players the game is symmetric (${\bf A}={\bf B}$ and $n_x=n_y=n$) and the game is well described by a single payoff matrix. An additional symmetry occurs in the partnership games \cite{hofbauer_98, sandholm_geb10} (or ''games with common interests'' \cite{monderer_geb96}) when the players share the utility equally, that is, if ${\bf A} = {\bf A}^T$.

For the two-player potential games we can introduce a potential matrix
\begin{equation}\label{eq:2ppot}
{\bf V}=\left(\matrix{ V_{11} & \cdots & V_{1n_y} \cr
                 \vdots & \ddots & \vdots \cr
                 V_{n_x1} & \cdots & V_{n_xn_y} \cr}\right).
\end{equation}
where the matrix components can also be expressed via the use of a matrix product, namely,
\begin{equation}\label{eq:Vij}
V_{ij}={\bf s}_{xi} \cdot {\bf V} {\bf s}_{yj} \,, \\
\end{equation}
that satisfies the conditions
\begin{eqnarray}\label{eq:2ppotcond}
V_{kj}-V_{ij}&=&A_{kj}-A_{ij} \,, \nonumber \\
V_{il}-V_{ij}&=&B_{il}-B_{ij} \,,
\end{eqnarray}
where $i,k =1, \ldots , n_x$ and $j,l =1, \ldots , n_y$.

In order to visualize the relevance of the above constraints we introduce now the dynamical graph representation of games where each possible strategy profile is denoted by a node arranged in the same way as in the bi-matrix formalism [see Eq.~(\ref{eq:bi-matrix})].

\begin{figure}[ht]
\centerline{\epsfig{file=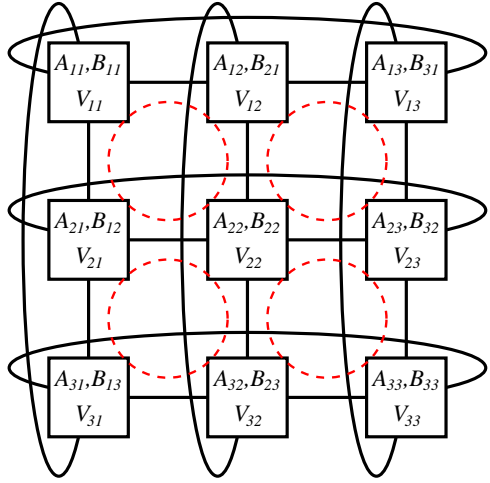,width=7cm}}
\caption{(Color online) Dynamical graph with payoff pairs and potential values for a two-player $3 \times 3$ game. Dashed (red) circles indicate a possible choice of the four independent four-edge loops to be considered for the existence of potential according to Kirchhoff's law.}
\label{fig:dg_g3x3}
\end{figure}

Figure \ref{fig:dg_g3x3} illustrates the dynamical graph for a two-player $3 \times 3$ game where the edges of the graph connect those pairs of strategy profiles where only one of the players changes her strategy.
This arrangement of nodes resembles a $3 \times 3$ square lattice with periodic boundary conditions. The payoff pairs and the elements of the potential matrix are indicated within the boxes representing strategy profiles. Notice, that the nodes of a column (or row) form a complete subgraph for any finite number of strategies. Along these edges we will consider the payoff variation of only that player who modifies her strategy.

The existence of potential implies that the summarized potential variation (or payoff variation of the active player) between any two strategy profiles [{\it e.g.} along a given series of unilateral strategy changes from $(s_{xi},s_{yj})$ to $(s_{xk},s_{yl})$] is independent of the path connecting the initial and final strategy profiles. In other words, the sum of potential variation is zero along each loop of the dynamical graph. More precisely, according to Kirchhoff's laws we should take into consideration only the independent loops, their number is the difference of the number of edges of the whole dynamical graph and those of its spanning tree \cite{kirchhoff_apc1847, desoer_69}.

For the $3 \times 3$ games the dynamical graph has 9 nodes and 18 edges as plotted in Fig.~\ref{fig:dg_g3x3} while the spanning tree has only 8 edges, {\it e.g.}, those forming a shape of $\exists$. The number of independent loops is 10 and the corresponding loops can be constructed by adding edges to the spanning tree consecutively. It is convenient to select the shortest loop containing the new edge. In the present case, however, the number of non-trivial independent loops is only 4 because along the three-edge loops (within a column or row) the conditions are satisfied as it is demonstrated by Eq.~(\ref{eq:loopV0}).

Accordingly, in the present case there are only four independent (internal) four-edge loops that can be selected as denoted by dashed (red) circles in Fig.~\ref{fig:dg_g3x3}. Kirchhoff's law ensures that the conditions of the existence of potential are satisfied for all the possible loops if it is satisfied for each independent four-edge loop.

\begin{figure}[ht]
\centerline{\epsfig{file=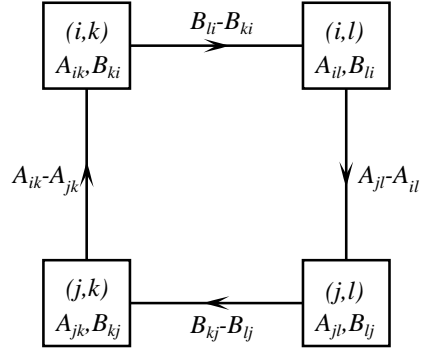,width=5.5cm}}
\caption{Payoff variations of the active player for a $2 \times 2$ sub-games along the directed four-edge loops of the dynamical graph. Within the boxes the upper parameters refer to the strategy profile and the corresponding payoffs are denoted in lower row.}
\label{fig:dg_2x2sg}
\end{figure}

For general treatment it is useful to discuss in detail the $2 \times 2$ sub-games where player $x$ can use her strategy $i$ or $j$ ($1 \le i <j \le n_x$) while player $y$ is restricted to select either her $k$th or $l$th strategy ($1 \le k <l \le n_y$). The relevant payoff variations for this sub-game are illustrated in Fig.~\ref{fig:dg_2x2sg}.
A potential exists if the sums of the payoff variations of the active player along these four-edge loops become zero, that is,
\begin{equation}\label{eq:loop2x2sgbmg}
B_{li}-B_{ki}+A_{jl}-A_{il}+B_{kj}-B_{lj}+A_{ik}-A_{jk}=0
\end{equation}
for all possible $2 \times 2$ sub-games \cite{monderer_geb96}. Conversely, all the sub-games of a potential game are potential games, too.

For a two-player game with $n_x$ and $n_y$ strategies we can distinguish $n_x(n_x-1)n_y(n_y-1)/4$ two-strategy sub-games whereas the number of independent four-edge loops is significantly less, {\it i.e.} $(n_x-1)(n_y-1)$, according to the Kirchhoff laws discussed above. The number of independent and relevant four-edge loops of the dynamical graph is reduced drastically for the symmetric matrix games. The reduction of the number of the relevant four-edge loops is related to the equivalence of payoff variations along the loops $(i,k) \to (i,l) \to (j,l) \to (j,k) \to (i,k)$ and $(k,i) \to (l,i) \to (l,j) \to (k,j) \to (k,i)$ if ${\bf A}={\bf B}$. As a result, if condition (\ref{eq:loop2x2sgbmg}) is satisfied along a loop, then it is satisfied also for its counter-loop.

Additionally, if ${\bf A}={\bf B}$, then we can distinguish three types of four-edge loops. In the first case, both players can choose the same two strategies ({\it e.g.}, $i=k$ and $j=l$) and then Eq. (\ref{eq:loop2x2sgbmg}) is always satisfied. That means that all the symmetric $2 \times 2$ games (and sub-games) are potential games. Section \ref{sec:proppot} is devoted to discuss the consequences of this inherent feature.

In the second case the players have a common strategy and their second strategies are distinct. For example, the upper right loop, indicated by the (red) dashed circle in Fig.~\ref{fig:dg_g3x3} represents such a situation when Eq.~(\ref{eq:loop2x2sgbmg}) becomes
\begin{equation}\label{eq:loop2x2w3s}
A_{12}-A_{21}+A_{23}-A_{32}+A_{31}-A_{13}=0\,.
\end{equation}
Due to the above-mentioned symmetries this is the only criterion for the existence of potential in the set of symmetric $3 \times 3$ matrix games.

The third type of the four-edge loops can appear only for $n \ge 4$ because here the players use different strategy pairs. For example, along the loop $(1,3) \to (1,4) \to (2,4) \to (2,3) \to (1,3)$ in a four-strategy game Eq.~(\ref{eq:loop2x2sgbmg}) obeys the following form:
\begin{equation}\label{eq:loop2x2w4s}
A_{13}-A_{31}+A_{24}-A_{42}+A_{32}-A_{23}+A_{41}-A_{41}=0\,.
\end{equation}
The structure of Eqs. (\ref{eq:loop2x2w3s}) and (\ref{eq:loop2x2w4s}) implies a relationship between the existence of potential and the absence of cyclic components of the basis games (see Sect.~\ref{sec:decomposition_smg}).

In summary, for an $n \times n$ matrix game, Kirchhoff's laws give $(n-1)^2$ conditions to be satisfied by the payoff matrix elements ($A_{ij}$ and $B_{ij}$) for the potential games. If ${\bf A}={\bf B}$ then the number of independent conditions is reduced to $(n-1)(n-2)/2$ in agreement with the fact that each symmetric $2 \times 2$ matrix game is a potential game and for the symmetric $3 \times 3$ potential games the nine payoff parameter $A_{ij}$ must satisfy only one simple linear relation given by (\ref{eq:loop2x2w3s}).

\subsection{Flow graphs}
\label{sec:flowdiagrams}

The flow graph is the directed version of the dynamical graph where each node represents a strategy profile and the directed edges connect those strategy pairs which differ in only one of the player's strategies. The arrows on these edges point towards the strategy profile resulting in higher income for the player modifying her strategy unilaterally. Notice that here the (bold) arrows differ from those we used in Fig.~\ref{fig:dg_2x2sg}. This approach limits the analysis to those cases where the preferences of these unilateral changes are defined.

Some of the properties of potential games are straightforward consequences of well-known results of the theory of graphs, especially of directed graphs \cite{harary_66}. The flow graphs are simple directed graphs that have neither parallel edges nor self-loops. Additionally, the flow graph of a potential game is free of directed loops as mentioned before.

\begin{figure}[ht]
\centerline{\epsfig{file=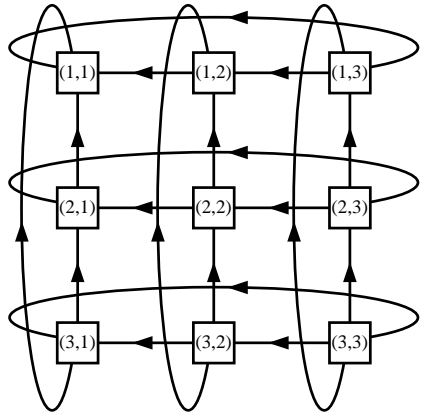,width=6cm}}
\caption{Flow graph for a three-strategy matrix game. Labels of vertices/boxes refer to the strategy profiles and the directed edges indicate the preferred strategy change for one of the players.}
\label{fig:fd_g3x3}
\end{figure}

Figure \ref{fig:fd_g3x3} represents a flow graph for a three-strategy game.
Due to the intimate relationship between the dynamical graphs and flow graphs here we use the same arrangement of strategy profiles as in Fig.~\ref{fig:dg_g3x3}. We emphasize, however, that for larger number of strategies ($n_x,n_y \ge 3$) the structure resembling the square lattice with periodic boundary conditions is preserved whereas the graph should be extended by additional directed edges connecting any pairs located within the same column or row. Remarkably, directed loops cannot occur within one row (and column) of the flow graph of a matrix game because of the conditions defined by Eq.~(\ref{eq:loopV0}). Furthermore, within each of these directed subgraphs there exists only one strategy providing the best income for the given player. As a strict Nash equilibrium is an optimal choice for both players, therefore the latter fact explains why the number of strict (pure) Nash equilibria is limited by $\min{(n_x,n_y)}$.

Figure \ref{fig:fd_g3x3} shows the flow graph of a three-strategy game having only one strict Nash equilibrium that can be found by starting the search for any strategy profile and allowing the players to choose unilaterally a better strategy from their own point of view. After some steps the walk in this directed graph ends in a Nash equilibrium represented by a node without outgoing edges. Evidently, a similar flow graph holds for games where the payoffs are modified by smaller components that are not capable of reversing the directions of arrows.

In the textbook of game theory the reader can find different methods for the determination of Nash equilibria. One of the standard methods is based on the iterated elimination of the strictly dominated strategies. In the notation of Sect.~\ref{sec:psp}, strategy $s_x$ of player $x$ is strictly dominated by strategy $s_x^{\prime}$ if $u_x(s_x,s_{-x})<u_x(s_x^{\prime},s_{-x})$ $\forall s_{-x}$. Such a dominance can be easily recognized in a flow graph. For example, Fig.~\ref{fig:fd_g3x3} represents a game where strategy 3 is dominated by strategy 2 as the arrows connecting the nodes of the second and third columns are pointed to left. Notice, that for this special case strategy 1 dominates the other strategies for both players.

As rational players do not play dominated strategies, these strategies can be eliminated. Once we have eliminated a dominated strategy it can happen that another strategy for one of the players becomes a dominated strategy. For the games illustrated in Fig.~\ref{fig:fd_g3x3} only the $(1,1)$ strategy profile survives after repeating this procedure.

The above method will not eliminate rows and columns including a strict Nash equilibrium. The iterated elimination of dominated strategies simplifies the game and shrinks the flow graph. In many cases the simplified model may be equivalent to a well-known game ({\it e.g.}, coordination game).

The number of Nash equilibria is a fundamental question in the theory of games. In the literature of graph theory \cite{bollobas_98, harary_66} the reader can find the discussion of finite directed acyclic graphs, {\it i.e.} finite directed graphs that have no directed loops. Such graphs always have at least one node without outgoing edges, otherwise a directed walk of arbitrary length could be constructed in the finite graph, as we could always leave any vertices of the loop, and by the finiteness of the graph the walk would contain a cycle, contradicting the assumption of acyclicity.

The latter statement specifies the Nash theorem for the case of potential games. Namely, at least one strict pure Nash equilibrium exists for the potential games. By reversing the direction of the edges, a similar argument shows, that a finite acyclic directed graph also has at least one node without incoming edges.  The corresponding node is represented by the strategy profile $(3,3)$ in Fig.~\ref{fig:fd_g3x3}.

Remarkably, the determination of flow graph for any matrix game can help us find the Nash equilibria as we only need to identify the strategy profiles without outgoing edge(s). In the rest of this work we discuss more complex flow graphs corresponding to multi-player games where many Nash equilibria can exist.

\subsection{General features of the potential}
\label{sec:generalfeatures}

The linear relationship between the individual payoff variation and potential variation implies general features. First we emphasize that a game as well as its potential is not influenced if all payoff elements (including $A_{ij}$, $B_{ij}$, and $V_{ij}$) are increased with a constant value because the player decisions depend only on the payoff differences.

If a bi-matrix game ${\bf G}=({\bf A},{\bf B}^T)$ has a potential ${\bf V}$ then the game ${\bf G}^{\prime}=(\alpha{\bf A},\alpha{\bf B}^T)$ also has a potential given as ${\bf V}^{\prime}= \alpha {\bf V}$. If $\alpha >0$ then the multiplication of the payoff elements can be interpreted as a choice of new unit. For negative $\alpha$, however, the game and its Nash equilibria can change drastically, whereas the existence of potential remains valid.

In addition, if we have two potential games, ${\bf G}=({\bf A},{\bf B}^T)$ and ${\bf G}^{\prime}=({\bf A}^{\prime},{\bf B}^{\prime T})$, with potentials $\bf V$ and ${\bf V}^{\prime}$ then the game obtained by the linear combination of payoffs ${\bf G}^{\star}=\alpha {\bf G} + \beta {\bf G}^{\prime} =(\alpha {\bf A}+ \beta {\bf A}^{\prime},{\alpha \bf B}^T+ \beta {\bf B}^{\prime T})$ is also a potential game with
${\bf V}^{\star}=\alpha {\bf V} + \beta {\bf V}^{\prime}$.

One can easily check that for symmetric matrix games (${\bf A}={\bf B}$) the potential is a symmetric matrix, {\it i.e.,} ${\bf V}={\bf V}^{T}$. This feature is the direct consequence of the fact that players exchange their payoff if they exchange their strategies. An additional symmetry occurs for the partnership games (${\bf A}={\bf B}$ and ${\bf A}={\bf A}^{T}$) when ${\bf V} = {\bf A}$. For such situations the players share their income equally, resembling the fraternal or egalitarian behavior \cite{monderer_geb96, hofbauer_98, sandholm_geb10}. In that case, the individual and common interests coincide and many intriguing events (e.g., social dilemmas) are dropped.

Within matrix games we can distinguish games with self- and cross-dependent payoffs \cite{szabo_pre14b}. For the cross-dependent payoffs the income of each player depends only on the co-player's strategy, that is the columns of the payoff matrices are composed of uniform values ($\gamma_j$ and $\delta_i$ for $j=1, \ldots , n_y$ and $i=1, \ldots , n_x$) as:
\begin{equation}
{\bf A}={\bf A}^{\rm (cr)}=\left(\matrix{ \gamma_1 & \cdots & \gamma_{n_y} \cr
                                  \vdots   & \ddots & \vdots \cr
                                  \gamma_1 & \cdots & \gamma_{n_y} \cr}\right),
                                  \,\,\,\,\,\,\,\,\,
{\bf B}={\bf B}^{\rm (cr)}=\left(\matrix{ \delta_1 & \cdots & \delta_{n_x} \cr
                                  \vdots   & \ddots & \vdots \cr
                                  \delta_1 & \cdots & \delta_{n_x} \cr}\right).
\label{eq:crossdependent}
\end{equation}
In these cases the unilateral strategy changes are not motivated by receiving a higher individual payoff and the corresponding potential is constant ({\it i.e.}, we can choose ${\bf V}^{\rm (cr)} = 0$). On the contrary, for the self-dependent payoffs the rows of the payoff matrices are constant, namely,
\begin{equation}
{\bf A}={\bf A}^{\rm (s)}=\left(\matrix{ \gamma_1 & \cdots & \gamma_1 \cr
                                  \vdots   & \ddots & \vdots \cr
                                  \gamma_{n_x} & \cdots & \gamma_{n_x} \cr}\right) , \,\,\,\,\,\,\,\,\,
{\bf B}={\bf B}^{\rm (s)}=\left(\matrix{ \delta_1 & \cdots & \delta_1 \cr
                                  \vdots   & \ddots & \vdots \cr
                                  \delta_{n_y} & \cdots & \delta_{n_y} \cr}\right) ,
\label{eq:selfdependent}
\end{equation}
and the components of the potential matrix obey the following form:
\begin{equation} \label{eq:Vijself}
V_{ij}^{\rm (s)} = \gamma_i + \delta_j,
\end{equation}
or in matrix notation
\begin{equation} \label{eq:Vmatrixself}
{\bf V}^{\rm (s)} = {\bf A} + {\bf B}^T.
\end{equation}

The linear relationship between the potential and the payoffs offers the possible use of decomposition when the matrix games are composed of elementary games reflecting basic symmetries as detailed in the next section.

\section{DECOMPOSITION OF TWO-PLAYER GAMES}
\label{sec:decomposition_i}

In the literature of game theory the concept of decomposition is used in different ways. For most of the cases the games with many payoff parameters are built from simpler games characterized by a significantly less number of parameters [see the works by \citet{szep_85}, \citet{kleinberg_mor86}, \citet{sandholm_geb10}, and \citet{candogan_mor11}]. This gives us a deeper insight into the general properties of interactions described by the tools of game theory. Relationship between payoffs, potential, flow graphs, and other types of interactions are discussed in the papers by \citet{candogan_mor11, candogan_geb13, hwang_sh_arx11} who used a different terminology. Now we follow the concept introduced recently in the papers by \citet{szabo_pre14b, szabo_pre15b}, which is based on the introduction of orthogonal elementary basis games representing proper features, {\it e.g.}, games with self- or cross-dependent payoffs. Our analysis will be focused on those decompositions that help us identify the components prohibiting the existence of potential. This way of decomposition is consistent with the stability analysis based on the systematic investigations of two-strategy sub-games \cite{adami_pre12, cui_pb_jtb14}.

\subsection{Decomposition of symmetric matrix games}
\label{sec:decomposition_smg}

Symmetric matrix games are defined by the $A_{ij}$ elements of the payoff matrices (${\bf B}={\bf A}$ and $n_x=n_y=n$). In the spirit of decomposition, the payoff matrix is a linear combination of four matrices:
\begin{equation}
{\bf A}=\left(\matrix{ A_{11} & A_{12} \cr
                                A_{21} & A_{22} \cr}\right)=
          A_{11}\left(\matrix{ 1 & 0 \cr
                                0 & 0 \cr}\right) +
          A_{12} \left(\matrix{ 0 & 1 \cr
                              0 & 0 \cr}\right) +
          A_{21}\left(\matrix{ 0 & 0 \cr
                              1 & 0 \cr}\right) +
          A_{22} \left(\matrix{ 0 & 0 \cr
                              0 & 1 \cr}\right)
 \label{eq:tradbasis}
 \end{equation}
with the suitable coefficients. For a linear arrangement of the  matrix components the above four matrices form a set of orthogonal "basis vectors" that are called basis matrices or basis games henceforth. We can choose, however, another set of orthogonal basis matrices representing fundamentally different games that involve the relevant symmetries. For example, a new set of orthogonal basis matrices can be defined as
\begin{equation}
{\bf f}(1)=\left(\matrix{ 1 & 1 \cr
                          1 & 1 \cr}\right) , \,\,\,\,
{\bf f}(2)= \left(\matrix{ 1 & -1 \cr
                           1 & -1 \cr}\right) , \,\,\,\,
{\bf f}(3)= \left(\matrix{ \phantom{-}1 & \phantom{-}1 \cr
                                     -1 & -1 \cr}\right) , \,\,\,\,
{\bf f}(4)= \left(\matrix{ \phantom{-}1 & -1 \cr
                             -1 & \phantom{-}1 \cr}\right).
\label{eq:newbasis2s}
\end{equation}
These basis games were used in many examples studied previously [see {\it e.g.} \cite{blume_l_geb93, sandholm_10}]. The orthogonality is defined by introducing the scalar product of two $n \times n$ matrices as
\begin{equation} \label{eq:scalarproduct}
{\bf A} \cdot {\bf B} = {\bf B} \cdot {\bf A} = \sum_{i,j=1}^n A_{ij} B_{ij},
\end{equation}
that is zero if ${\bf A}$ and ${\bf B}$ are orthogonal to each other. Notice that the basis games given by Eqs. (\ref{eq:tradbasis}) and (\ref{eq:newbasis2s}) form two sets of orthogonal basis matrices as ${\bf f}(m) \cdot {\bf f}(m^{\prime}) =0$ if $m \ne m^{\prime}$ ($m,m^{\prime}=1, \ldots, n^2$) and otherwise ${\bf f}(m) \cdot {\bf f}(m) ={\cal N}(m) > 0$.

Using the orthogonal basis matrices one can express any $n \times n$ payoff matrix as
\begin{equation}
{\bf A}=\sum_{m=1}^{n^2} \alpha(m) {\bf f}(m) \,,
\label{eq:nob2x2}
\end{equation}
where the coefficients $\alpha(m)$ are given by the expression
\begin{equation}
\alpha(m)= {1 \over {\cal N}(m)} {\bf A} \cdot {\bf f}(m) =
{1 \over {\cal N}(m)} \sum_{j,k=1}^{n} A_{jk} f_{jk}(m).
\label{eq:coef}
\end{equation}

The first basis matrix ${\bf f}(1)$ given by (\ref{eq:newbasis2s}) defines a game with constant payoffs. In fact this is the irrelevant component of the game that does not influence the decision of selfish players and $\alpha(1)$ [defined by Eq.~(\ref{eq:nob2x2})] is equal to the average value of $A_{ij}$. The linear combinations of the first and second components ({\it e.g.}, ${\bf A} ={\bf A}^{\rm (cr)} = \alpha(1){\bf f}(1) + \alpha(2){\bf f}(2)$) describe all games with cross-dependent payoffs. Similarly, the matrix ${\bf A} = {\bf A}^{\rm (s)} = \alpha(1){\bf f}(1) + \alpha(3){\bf f}(3)$ corresponds to a game with self-dependent payoffs.

In game theory ${\bf f}(4)$ represents the coordination game when players are enforced to choose the same strategy. At the same time, the game with payoff matrix ${\bf A} =-{\bf f}(4)$ is an anti-coordination game where the best results can be achieved by the players if they choose opposite strategies. Thus the sign of $\alpha(4)$ defines whether coordination or anti-coordination type interaction is built into the game and its strength is measured by $|\alpha(4)|$.

In the light of the above results, a symmetric two-strategy game is composed of three types of components representing the self-dependent and cross-dependent payoffs, and the coordination type interaction. Due to the general features discussed in Sec.~\ref{sec:generalfeatures} only two of these components contribute to the potential matrix that obeys the following form:
\begin{equation}
\label{eq:pot_2x2s}
{\bf V}= \alpha(3)[{\bf f}(3)+{\bf f}^{T}(3)] + \alpha(4) {\bf f}(4)
\end{equation}
omitting the arbitrary constant [proportional to ${\bf f}(1)$].

The mentioned general features are preserved when the number of strategies is increased, whereas a fundamentally new type of interaction emerges if $n>2$ (or even if ${\bf B} \ne {\bf A}$). This type of interaction represents the cyclic dominance that prevents the existence of potential.

For the illustration what happens for $n>2$, we briefly discuss now a possible decomposition of the symmetric $3 \times 3$ matrix games. A more detailed analysis is available in a recent paper by \citet{szabo_pre14b} who suggested the introduction of a set basis matrices representing the two-dimensional Fourier components of ${\bf A}$. Instead of it, now we use another notation based on the dyadic products \cite{szabo_pre15b}. In this notation the basis matrices ${\bf g}(m)$ ($m=1, \ldots , 9$) are expressed with the help of dyadic products ${\bf e}(k) \otimes {\bf e}(l)$ ($k,l=1,2,3$) of the following three three-dimensional orthogonal basis vectors
\begin{equation}
\label{eq:e123}
{\bf e}(1)= \left( \matrix{1 \cr 1 \cr 1 \cr}\right),\,\,\,\,\,\,
{\bf e}(2)= \left( \matrix{-1 \cr -1 \cr \phantom{-}2}\right),\,\,\,\,\,\,
{\bf e}(3)= \left( \matrix{\phantom{-}1 \cr -1 \cr \phantom{-}0 \cr}\right) .
\end{equation}
The relationship between the matrix label $m$ and the vector labels $k$ and $l$ is given by the following definitions:
\begin{eqnarray}
{\bf g}(1)&=&{\bf e}(1) \otimes {\bf e}(1)= \left(\matrix{ 1  &  1  &  1 \cr
                                               1  &  1  &  1 \cr
                                               1  &  1  &  1 \cr} \right),
                                               \label{eq:dyad3x3m1}  \\
{\bf g}(2)&=&{\bf e}(1) \otimes {\bf e}(2)= \left(\matrix{-1  & -1  &  2 \cr
                                               -1  & -1  &  2 \cr
                                               -1  & -1  &  2 \cr} \right),
                                               \label{eq:dyad3x3m2}  \\
{\bf g}(3)&=&{\bf e}(1) \otimes {\bf e}(3)= \left(\matrix{ 1  & -1  &  0 \cr
                                               1  & -1  &  0 \cr
                                               1  & -1  &  0 \cr} \right),
                                               \label{eq:dyad3x3m3}\\
{\bf g}(4)&=&{\bf e}(2) \otimes {\bf e}(1)= \left(\matrix{-1  & -1  & -1 \cr
                                               -1  & -1  & -1 \cr
                                                \phantom{-}2  &  \phantom{-}2  &  \phantom{-}2 \cr} \right),
                                                \label{eq:dyad3x3m4}  \\
{\bf g}(5)&=&{\bf e}(3) \otimes {\bf e}(1)= \left(\matrix{ \phantom{-}1  &  \phantom{-}1  &  \phantom{-}1 \cr
                                              -1  & -1  & -1 \cr
                                               \phantom{-}0  &  \phantom{-}0  &  \phantom{-}0 \cr} \right),
                                               \label{eq:dyad3x3m5}\\
{\bf g}(6)&=&{\bf e}(2) \otimes {\bf e}(2)= \left(\matrix{\phantom{-}1  & \phantom{-}1  & -2 \cr
                                                \phantom{-}1  & \phantom{-}1 & -2 \cr
                                               -2  & -2 &  \phantom{-}4 \cr} \right),
                                               \label{eq:dyad3x3m6}\\
{\bf g}(7)&=&{\bf e}(3) \otimes {\bf e}(3)= \left(\matrix{\phantom{-}1  &  -1  &  0 \cr
                                                  -1  &  \phantom{-}1  & 0 \cr
                                                 \phantom{-}0  & \phantom{-}0  &  0 \cr} \right),  \label{eq:dyad3x3m7} \\
{\bf g}(8)&=&{1 \over 2}\left[{\bf e}(2) \otimes {\bf e}(3)+{\bf e}(3) \otimes {\bf e}(2)\right] = \left(\matrix{     -1  & \phantom{-}0 & \phantom{-}1 \cr
                                               \phantom{-}0  &  \phantom{-}1 & -1 \cr
                                               \phantom{-}1  & -1 &  \phantom{-}0 \cr} \right),  \label{eq:dyad3x3m8}       \\
{\bf g}(9)&=&{1 \over 2}\left[{\bf e}(2) \otimes {\bf e}(3)-{\bf e}(3) \otimes {\bf e}(2)\right] = \left(\matrix{ \phantom{-}0  & \phantom{-}1 & -1 \cr
                                               -1  &  \phantom{-}0 & \phantom{-}1 \cr
                                               \phantom{-}1  & -1 &  \phantom{-}0 \cr} \right).
\label{eq:dyad3x3m9}
\end{eqnarray}

Accordingly, the $3 \times 3$ payoff matrix is expressed by the linear combinations of these ${\bf g}(m)$s as
\begin{equation}
{\bf A}=\sum_{m=1}^{9} \beta(m) {\bf g}(m) \,,
\label{eq:nob3x3}
\end{equation}
where the coefficients $\beta(m)$ are given by the expression
\begin{equation}
\beta(m)= {1 \over {\cal N}(m)} {\bf A} \cdot {\bf g}(m)
= {1 \over {\cal N}(m)} \sum_{j,k=1}^{3} A_{jk} g_{jk}(m),
\label{eq:coef3x3}
\end{equation}
and the normalization factors are defined as above, that is, ${\cal N}(m)= {\bf g}(m) \cdot {\bf g}(m)$ .

The present set of basis vectors reflects similar features and symmetries we found for the $2 \times 2$ payoff matrices. Namely, the first component ${\bf g}(1)$ defines a constant (or average) contribution to the payoffs and for later convenience its contribution will be denoted for $n > 2 $ as
\begin{equation}
{\bf A}^{\rm (av)}=\beta(1) {\bf g}(1) \,
\label{eq:A_av}
\end{equation}
where ${\bf g}(1)$ denotes the all-ones matrix and
\begin{equation}
\beta(1) = {1 \over n^2}\sum_{i,j=1}^{n} A_{ij} .
\label{eq:beta_av}
\end{equation}

The linear combination of the first three terms [${\bf A}^{\rm (cr)}={\beta(1)\bf g}(1)+{\beta(2)\bf g}(2)+{\beta(3)\bf g}(3)$] spans the whole set of symmetric $3 \times 3$ games with cross-dependent payoffs. On the other hand, the self-dependent payoff component of a game ${\bf A}$ can be given as ${\bf A}^{\rm (s)}={\beta(1)\bf g}(1)+{\beta(4)\bf g}(4)+{\beta(5)\bf g}(5)$ where values of $\beta(1)$, $\beta(4)$, and $\beta(5)$ are given by Eqs.~(\ref{eq:coef3x3}). Straightforward calculations yield that these portions of payoffs obey a simple form:
\begin{equation}\label{eq:self3x3}
{\bf A}^{\rm (s)}=\left(\matrix{ \gamma_1 & \gamma_1 & \gamma_1 \cr
                                 \gamma_2 & \gamma_2 & \gamma_2 \cr
                                 \gamma_3 & \gamma_3 & \gamma_3 \cr}\right) ,
\end{equation}
where $\gamma_i$ is equal to the average values of payoffs in the $i$th ($i=1,2$ and 3) row of ${\bf A}$, that is,
\begin{equation}\label{eq:self3x3g}
\gamma_i = {1 \over 3}\sum_{j=1}^3 A_{ij} \,.
\end{equation}
The values of $\gamma_i$ are equivalent to the average income of player $x$ when she chooses her $i$th strategy, while her co-player follows a mixed strategy by choosing the three options with equal probabilities. For games with more than one Nash equilibria the concept of risk dominance, suggested by \citet{harsanyi_88}, gives us an additional criterion to select one of them. The risk dominance dictates to choose those strategies that provide the highest $\gamma_i$ value. Accordingly, ${\bf A}^{\rm (s)}$ quantifies the risk dominance, too. Figure \ref{fig:fd_g3x3} illustrates the flow graph of ${\bf A}^{\rm (s)}$ for $\gamma_1> \gamma_2 > \gamma_3$. For games ${\bf A}^{\rm (s)}$ a potential exists with a potential matrix ${\bf V}^{\rm (s)}={\bf A}^{\rm (s)}+{\bf A}^{{\rm (s)}T}$ as it is detailed in Sect.~\ref{sec:generalfeatures}.

Evidently, one can perform a similar calculation for the evaluation of the component ${\bf A}^{\rm (cr)}= \beta (1){\bf g}(1)+\beta(2) {\bf g}(2) + \beta(3) {\bf g}(3)$ with cross-dependent payoffs that can be expressed as $A^{\rm (cr)}_{ij}=\sum_{i=1}^3 A_{ij}/3$. These terms do not contribute to the potential matrix (${\bf V}^{(cr)}=0$).

The component ${\bf g}(7)$ [see (\ref{eq:dyad3x3m7})] corresponds to an elementary coordination type game where the players achieve the best if both choose either the first or the second strategy simultaneously. Due to its relevance this payoff matrix is denoted henceforth as ${\bf d}(1,2)$. Additionally, we can introduce further elementary coordination games ${\bf d}(p,q)$ defined as
\begin{equation}
d_{ij}(p,q)=\cases{\phantom{-}1,  &if $i=j=p$, \cr
                  \phantom{-}1,  &if $i=j=q$, \cr
                  -1, &if $i=p$ and $j=q$, \cr
                  -1, &if $i=q$ and $j=p$, \cr
                  \phantom{-}0,  &otherwise, }
\label{eq:isingsubgames}
\end{equation}
even for $n>3$ when $1 \le p < q \le n$ that describes similar relationship between the strategies $p$ and $q$ (for simplicity the $n$-dependence is not denoted). Notice that ${\bf d}(1,3)$ can be obtained from ${\bf d}(1,2)$ by exchanging the second and third rows and columns simultaneously. Although the matrices ${\bf d}(1,2)$, ${\bf d}(1,3)$, and ${\bf d}(2,3)$ are not orthogonal to each other, these basis games span the three-dimensional subspaces of the coordination type interactions defined as ${\bf A}^{\rm (coord)} = \beta (6){\bf g}(6)+\beta(7) {\bf g}(7) + \beta(8){\bf g}(8)$. Applying the expressions (\ref{eq:nob3x3}) and (\ref{eq:coef3x3}) one can easily check that, for example,
\begin{equation}
{\bf d}(1,3)=
\left(\matrix{ \phantom{-}1  &  \phantom{-}0  &            -1 \cr
               \phantom{-}0  &  \phantom{-}0  &  \phantom{-}0 \cr
                         -1  &  \phantom{-}0  &  \phantom{-}1 \cr} \right) =
 {1 \over 4}\left[{\bf g}(6)+{\bf g}(7)- 2{\bf g}(8)\right].
\label{eq:d13}
\end{equation}

The anti-coordination type interactions between any strategy pairs ($p < q$) are also located along the half-lines ($\beta {\bf d}(p,q)$ with $\beta <0$) in the space of ${\bf A}^{\rm (coord)}$. For example, the game with a payoff matrix ${\bf A}^{\rm (coord)} = - {\bf g}(7)$ corresponds to a situation where the players have two equivalent Nash equilibria, namely the strategy pairs $(1,2)$ and $(2,1)$.

The coordination type games include curious cases when the game has three equivalent Nash equilibria. For instance, the strategy pairs $(1,1)$, $(2,2)$, and $(3,3)$ are equivalent Nash equilibria in a game defined as ${\bf A}= {\bf d}(1,2)+{\bf d}(1,3)+{\bf d}(2,3)$. If one exchanges the first and second rows and columns simultaneously in the latter payoff matrix, then the resultant game is also a coordination type game with three equivalent Nash equilibria: $(1,2)$, $(2,1)$, and $(3,3)$.

Notice that ${\bf A}^{\rm (coord) T}={\bf A}^{\rm (coord)}$ indicating the coincidence of individual and common interests within this type of games and the existence of a potential matrix that is identical to the symmetric payoff matrix [${\bf V}^{\rm (coord)}={\bf A}^{\rm (coord)}$].

The last component (${\bf g}(9) = {\bf A}^{\rm (rps)}$) represents the well investigated rock-paper-scissors game that has only one mixed Nash equilibrium when the three strategies are selected with the same probability. The rock-paper-scissors games exemplify those systems from ecology to non-equilibrium physics, where three states (strategies or species) cyclically dominate each other \cite{may_siam75, hofbauer_88, tainaka_lncs01, frey_pa10, szolnoki_jrsi14c}.

\begin{figure}[ht]
\centerline{\epsfig{file=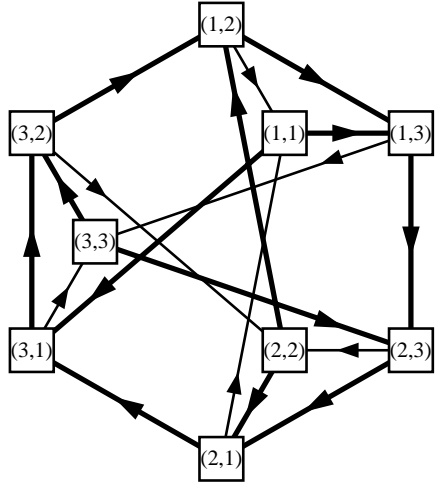,width=6cm}}
\caption{Flow graph of the rock-scissors-paper game. Thick edges represent the best responses.}
\label{fig:rspfg}
\end{figure}

The rock-paper-scissors basis game (${\bf g}(9)$) has no potential because of the existence of directed loops in the flow graph shown in Fig.~\ref{fig:rspfg}. For the illustration of the three-fold symmetries within this flow graph, here the strategy profiles (nodes) are rearranged. This flow graph illustrates clearly the presence of a directed six-edge loop along the hexagonal periphery where the thick directed edges represent the "best response".
Besides it, this flow graph contains six directed four-edge loops [ see {\it e.g.}, the loop $(1,2) \to (1,3) \to (2,3) \to (2,2) \to (1,2)$].

As potential exists for all the linear combinations of the first eight basis games, therefore the absence of the rock-paper-scissors component ($\beta(9)=0$ ) can be interpreted as a necessary condition for the existence of potential. The latter condition, more precisely the mathematical expression of ${\bf A} \cdot {\bf A}^{\rm (rps)} = 0$, is equivalent to (\ref{eq:loop2x2w3s}) we derived previously with the application of the Kirchhoff laws.

\begin{figure}[ht]
\centerline{\epsfig{file=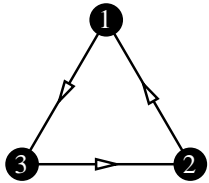,width=3.1cm}}
\caption{The adjacency matrix of this three-node directed graph is equivalent to payoff matrix of the rock-paper-scissors game. The directed three-edge loop refers to cyclic dominance among the three strategies denoted by labeled black bullets.}
\label{fig:dirgw3s}
\end{figure}

The payoff matrix ${\bf g}(9) = {\bf A}^{\rm (rps)}$ can be considered as the adjacency matrix of a simple directed graph (shown in Fig.~\ref{fig:dirgw3s}) that illustrates graphically the cyclic dominance among the three strategies represented by the nodes of this dominance graph.
In order to avoid confusion with the dynamical and flow graphs here we used a third way for the graphical illustration of a simple directed graph. The conventions of the arrow directions in the flow and dominance graphs become identical if the adjacency matrix elements of a directed graph are defined as $A_{ij}=-A_{ji}=1$ if there is a directed link from the strategy (node) $j$ to $i$ and otherwise $A_{ij}=0$. In words, the dominance graph in Fig.~\ref{fig:dirgw3s} corresponds to a game where strategy 1 dominates 2, 2 dominates 3, and 3 dominates 1. This direction convention ensures that the arrows will show the direction of probability current in evolutionary systems discussed later.

It is noteworthy, that the cyclic dominance graph is analogous to the cyclic food web used in ecological models to describe the predation-prey relationship. The adjacency matrix of the tree-like dominance graph (trophic food web) can be related to some anti-symmetric components in the sub-space of games with the self- and cross-dependent payoffs.

Evidently, one can choose another set of basis matrices with features suitable for the problems to be studied. For example we can separate the symmetric and anti-symmetric parts of the self- and cross-dependent payoffs which allow the analysis of the zero-sum components or the strength of social dilemmas.

Most of the general properties of the symmetric three-strategy games are preserved for $n>3$. The decomposition of a symmetric matrix game into elementary components becomes impressive for $n=2^k$ ($k$ is an integer) when the columns of the Walsh-Hadamard matrices \cite{ahmed_75} are used as basis vectors in the dyadic decomposition \cite{szabo_pre15b}. For this choice both the basis vectors and the derived basis matrices are composed of only $+1$s and $-1$s that simplify the calculations [for an example see Eq.~(\ref{eq:newbasis2s})]. These calculations have confirmed the relevance of the mentioned four types of interactions for $n=4$ \cite{szabo_pre15b}. Similar results are concluded by \citet{candogan_mor11, hwang_sh_arx11} who studied the decomposition of $n$-strategy games without introducing a definite set of basis games.

According to the analyses mentioned, the games with self-dependent payoffs are spanned by $n$ dyadic products, as ${\bf e}(1) \otimes {\bf e}(j)$ for $j=1, \ldots, n$, of an orthogonal set of basis vectors (${\bf e}(1), \ldots , {\bf e}(n)$) when the first one is composed of only $+1$s, that is, $e_i(1)=1$ for $i=1, \ldots, n$. In this notation the basis games with cross-dependent payoffs are given as ${\bf e}(j) \otimes {\bf e}(1)$. The all-ones matrix (${\bf g}(1)={\bf e}(1) \otimes {\bf e}(1)$) belongs to both types and plays the role of irrelevant term. Its coefficient $\beta(1)$ quantifies the average payoff. The direct pair interactions are missing within their unified parameter space spanned by the $(2n-1)$ orthogonal basis matrices that may even be divided into the sum of symmetric (${\bf e}(1) \otimes {\bf e}(j)+{\bf e}(j) \otimes {\bf e}(1)$) and anti-symmetric (${\bf e}(1) \otimes {\bf e}(j)-{\bf e}(j) \otimes {\bf e}(1)$) basis matrices. Thus the unified parameter space of ${\bf A}^{\rm (cr)}$ and ${\bf A}^{\rm (s)}$ can also be spanned by the mentioned $n$ symmetric and $(n-1)$ anti-symmetric basis games.

The rest of the symmetric parameter space of games is equivalent to the linear combinations of the additional symmetric basis matrices defined by the dyadic products as ${\bf e}(k) \otimes {\bf e}(j)+{\bf e}(j) \otimes {\bf e}(k)$ if $1 < k \le j \le n$. \citet{hwang_sh_arx11} have shown that this set of games (denoted as ${\bf A}^{\rm (coord)}$) are also spanned by the $n(n-1)/2$ ${\bf d}(p,q)$ matrices defined by Eq.~(\ref{eq:isingsubgames}) (for $1 \le p < q \le n$). Notice that all the mentioned basis matrices of ${\bf A}^{\rm (coord)}$ are orthogonal to both ${\bf A}^{\rm (s)}$ and ${\bf A}^{\rm (cr)}$ because the sums of the matrix elements are zero in the rows and columns separately. The number of these orthogonal basis games, as well as the number of the independent ${\bf d}(p,q)$ matrices, is $n(n-1)/2$.

The rest of the anti-symmetric parameter space of the payoff matrix is spanned by the orthogonal basis matrices ${\bf e}(k) \otimes {\bf e}(j)-{\bf e}(j) \otimes {\bf e}(k)$ where $1 < k < j \le n$. The number of these orthogonal basis games is $(n-1)(n-2)/2$ and their linear combinations are also denoted by ${\bf A}^{(cycl)}$ because of the presence of cyclic dominance.

\begin{figure}[ht]
\centerline{\epsfig{file=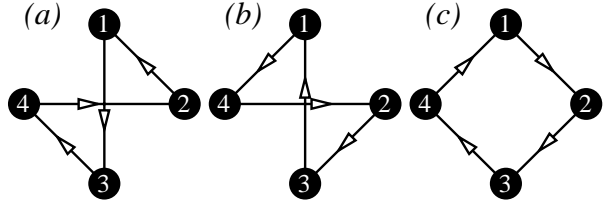,width=8cm}}
\caption{The adjacency matrices of these four-edge directed graphs are orthogonal to each other and represent the three cyclic basis games for $n=4$.}
\label{fig:dirgw4s}
\end{figure}

Calculations \cite{szabo_pre15b} for $n=4$ have indicated that ${\bf A}^{\rm (cycl)}$ is spanned by three orthogonal basis games defined by the adjacency matrices of the four-edge directed graphs plotted in Fig.~\ref{fig:dirgw4s}.
Additionally, ${\bf A}^{\rm (cycl)}$ contains games where the cyclic dominance is limited to three strategies as in the rock-paper-scissors game. For example, the sum of these three cyclic orthogonal basis games is equivalent to a rock-paper-scissors type sub-game (with strategies 2, 3, and 4). The knowledge of these three cyclic basis games simplifies the identification of potential games, because it is enough to check whether the scalar products of ${\bf A}$ by the three cyclic basis games become zero simultaneously, or not. Potential exists if all these scalar products are zero.

\begin{figure}[ht]
\centerline{\epsfig{file=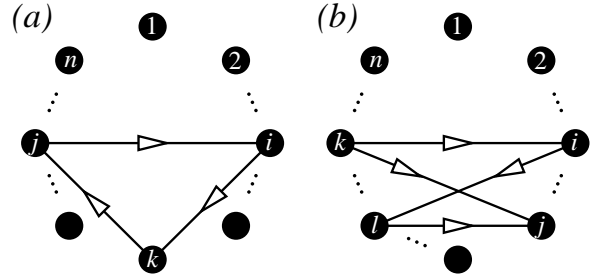,width=7.8cm}}
\caption{Directed graphs with $n$ nodes and a single three- (a) or four-edge (b) directed loop.}
\label{fig:dirgwns}
\end{figure}

For $n>4$, however, determining the existence of potential becomes difficult because of the large number of the possible four-edge and three-edge loops illustrated in Fig.~\ref{fig:dirgwns}.
It is emphasized, that the orthogonality between ${\bf A}$ and the adjacency matrices of these types of the directed graphs reproduces the same conditions for the existence of potential that we have derived in Sec.~\ref{sec:2pg}. The application of the Kirchhoff laws gives us a method to select $(n-1)(n-2)/2$ independent loops to be checked.

In sum, the payoff matrix of a two-player symmetric game can be decomposed into the sum of four classes of payoff matrices representing the self-dependent, the cross-dependent, the coordination, and the cyclic dominance type interactions. More precisely,
\begin{equation}\label{eq:classes}
{\bf A}= {\bf A}^{\rm (s)} + {\bf A}^{\rm (cr)} - {\bf A}^{\rm (av)}
+ {\bf A}^{\rm (coord)} + {\bf A}^{\rm (cycl)}\,,
\end{equation}
that takes into consideration that ${\bf A}^{\rm (av)}$ is involved in ${\bf A}^{\rm (s)}$ and ${\bf A}^{\rm (cr)}$. Potential exists if ${\bf A}^{\rm (cycl)}=0$ and the potential matrix can be given as
\begin{equation}\label{eq:potmatrix_ns}
{\bf V}= {\bf A}^{\rm (s)} + {\bf A}^{{\rm (s)}T} + {\bf A}^{\rm (coord)}.
\end{equation}

\subsection{Decomposition of two-strategy bi-matrix games}
\label{sec:decomposition_bmg}

The decomposition of the bi-matrix games requires the straightforward extension of the scalar product of two bi-matrix games analogously to the traditional scalar product of two vectors. As the bi-matrix game ${\bf G}$ is described by the elements of two matrices (${\bf A}$ and ${\bf B}$), therefore the scalar product of the games ${\bf G}$ and ${\bf G}^{\prime}$ (determined by (${\bf A}^{\prime}$ and ${\bf B}^{\prime}$)) is defined as
\begin{equation}\label{eq:GxG}
{\bf G} \cdot {\bf G}^{\prime} =  \sum_{i,j=1}^{n} [A_{ij}A_{ij}^{\prime} + B_{ij}B_{ij}^{\prime}].
\end{equation}
The decomposition of the two-strategy games becomes impressive if we use the following orthogonal basis games:
\begin{eqnarray}
{\bf f}^{\prime}(1)&=&\left(\matrix{(\phantom{-}1,\phantom{-}1) & (\phantom{-}1,\phantom{-}1) \cr
                                    (\phantom{-}1,\phantom{-}1) & (\phantom{-}1,\phantom{-}1) \cr}\right) \\
{\bf f}^{\prime}(2)&=&\left(\matrix{(\phantom{-}1,\phantom{-}1) & (          -1,\phantom{-}1) \cr
                                    (\phantom{-}1,          -1) & (          -1,          -1) \cr}\right) \\
{\bf f}^{\prime}(3)&=&\left(\matrix{(\phantom{-}1,\phantom{-}1) & (\phantom{-}1,          -1) \cr
                                    (          -1,\phantom{-}1) & (          -1,          -1) \cr}\right) \\
{\bf f}^{\prime}(4)&=&\left(\matrix{(\phantom{-}1,\phantom{-}1) & (          -1,          -1) \cr
                                    (          -1,          -1) & (\phantom{-}1,\phantom{-}1) \cr}\right) \\
{\bf f}^{\prime}(5)&=&\left(\matrix{(\phantom{-}1,          -1) & (\phantom{-}1,          -1) \cr
                                    (\phantom{-}1,          -1) & (\phantom{-}1,          -1) \cr}\right) \\
{\bf f}^{\prime}(6)&=&\left(\matrix{(\phantom{-}1,          -1) & (          -1,          -1) \cr
                                    (\phantom{-}1,\phantom{-}1) & (          -1,\phantom{-}1) \cr}\right) \\
{\bf f}^{\prime}(7)&=&\left(\matrix{(\phantom{-}1,          -1) & (\phantom{-}1,\phantom{-}1) \cr
                                    (          -1,          -1) & (          -1,\phantom{-}1) \cr}\right) \\
{\bf f}^{\prime}(8)&=&\left(\matrix{(\phantom{-}1,          -1) & (          -1,\phantom{-}1) \cr
                                    (          -1,\phantom{-}1) & (\phantom{-}1,          -1) \cr}\right)
\label{eq:newbasis2a}
\end{eqnarray}
that are composed of $+1$s and $-1$s. Using these orthogonal basis games, all the two-strategy bi-matrix games can be given as
\begin{equation}
{\bf G}=\sum_{m=1}^{8} \alpha^{\prime}(m){\bf f}^{\prime}(m)
\label{eq:bmgdecomp}
\end{equation}
where the coefficients $\alpha^{\prime}(n)$ are expressed as before, {\it i.e.},
\begin{equation}
\alpha^{\prime}(m)={1 \over 8} {\bf G} \cdot {\bf f}^{\prime}(m).
\label{eq:coef_bmg}
\end{equation}

Notice that the first four basis games span the parameter space of the symmetric games when ${\bf B}={\bf A}$, that have been discussed previously. In other words, ${\bf f}^{\prime}(1), \ldots , {\bf f}^{\prime}(4)$ denote the bi-matrix version of the symmetric two-strategy orthogonal basis games given by Eqs. (\ref{eq:newbasis2s}).

The linear combinations of the additional second four basis games describe the anti-symmetric games where ${\bf B}=-{\bf A}$. The resulting four basis games [${\bf f}^{\prime}(5), \ldots , {\bf f}^{\prime}(8)$] are obtained from the first four ones by reversing the sign of payoffs received by the second player. All these basis games represent some properties. For example, the coefficient of ${\bf f}^{\prime}(5)$ quantifies the difference in the average payoffs between the players $x$ and $y$.

The games with self-dependent payoffs are given as
\begin{equation}
{\bf G}^{\rm (s)} =\alpha^{\prime}(1) {\bf f}^{\prime}(1) + \alpha^{\prime}(3) {\bf f}^{\prime}(3)+ \alpha^{\prime}(5) {\bf f}^{\prime}(5)+\alpha^{\prime}(7) {\bf f}^{\prime}(7),
\label{eq:self_bmg}
\end{equation}
and the resultant payoff matrices ${\bf A}^{\rm (s)}$ and ${\bf B}^{\rm (s)}$ are composed of uniform rows with payoff parameters representing averages values as for the symmetric games ({\it e.g.}, $A_{ij}^{\rm (s)}=(A_{i1}+A_{i2})/2$ and $B_{ij}^{\rm (s)}=(B_{i1}+B_{i2})/2$).

Similar expressions define the components with cross-dependent payoffs:
\begin{equation}
{\bf G}^{\rm (cr)} =\alpha^{\prime}(1) {\bf f}^{\prime}(1) + \alpha^{\prime}(2) {\bf f}^{\prime}(2)+ \alpha^{\prime}(5) {\bf f}^{\prime}(5)+\alpha^{\prime}(6) {\bf f}^{\prime}(6).
\label{eq:cross_bmg}
\end{equation}

The basis game ${\bf f}^{\prime}(4)$ defines the symmetric coordination type interaction with payoffs ${\bf A}^{\rm (coord)}={\bf B}^{\rm (coord)}=\alpha^{\prime}(4){\bf f}(4)$ where ${\bf f}(4)$ is defined by Eq.~(\ref{eq:newbasis2s}).

The bi-matrix game ${\bf f}^{\prime}(8)$ is equivalent to the matching pennies game, which is a well-known zero-sum game where the first player wins a payoff unit from the co-player if they both choose the same strategy. For opposite choices the second player wins 1 from the first player. This game has a mixed Nash equilibrium, where the players choose their strategy at random.

\begin{figure}[ht]
\centerline{\epsfig{file=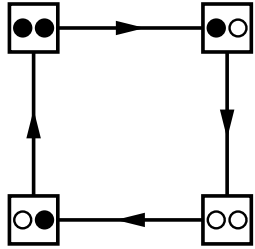,width=3.5cm}}
\caption{For the matching pennies game one of the players always wishes to reverse her strategy in a way maintaining cyclic strategy modifications.}
\label{fig:matpen}
\end{figure}

The flow graph of the matching pennies game (see Fig.~\ref{fig:matpen}) illustrates the appearance of a directed loop because for each strategy profile the unsatisfied player can increase her own payoff by 2 by reversing her strategy.
As a result, if in an iterated game the randomly selected player is allowed to modify her own strategy unilaterally then the actual strategy profile will evolve cyclically. Such situations were reported previously by \citet{van_valen_et73, van_valen_et80} who suggested the concept of Red Queen mechanism to explain biological evolution via a constant arm race between co-evolving species. The interactions between buyers and sellers \cite{friedman_d_e91}, property owners and criminals \cite{cressman_cje98}, or conformists and rebels \cite{cao_zg_tcs14} exhibit similar features. For very recent investigations of the Red Queen mechanism at the level of population dynamics we suggest reading the papers by \citet{sardanyes_14} and \citet{juul_pre13b}. The cyclic variation in the strategy profiles has been observed by \citet{xu_b_epjb14} in human experiments.

Evidently, potential does not exist for the matching pennies game and also for those two-strategy bi-matrix games that include this component. It is needless to emphasize that the bi-matrix game ${\bf G}= {\bf G}^{\rm (pot)}$ is a potential game, if $\alpha^{\prime}(8)={\bf G} \cdot {\bf g}(8)=0$, and the latter existence criterion for the potential is equivalent mathematically to Eq.~(\ref{eq:loop2x2sgbmg}) for the particular case when $i=k=1$ and $j=l=2$.

For the rest of (seven) basis games the potential of a non-symmetric bi-matrix game can be easily determined. According to this calculation, the potential matrix ${\bf V}^{\rm (bmg)}$ can be built up only from three components as
\begin{equation}
{\bf V}^{\rm (bmg)} = \alpha^{\prime}(3)[{\bf f}(3)+{\bf f}^{T}(3)] + \alpha^{\prime}(4) {\bf f}(4) + \alpha^{\prime}(7)[{\bf f}(3)-{\bf f}^{T}(3)].
\label{eq:pot_2x2bmg}
\end{equation}

It is worth mentioning that the game ${\bf G}={\bf G}^{\rm (pot)}+\alpha^{\prime} (8) {\bf f}^{\prime}(8)$ can even be an ordinal potential game with the potential of ${\bf G}^{\rm (pot)}$ [given by Eqs.~(\ref{eq:pot_2x2bmg})] if $|\alpha^{\prime}(8)|$ does not exceed a threshold value dependent on the payoff differences in ${\bf G}^{\rm (pot)}$, more precisely, if the contributions of the matching pennies game cannot reverse the arrow directions in the flow graph that ensures the existence of at least one pure Nash equilibrium. Conversely, if the cyclic component dominates the system behavior, then the game has only a mixed Nash equilibrium until ${\bf G}^{\rm (pot)}$ is weak enough to reverse the flow direction along at least one edge in the flow graph.

\subsection{Properties of two-player two-strategy games}
\label{sec:proppot}

In the previous sections we have shown a general method to evaluate the potential matrix in the absence of cyclic (or matching pennies) components. This method is based on the concept of decomposition and exploited the symmetries simplifying the calculations. Now we introduce another method and notation to discuss the general properties of the two-player two-strategy games.

For a deeper discussion of the symmetric two-strategy games (${\bf A}= {\bf B}$) we use the traditional notation introduced for the investigation of social dilemmas \cite{macy_pnas02, santos_pnas06, sigmund_10} where the strategies are called defection and cooperation (in short $s_{x1}, s_{y1}=D$ and $s_{x2}, s_{y2}=C$). Within this terminology both players receive $P$ (punishment) or $R$ (reward) for mutual defection or cooperation meanwhile the defector receives $T$ (temptation) and her cooperative co-player gets $S$ (sucker's payoff) if they choose opposite strategies. For the case of prisoner's dilemma ($S<P<R<T$) both selfish players are enforced to choose defection (this is the pure Nash equilibrium) because the cooperation is a dominated strategy, that is, the players receive higher payoff when choosing $D$ irrespectively of the co-players decision. The curiosity of the prisoner's dilemma game is that the players' rational (selfish) choices yield a suboptimal outcome.

There are two other (weaker) social dilemmas, namely the hawk-dove \cite{maynard_82} (called also snowdrift \cite{hauert_n04} or chicken \cite{morris_94}) game (with a payoff rank of $P<S<R<T$) and the stag hunt game ($S<P<T<R$). Most of the previous analyses are performed when setting $P=0$ and $R=1$ without loss of generality.

\begin{figure}[ht]
\centerline{\epsfig{file=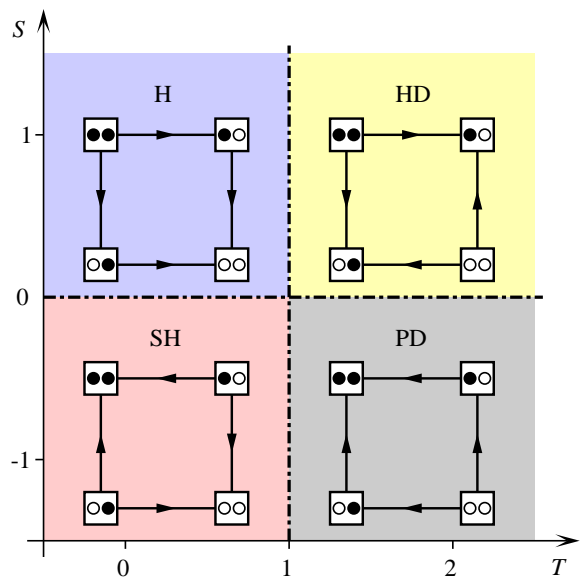,width=7.5cm}}
\caption{(Color online) Four types of symmetric two-strategy games can be distinguished as a function of $T$ and $S$ for the notation of social dilemmas when the black and white bullets represent $D$ and $C$ strategies. The four segments are characteristic to games abbreviated as:  H = harmony; HD = hawk-dove; SH = stag hunt; and PD = prisoner's dilemma. The pure Nash equilibria are represented by nodes without outgoing edges in the flow graphs.}
\label{fig:socdil_map}
\end{figure}

In the terminology of social dilemmas the bi-matrix of the symmetric two-strategy game is usually given as
\begin{equation}\label{eq:socdilG}
{\bf G}={\bf G}^{\rm (sd)}=\left(\matrix{ (0,0) & (T,S) \cr
                 (S,T) &(1,1) \cr} \right) .
\end{equation}
For these payoff parameters the three social dilemmas are positioned on the $S-T$ plane as illustrated in Fig.~\ref{fig:socdil_map} where four regions are distinguished.

Within each region the two-person game can be characterized by the same flow graph denoted. It is emphasized that these notations of strategies and payoffs are related to the case of the prisoner's dilemma game. For other types of games one can find more suitable names for the strategies and payoffs, {\it e.g.}, for the hawk-dove game the players are to choose between the aggressive and peaceful (conflict avoiding) behaviors and the resultant payoff can be well described by using the terms of equal share of benefit, exploitation, and cost of injury. Anyway, the hawk-dove game has an additional mixed Nash equilibrium when the players choose one of their two strategies with a probability dependent on the payoffs.

The absence of additional types of flow graphs is related to the symmetries that prescribe equivalent payoff variations for the active player deviating from the $(D,D)$ or $(C,C)$ strategy profiles. Due to these symmetries all the symmetric two-strategy games are potential games and the potential matrix is formally given by Eq.~(\ref{eq:pot_2x2s}) with the decomposition of the payoff matrix
\begin{equation}\label{eq:socdilApayoff}
{\bf A}={\bf A}^{\rm (sd)}=\left(\matrix{ 0 & T \cr
                       S & 1 \cr} \right)
\end{equation}
into the sum of four components [as defined by Eqs.(\ref{eq:nob2x2}) and (\ref{eq:newbasis2s})] with coefficients
\begin{equation}
\alpha (1)= {T+S+1 \over 4}, \,\,\,\,\,
\alpha (2)={S-T-1 \over 4},  \,\,\,\,\,
\alpha (3)= {T-S-1 \over 4}, \,\,\,\,\,
\alpha (4)={1-T-S \over 4}. \nonumber
\label{eq:socdilAcoef}
\end{equation}
According to Eq. (\ref{eq:pot_2x2s}) the potential obeys the following form:
\begin{equation}\label{eq:socdilV}
{\bf V}^{\rm (sd)}=\alpha (4) \left(\matrix{ \phantom{-}1 & -1 \cr
                                        -1 & \phantom{-}1 \cr} \right) +
\alpha (3) \left(\matrix{ 2 & \phantom{-}0 \cr
                          0 & -2 \cr} \right) .
\end{equation}
This expression will be used in Sec. \ref{sec:mtoI} to illuminate the relationship between the Ising models and the multi-agent evolutionary games, if the interaction is defined by symmetric two-strategy games. It is emphasized, however, that the above potential matrix can be expressed by an even simpler expression,
\begin{equation}\label{eq:socdilVsimple}
{\bf V}^{\rm (sd)}=\left(\matrix{ 0 & S \cr
                       S & 1-T+S \cr} \right)\,,
\end{equation}
that is obtained from (\ref{eq:socdilV}) by adding a suitable constant to $V_{ij}$s. The second version of the potential can be evaluated by using a simple algorithm. First we choose $V_{11}=A_{11}$, subsequently $V_{12}=V_{21}=A_{21}$, and finally $V_{22}=V_{12}+A_{22}-A_{12}$ in the spirit of Eqs.~(\ref{eq:potential}). This method can be adopted for larger number of strategies if the existence of potential is already justified. The resultant potential matrices are convenient when deriving phase diagrams for some types of multi-agent evolutionary games.

It is worth mentioning that the values of the potential are equal for the two pure Nash equilibria [$(C,D)$ and $(D,C)$] within the region of the hawk-dove game (see Fig.~\ref{fig:socdil_map}). On the contrary, the potential values differ between the $(D,D)$ and $(C,C)$ Nash equilibria within the region of stag hunt games. To be more quantitative, $(C,C)$ is the preferred Nash equilibrium if $S>(T-1)$ when $T<1$ and the strategy pair $(D,D)$ is preferred if $S < \min{[(T-1),0]}$. The location of the preferred Nash equilibria is illustrated graphically in the next section (see Fig.~\ref{fig:pne_st}) where it is contrasted with the case of collaborating players.

Within the region of the hawk-dove (or anti-coordination) games the equivalence between $(C,D)$ and $(D,C)$ Nash equilibria is evidently broken if ${\bf A} \ne {\bf B}$. For these bi-matrix games the potential (\ref{eq:pot_2x2bmg}) obeys the form:
\begin{equation}
{\bf V}^{\rm (bmg)}=\alpha^{\prime} (4) \left(\matrix{ \phantom{-}1 & -1 \cr
                                        -1 & \phantom{-}1 \cr} \right) +
\alpha^{\prime} (3) \left(\matrix{ 2 & \phantom{-}0 \cr
                          0 & -2 \cr} \right) +
\alpha^{\prime} (7) \left(\matrix{ \phantom{-}0 & \phantom{-}2 \cr
                          -2 & \phantom{-}0 \cr} \right),
\label{eq:pot_bmg_decomp}
\end{equation}
that we use later when drawing a parallel between the multi-agent evolutionary games and Ising models. Now we emphasize that the first components in Eqs.~(\ref{eq:socdilV}) and (\ref{eq:pot_bmg_decomp}) measure the strength of coordination type interaction, the second ones quantify the potential difference between the $(1,1)$ and $(2,2)$ strategy pairs whereas the third term distinguishes the potential value between the states $(1,2)$ and $(2,1)$.

\subsection{Fraternal collaboration versus individualism}
\label{sec:coll_vs_individ}

In traditional game theory \cite{neumann_44} the formation of coalition is permitted for the so-called cooperative games. Some of the essential elements of the games are dropped if the players are allowed to agree in how they wish to increase the sum of their income for two-player potential games. On the other hand, such situations can occur frequently in our everyday life. Thus we analyze now what happens if fraternal players try to maximize their total income shared equally. The results can serve as a reference when comparing actions motivated by individual or common interest. For this goal we introduce a reevaluated payoff matrix, ${\bf A}^{\rm (fr)}=({\bf A}+{\bf A}^{T})/2$, defining the individual payoffs for symmetric games with $n$ strategies. For these games the individual and common interests coincide and the preferred Nash equilibrium may differ from those proposed for a potential game ${\bf A}$.

\begin{figure}[ht]
\centerline{\epsfig{file=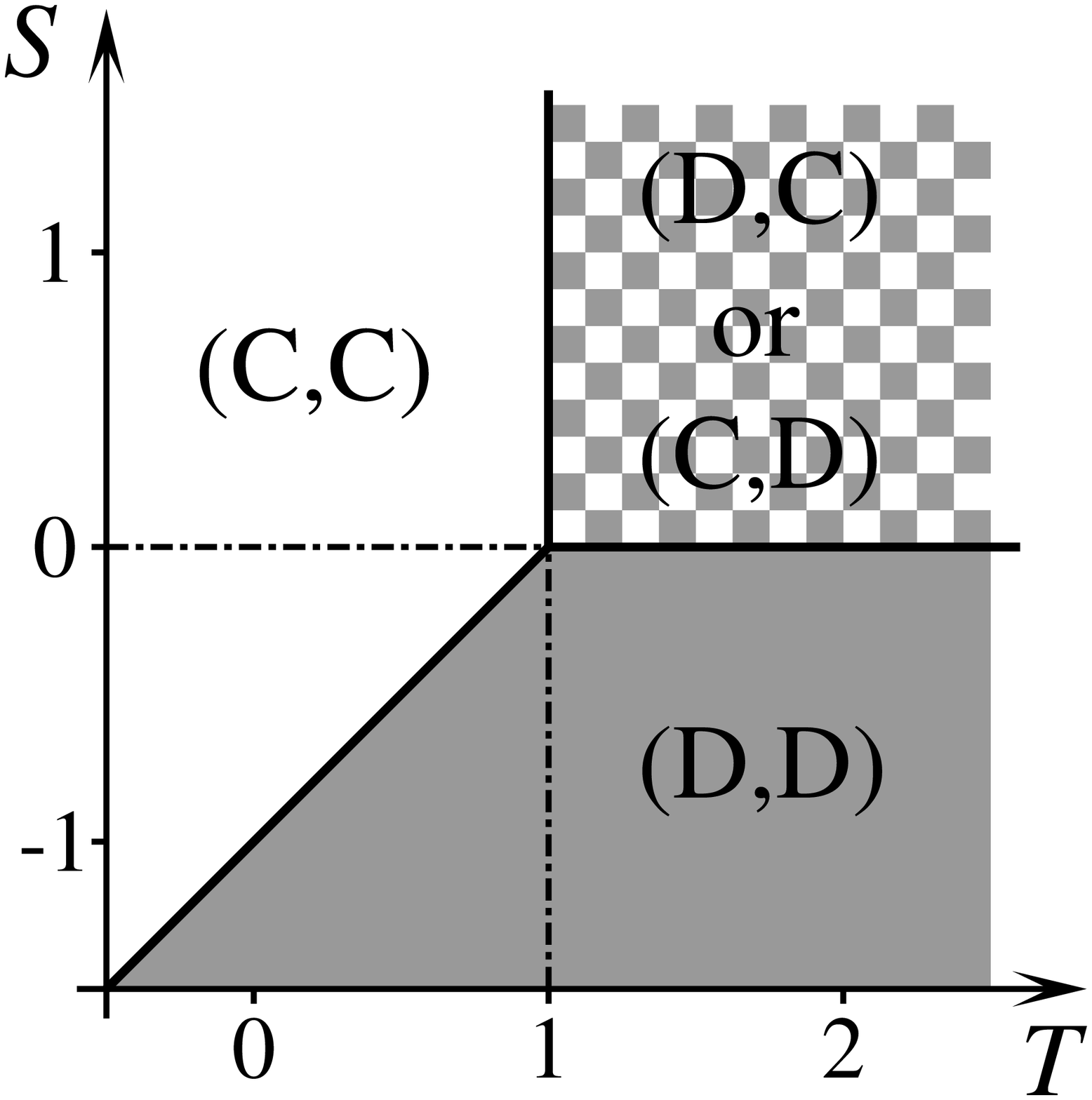,width=4cm} \epsfig{file=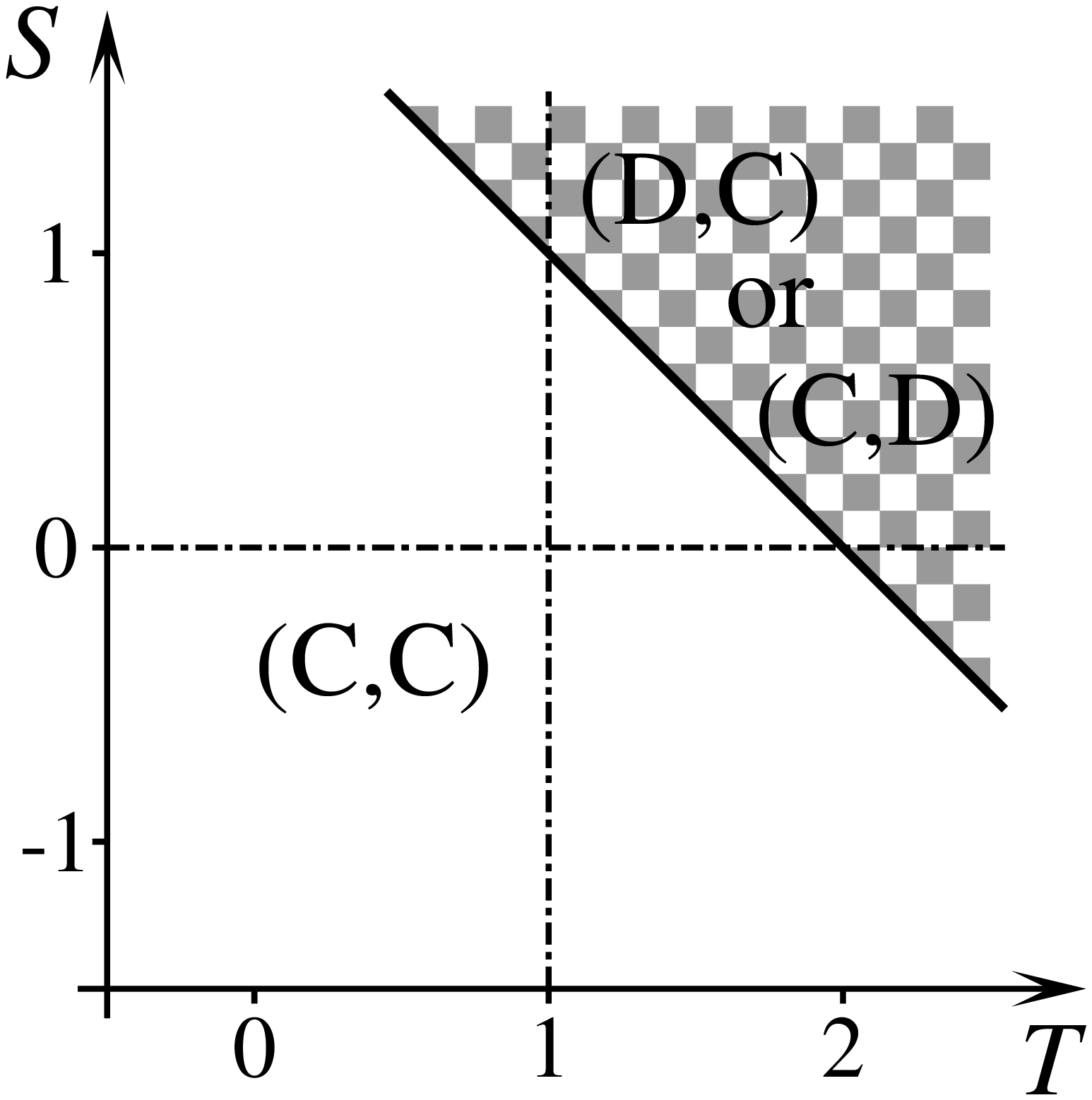,width=4cm}}
\caption{Preferred Nash equilibria for selfish  (left) and fraternal (right) players. Dash-dotted lines divide the $S-T$ plane into four regions nominated in Fig.~\ref{fig:socdil_map}}
\label{fig:pne_st}
\end{figure}

The equal share of payoffs eliminates the anti-symmetric components of ${\bf A}$ and the corresponding potential matrix is given as ${\bf V}^{\rm (fr)}={\bf A}^{\rm (fr)}$. In the two-strategy social dilemma notation discussed above
\begin{equation}
{\bf V}^{\rm (fr)}={\bf A}^{\rm (fr)}=\left(\matrix{ 0 & (T+S)/2 \cr
                       (T+S)/2 & 1 \cr} \right)
\label{eq:socdil_fr_payoff}
\end{equation}
which predicts two distinct regions on the $T-S$ plane when considering the preferred Nash equilibrium that may be either $(C,C)$ (if $(T+S)<2$) or the $(C,D)$ or $(D,C)$ strategy profiles (if $(T+S)>2$). The latter results differ strikingly from those we obtained above as it is illustrated in Fig.~\ref{fig:pne_st}. The differences in the location of the maximum matrix elements of ${\bf V}$ [see Eq.~(\ref{eq:socdilVsimple})] and ${\bf V}^{\rm (fr)}$ imply social dilemmas for the two-strategy games.

A similar comparison can be performed for other potential games even for $n>2$. In that case we use the notation of Eq.~(\ref{eq:classes}) when potential exists if ${\bf A}^{\rm (cycl)}=0$ and the resulting potential matrix is determined by only two components (${\bf A}^{\rm (s)}$ and ${\bf A}^{\rm (coord)}$) as expressed in (\ref{eq:potmatrix_ns}). For these games the effective payoff and potential matrices for fraternal players involve the contribution of ${\bf A}^{\rm (cr)}$, too, as
\begin{equation}
{\bf V}^{\rm (fr)}= {\bf A}^{\rm (fr)}={\bf A}^{\rm (coord)}-{\bf A}^{\rm (av)}+
{1 \over 2} [{\bf A}^{\rm (s)}+{\bf A}^{{\rm (s)}T} + {\bf A}^{\rm (cr)}+{\bf A}^{{\rm (cr)}T}].
\label{eq:potmatrix_fr}
\end{equation}
We face ''social dilemma'' if the preferred Nash equilibria are distinct, that is, when  $(i,j) \ne (i^{\prime},j^{\prime})$ where $\max{[V_{kl}]}=V_{ij}$ and $\max{[V_{kl}^{\rm (fr)}]}=V_{i^{\prime}j^{\prime}}^{\rm (fr)}$. The conflicting situations involve cases when the selfishness results in lower payoffs for both players (as it happens for the prisoner's dilemma) or also those ones when the loss of the sucker exceeds the extra profit of the winner. The discrepancy between the individual and common interests can be quantified by
\begin{equation}\label{eq:potmatrix_dif}
{\bf V}- {\bf V}^{\rm (fr)}={1 \over 2} [{\bf A}^{\rm (s)}+{\bf A}^{{\rm (s)}T} - {\bf A}^{\rm (cr)} - {\bf A}^{{\rm (cr)}T}] \nonumber
\end{equation}
depending only on the self- and cross-dependent components. More precisely, ${\bf V}-{\bf V}^{\rm (fr)}$ is the potential of the antisymmetric portion of the self- and cross-dependent payoff components defined as $({\bf A}^{\rm (s)}+ {\bf A}^{\rm (cr)}- {\bf A}^{{\rm (s)}T}-{\bf A}^{{\rm (cr)}T})/2 = {\bf A}-{\bf A}^T$. Accordingly, no dilemmas arise if ${\bf A}^{\rm (s)}={\bf A}^{{\rm (cr)}T}$. In the light of this result for a symmetric $n$-strategy potential game we have $(n-1)$ independent parameters to control the absence or presence of the social dilemma in the above sense.

The social dilemmas and their consequences are preserved for spatial multi-agent evolutionary games with equivalent players if logit rules control the strategy updates as it is demonstrated by \citet{szabo_jtb12} who studied cases of non-equal sharing for $n=2$.

\section{MULTI-PLAYER POTENTIAL GAMES}
\label{sec:mppg}

Up to this point we have discussed only two-player games. Evidently the sum or any linear combinations of independent two-player potential games are also potential games as the original conditions (\ref{eq:potential}) are satisfied. This situation remains valid even for the cases when some of the players participate in more than one potential games using the same strategies in each action. Remarkably, if a player uses two (or more) strategies against her co-players then she can be considered as two (or more) players. In many biological and economical systems the strategy represents the type of species or behavior that is the same for all the interactions for a given individual. For human systems, however, one can assume that the players use different strategies in the games they participate. Furthermore, human players can use mixed (stochastic) strategies in the repeated games.

Henceforth we assume that the players use pure strategies expressed by unit vectors [see Eq.~(\ref{eq:purestrats})] for each player $x=1, \ldots , N$ where $N$ denotes the number of players. In the present game theoretical models the players wish to maximize their own payoff by consecutive strategy refreshments disregarding the payoff variation for all the other co-players. If the players play two-person potential games with a subset of the other players, then their individual motivation can be well quantified by the variation of potential
\begin{equation}\label{eq:Nppot}
U({\bf s})= \sum_{\langle x,y \rangle }{\bf s}_x \cdot {\bf V}(x,y)
{\bf s}_{y}\,
\end{equation}
where the summation runs over all interacting pairs (denoted as $\langle x,y \rangle$) defined by the edges of the connectivity graph where each node represents a player.

In real social systems the pair potential may be different for all interacting pairs as indicated by the expression ${\bf V}(x,y)$. In most of the cases, however, we assume that the players are equivalent and the games are uniform. For the latter discussions the arguments of the potential matrix ${\bf V}$ are omitted. Many relevant features of spatial evolutionary games can be investigated where the equivalent players are located on the sites of lattice with periodic boundary conditions. The investigation of these systems can be efficiently supported by the methods developed for studying models of solid state physics where all the relevant symmetries are satisfied.

The latter simplifications exclude the analysis of systems with players interacting via non-symmetric games. At the same time, the systematic investigation of the multi-agent systems with non-symmetric games can be performed on bipartite networks that give us a convenient connectivity structure. For these multi-agent systems we distinguish two types of players ({\it e.g.} males and females, buyers and sellers, {\it etc.}) who are distributed on a bipartite graph providing that each player plays games with players of opposite type. For these systems the connectivity network is divided into two subgraphs with sites $x \in X$ and $y \in Y$. The bipartite connectivity graph ensures that players at sites $x \in X$ interact exclusively with those at sites $y \in Y$, and {\it vice versa}.

Such a situation can even be realized in lattice systems. For example, if the players are located at the sites of a square lattice, then the sites are divided into two equivalent sublattices corresponding to the white and dark squares of the chessboard. These systems will be discussed later in Sec. \ref{sec:mtoI} and \ref{sec:emp2x2}.

The first investigations of the multi-agent evolutionary games were performed for well-mixed populations corresponding to a complete connectivity graph where interactions exist between all the possible pairs. Since the pioneering work of \citet{nowak_n92b,nowak_ijbc93} the games are studied progressively on different lattices. For these spatial systems the sites are equivalent if periodic boundary conditions are applied for a finite square box of $L \times L = N$ sites and the interactions are restricted to the same size of neighborhood for each player ({\it e.g.}, nearest or nearest and next-nearest neighbors). The main advantage of the lattice structured population is based on the translation symmetry providing good conditions for studying the consequences of some characteristic features in the structure of neighborhood.

For real social systems, however, the irregular networks give a more adequate background for the investigation of multi-agent systems. This is the reason why the analysis of lattice systems was extended to different graphs representing diluted lattices \cite{nowak_ijbc94,vainstein_pre01}, real social networks \cite{holme_pre03} random graphs \cite{kim_bj_pre02,masuda_pla03,duran_pd05,vukov_pre06,de-santis_jpa07}
including different small-world \cite{wu_zx_pre05,tomassini_pre06} and  scale-free networks \cite{santos_prl05,santos_prsb06}.

\subsection{Pure Nash equilibia in multi-player games}
\label{sec:searchNE}

\begin{figure}[ht]
\centerline{\epsfig{file=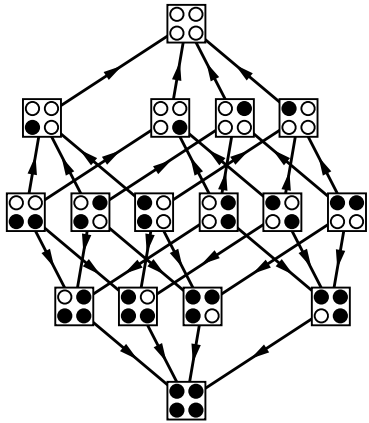,width=5cm}}
\caption{The flow graph in a four player potential game if the potential is composed of two-player coordination games when players benefit from choosing the same strategy.}
\label{fig:dg4p2sg}
\end{figure}

In this section we briefly survey the relevant variations in the number of Nash equilibria when an $N$-player game on a lattice (or network) is built from pair interactions between the neighbors. Regardless of the connectivity structure, the multi-agent systems have $2^N$ microscopic states (${\bf s}$) if all the pair interactions are characterized by a $2 \times 2$ game. Furthermore, if the possible transitions between two microscopic states are limited to those ${\bf s} \to {\bf s}^{\prime}$ where only one player modifies her strategy ({\it e.g.}, $s_x \to s_x^{\prime}$) then the dynamical graph can be represented by an $N$-dimensional hypercube \cite{schnakenberg_rmp76, szabo_pre10b}. Such a dynamical graph is shown (for $N=4$) in Fig.~\ref{fig:dg4p2sg} where the nodes and edges of the four-dimensional hypercube are projected onto a two-dimensional plane from a suitable direction in such a way that the number of players playing the first strategy differs by 1 from level to level.
Notice that the parallel edges represent strategy changes for the same player while the strategy profiles are different on the neighborhood. If the strategy profile evolves along the directed edges then the system evolves into one of the two Nash equilibria. Such a behavior occurs when randomly chosen players are allowed to increase their income by choosing another strategy in the knowledge of their other possible payoff(s). The iteration of this process ends in one of the Nash equilibria that are absorbing points. Evidently the final state depends on the initial state. The latter method can be considered as a way to find pure Nash equilibria \cite{brown_gw_51, rosenthal_ijgt73, monderer_jet96, candogan_mor11}.

This method can be utilized for all the potential (and ordinal potential) games to find pure Nash equilibria due to the absence of directed loops in their flow graph. The $N$-player systems have only one Nash equilibrium if the interaction is composed of pair interactions belonging uniformly to either the harmony or prisoner's dilemma games.

In the example illustrated in Fig.~\ref{fig:dg4p2sg} there are only two Nash equilibria and each has a basin of attraction in the space of strategy profiles. A similar behavior occurs for larger number of players if all the possible pairs interact via a coordination game.

\begin{figure}[ht]
\centerline{\epsfig{file=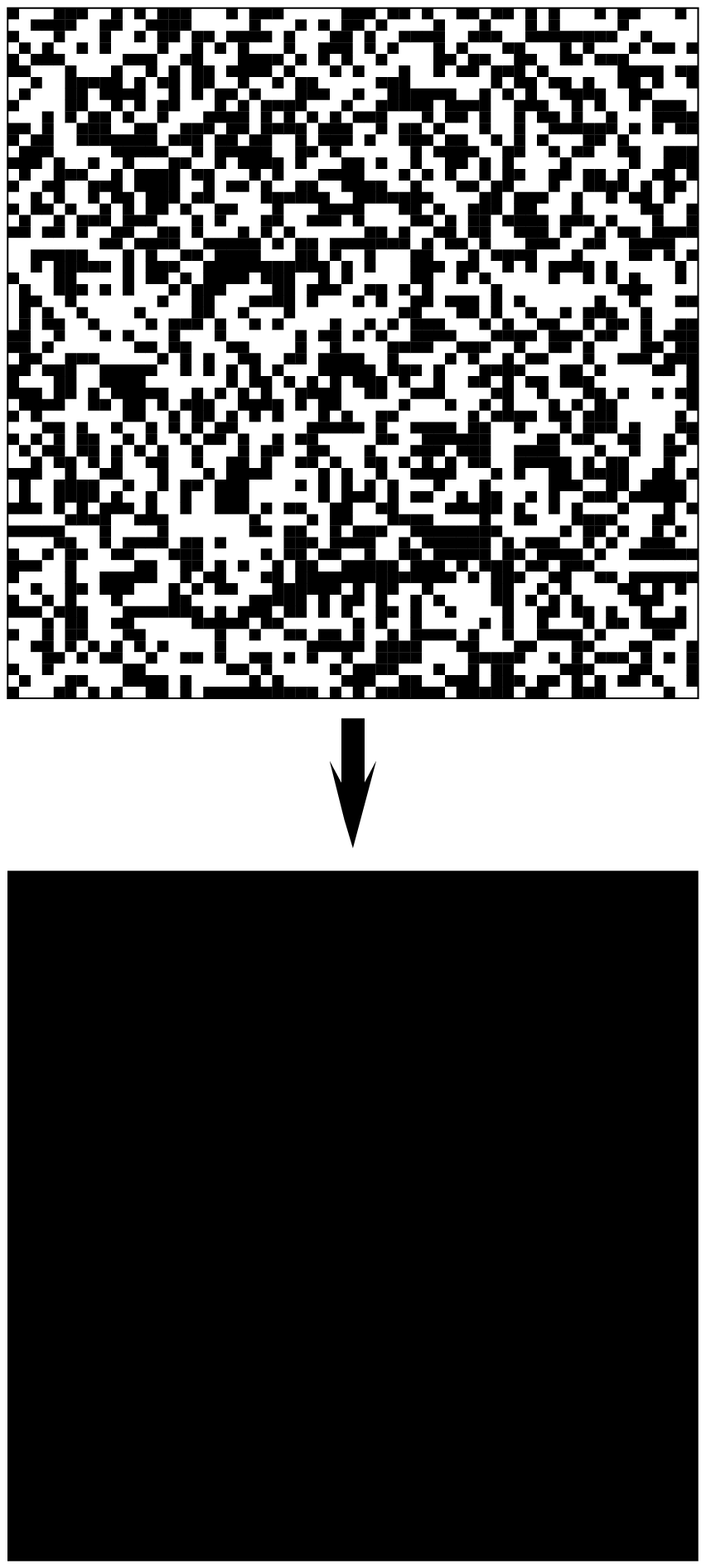,width=3cm}
\epsfig{file=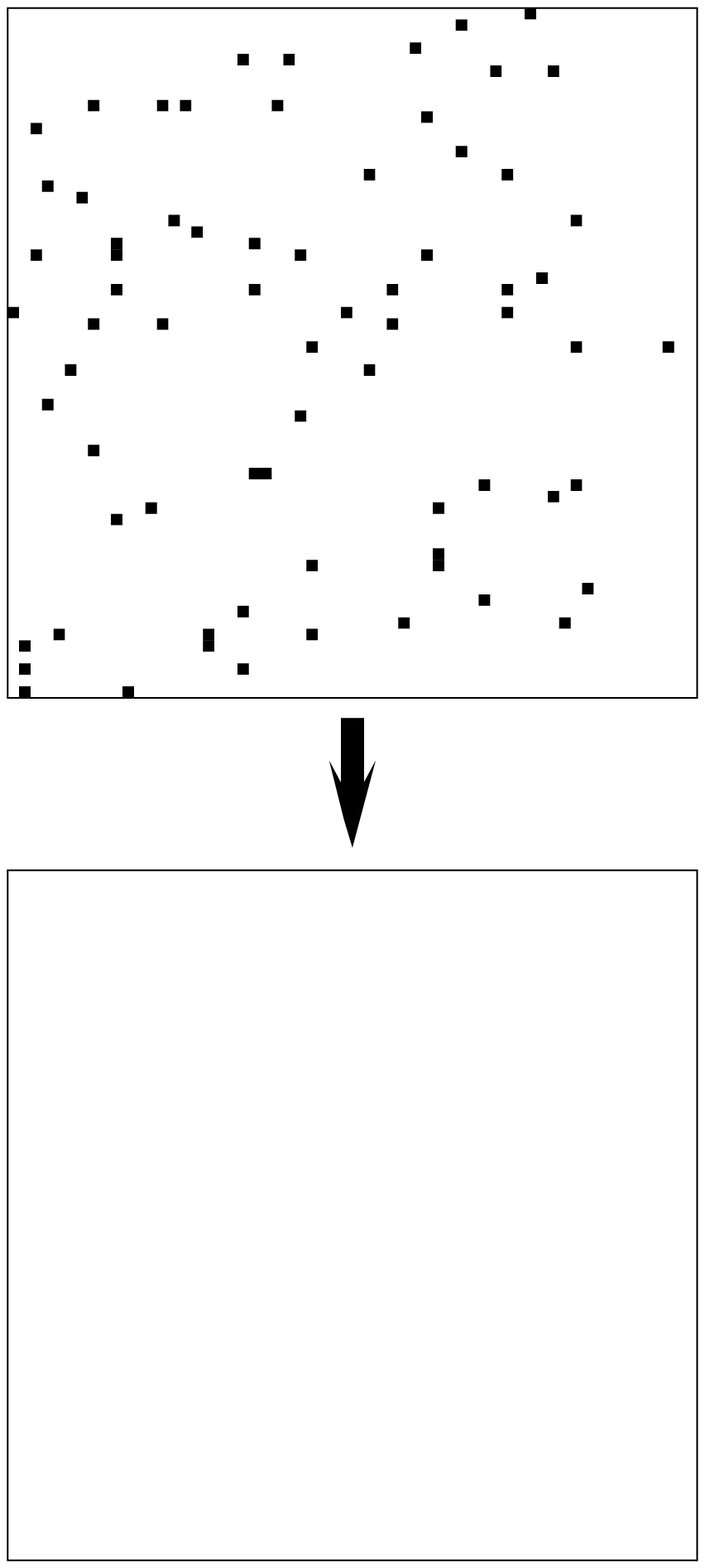,width=3cm}
\epsfig{file=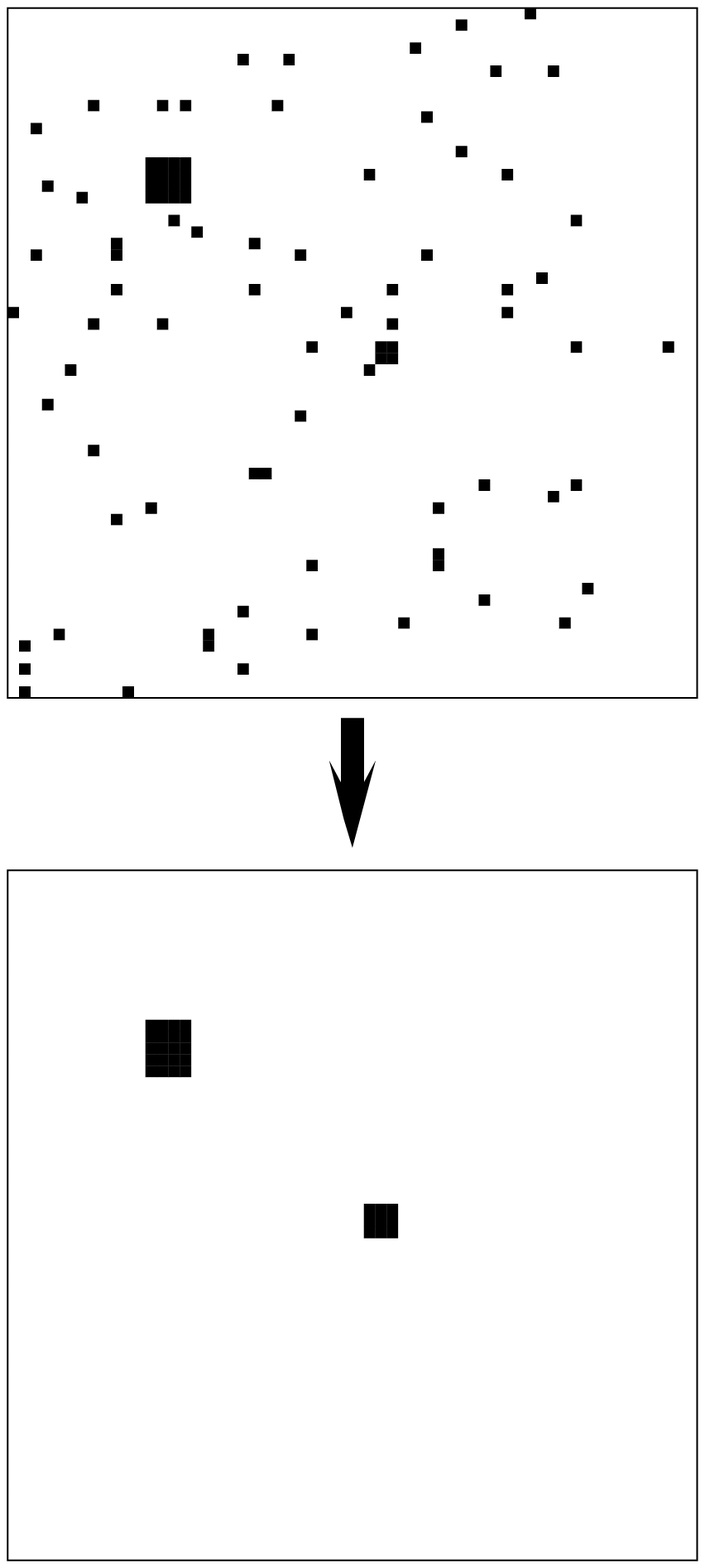,width=3cm}}
\caption{Final spatial distributions of strategies (Nash equilibria) (lower snapshots) if the evolution is started from three different initial states (upper snapshots) when the players play stag hunt games ($T=0.5$, S=-0.6) with the four neighbors on the square lattice. In the left case the system is started from a random initial state where $D$ and $C$ strategies are present with the same probability. In the initial state of the middle case $D$s are created randomly with a probability of $0.02\,$. For the right case the latter initial state is modified by adding two rectangular blocks of $D$s.}
\label{fig:tendtoNE}
\end{figure}

More sophisticated processes occur in a spatial system where the players are located on a square lattice and play stag hunt games with their four nearest neighbors. In that case if we allow a randomly chosen player to increase her payoff by choosing the opposite strategy then the iteration of these steps drives the system into the homogeneous $D$ state ($s_x=D$ $\forall x$) if the sufficiently large system is started from a random initial strategy profile for $S<T-1$ and $T<1$ (see the left plots in Fig.~\ref{fig:tendtoNE}). This is the preferred Nash equilibrium where $U({\bf s})$ reaches its maximum. This system, however, has many other Nash equilibria that may be achieved by the application of the above evolutionary dynamics if we choose different initial states as it is illustrated in Fig.~\ref{fig:tendtoNE}.
The reader can easily check that in the homogeneous $C$ state ($s_x=C$ $\forall x$) nobody can benefit from changing her strategy unilaterally. This feature implies that the latter Nash equilibrium has a finite basin of attraction if the formation of a $D-D$ pair on two neighboring sites is not favored either. Consequently, the second Nash equilibrium can be achieved from such a random initial state where the ratio of $D$ and $C$ strategies is small and typically only isolated $D$ strategies or small cluster of $D$s are present in the sea of $C$s. Such a situation is illustrated by the middle pair of snapshots of Fig.~\ref{fig:tendtoNE}.

Despite the small ratio of $D$ strategies, larger clusters of $D$s may also appear in the random initial state, particularly if the system is sufficiently large [for a more quantitative analysis we suggest consulting surveys of percolation theory \citep{essam_rpp80, stauffer_92}]. If the size of a rectangular cluster of $D$s exceeds a threshold value, then this colony can remain alive or it can even expand if this process is supported by the presence of solitary $D$s along the periphery as demonstrated in the plots at the right of Fig.~\ref{fig:tendtoNE}. Consequently, the selected Nash equilibrium depends on the initial state and also on the random strategy refreshment of players.

\begin{figure}[ht]
\centerline{\epsfig{file=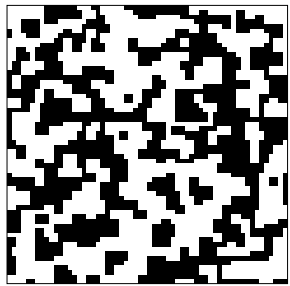,width=4cm}}
\caption{Typical distribution of $D$ and $C$ strategies on a square lattice in a Nash equilibrium occurring in the stag hunt region for  $S=-0.5$ and $T=0.5$.}
\label{fig:sh_polydom}
\end{figure}

Similar phenomena can be observed for the spatial stag hunt games if $S>T-1$. The only difference is that the roles of the $D$ and $C$ strategies are exchanged in the evolutionary processes. On the contrary, a basically different behavior occurs along the boundary $S=T-1$ separating the above discussed regions of parameters where the homogeneous states form a frozen poly-domain structure as demonstrated in Fig.~\ref{fig:sh_polydom}.
As this is the typical behavior in the whole region of the hawk-dove games, we will not discuss the resulting processes separately.

\begin{figure}[ht]
\centerline{\epsfig{file=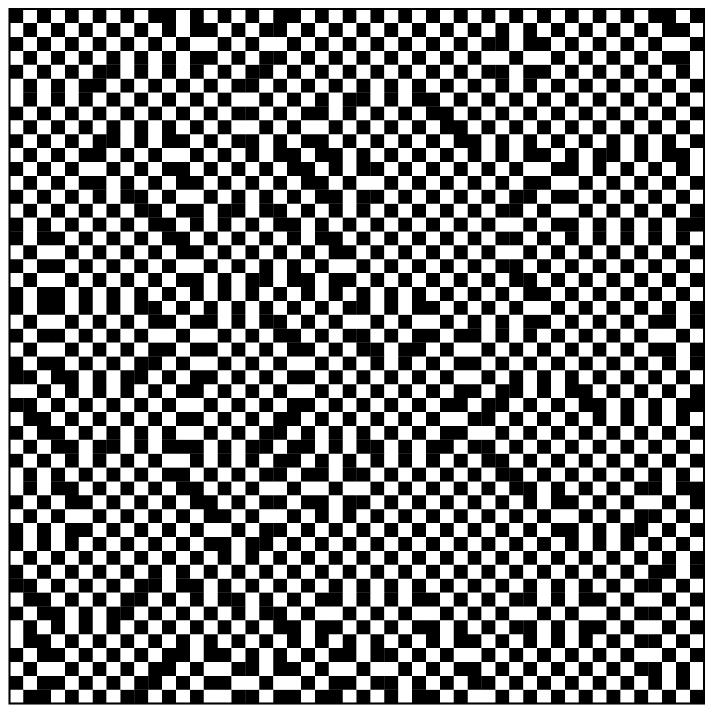,width=4.5cm}
\epsfig{file=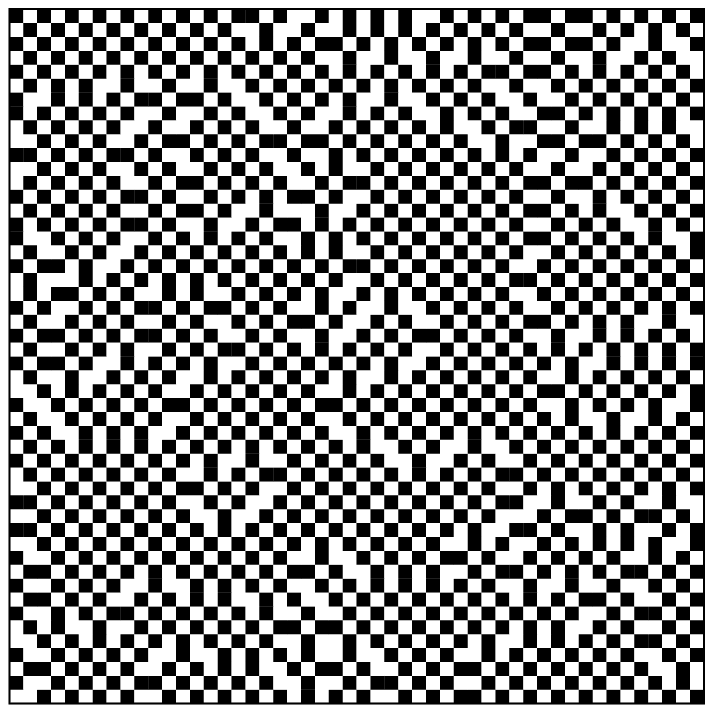,width=4.5cm}}
\caption{Typical strategy distributions on a $50 \times 50$ box of sites for Nash equilibria in a HD game on a square lattice for nearest neighbor interactions if $S=0.2$ (left panel) and $S=0.4$ (right panel) for $T=1.3$.}
\label{fig:NE-sHD}
\end{figure}

Within the region of hawk-dove games ($T>1$ and $S>0$) the lattice system has two equivalent sublattice ordered states where the potential reaches its maximum. The analysis of these states requires the introduction of two sublattices ($X$ and $Y$) as described above. In the first ordered structure the sites of sublattice $X$ are occupied by $D$s, and $C$s are in the other sublattice. For the second ordered structure the sublattice occupancies are reversed. These long range ordered structures are equivalent (because of the translation symmetry) and can be observed in finite systems if the linear size $L$ is even for periodic boundary conditions. If such a system is started from a random initial strategy distribution, then the random strategy update of players favoring the increase of their own payoff results in a frozen poly-domain structure as illustrated in Fig.~\ref{fig:NE-sHD}. Similar patterns are reported by \citet{sysiaho_epjb05}, who studied additionally the cases where second neighbor interactions are also taken into consideration.

In the frozen poly-domain structures both types of ordered strategy arrangements are recognizable within the domains. Notice the absence of point defects inside the ordered spatial regions where the players are satisfied and will not modify their strategy. The opposite ordered structures are separated by boundary layers. The average composition of these boundary layers, however, depends on the sign of $S-T+1$ as demonstrated by two opposite examples in Fig.~\ref{fig:NE-sHD}.

\begin{figure}[ht]
\centerline{\epsfig{file=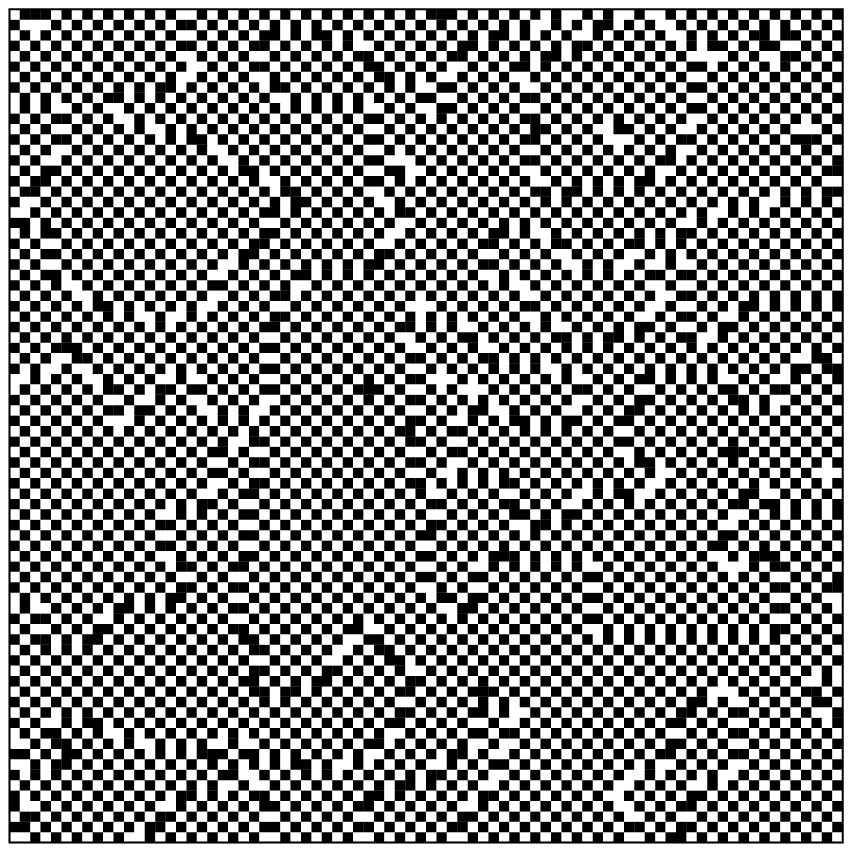,width=3.5cm}
\epsfig{file=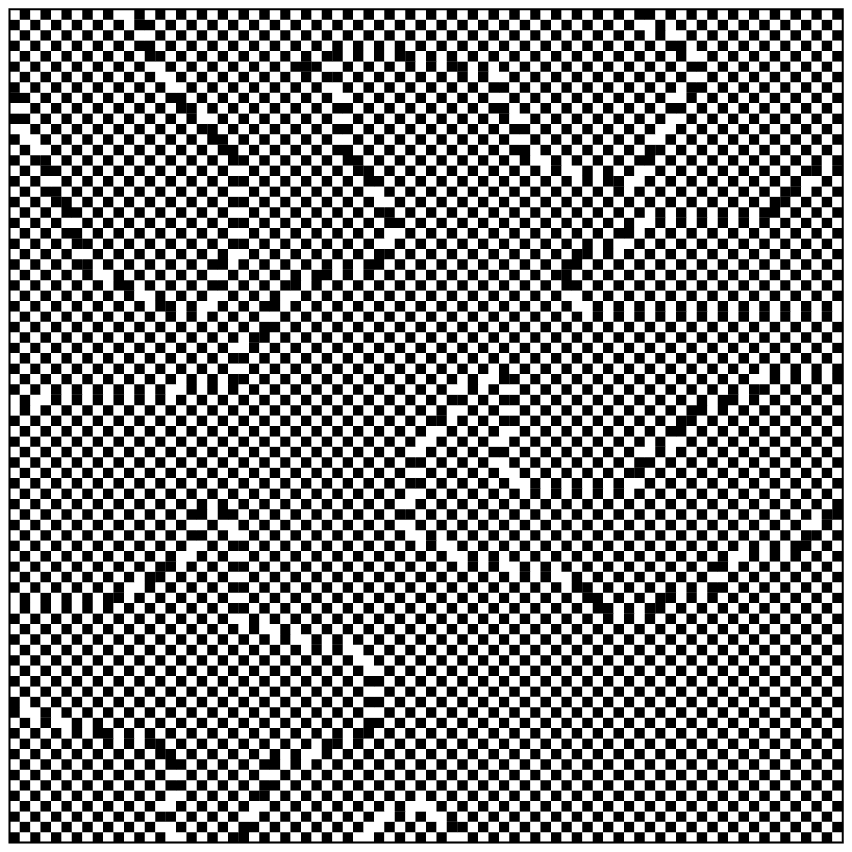,width=3.5cm}
\epsfig{file=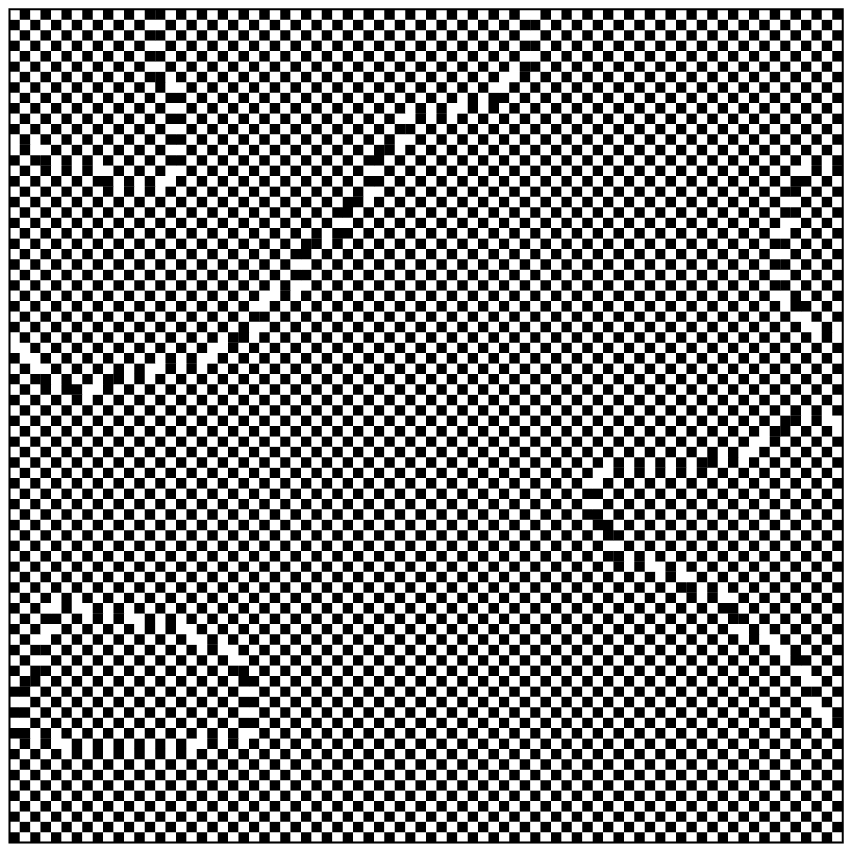,width=3.5cm}}
\caption{Three snapshots at times $t=10$, 40, and 160 MCS (from left to right) illustrating the evolution of strategy distribution on a $80 \times 80$ box of a larger system for $S=0.4$ and $T=1.4$.}
\label{fig:myothd}
\end{figure}

Evidently, the above-mentioned difference in the boundary layers vanishes if $S-T+1=0$ because the interface becomes a mixture of two types illustrated in Fig.~\ref{fig:NE-sHD}. In this special case there are many players along the interfaces who have the same payoff for the opposite strategies. If we additionally allow these players also to modify their strategy, then these additional stochastic events in the strategy update yield a domain growing process that drives the system into one of the homogeneous sublattice ordered strategy arrangements. This evolutionary process is demonstrated in Fig.~\ref{fig:myothd} where the time is measured in the unit of Monte Carlo steps (MCS) [within one MCS each player receives a chance once on the average to change her strategy].

Finally we mention that the latter additional options result in a similar domain growing process in the stag hunt region, too, if $S-T+1=0$. These ordering processes will be discussed later when studying the consequences of a more general class of stochastic evolutionary rules. Finally we mention that similar phenomena occur on bipartite networks.

\section{EVOLUTIONARY POTENTIAL GAMES}
\label{sec:epg}

Evolutionary games are devoted to explore the consequences of different evolutionary rules defining the ways how individuals can modify their own strategies in order to have a higher payoff. The first efforts were mainly focused on the maintenance of cooperative behavior in the social dilemmas where the results strongly depend on the set of strategies, the connectivity structure of players, and the dynamical rules that may involve variations in all ingredients of the mathematical models (including the current strategy, interaction or learning networks, dynamical rules, personal features, and their combinations). As mentioned in the Introduction the concept of evolutionary games was originally suggested and applied by  \citet{maynard_n73,maynard_82,axelrod_s81,axelrod_84,hofbauer_88} to provide a mathematical background for the description of the Darwinian evolution for biological and social systems. This is the reason why most of the investigations during the last decades used dynamical rules based on the adoption/imitation of the more successful strategy (or biological species). As a result different dynamical rules are introduced and studied systematically for a wide scale of conditions [for a survey see \cite{nowak_06, szabo_pr07, roca_plr09}].

For example, in lattice systems the "imitating the best" rule \cite{nowak_n92b} dictates the deterministic adoption of that player's strategy who received the higher payoff in the neighborhood; the "imitating the better"  rules allow to adopt any better strategy from the local neighborhood with a probability dependent on the neighbors's payoff \cite{helbing_incoll98,schlag_jet98}. Besides it there are other imitation rules [introduced and studied e.g., by \citet{nowak_ijbc94}; \citet{szabo_pre98}; \citet{alonso-sanz_ijbc01c};  \citet{masuda_pla03}; \citet{ohtsuki_prslb06}, \citet{wild_n09}; \citet{wu_b_pre10}; and many others] when the players can adopt the strategy even from a worse neighbor though the more successful strategies continue to be preferred.

For all types of imitation rules the system has absorbing states because this rule cannot recreate a strategy that has become extinct previously. For example, if one of the homogeneous states is approached then the system remains there forever. On the contrary, the introduction of mutation [see \cite{willensdorfer_jtb05}, \cite{antal_jtb09}, and \cite{tarnita_jtb09b}] is capable of reviving strategies that have vanished previously.

For the rules based on fictitious game (as detailed in the Section \ref{sec:searchNE}) each player is capable of determining her own payoff for her strategies with an assumption that their co-player will not change their strategies. For the "best response" version of these rules the player chooses those strategies that provide the highest payoff under the mentioned conditions. If more than one "best response" exists, then the player selects one of them at random. The smoothed versions of this rule allow the player to select all her possible strategies with a probability dependent on the payoff variation. The concept of potential games becomes fruitful for those special versions of these dynamical rules which drive the system into the so-called Gibbs state characterized by the Boltzmann distribution [for an English translation of Boltzmann's original work see \cite{sharp_e15}]. For these special cases we can utilize the enormous amount of results obtained in statistical and solid state physics to explain phenomena emerging in social and biological systems, too.

Evidently, the latter evolutionary processes based on fictitious games are relevant in human systems. Similar situations can occur in biological systems if the species creates a large number of mutants with a survival probability dependent on the neighborhood. We have to emphasize, however, that the quantitative analysis of the social and biological systems requires relevant extensions of the possible rules opening new challenges for both the non-equilibrium statistical physics and super-statistics.

\subsection{Evolutionary rules leading to Boltzmann distribution}
\label{sec:evoldyn}

In the field of game theory \citet{blume_l_geb93} described first that for the application of the so-called log-linear or logit evolutionary rules the potential games evolve into the Gibbs state assuring the validity of the direct application of the concepts and tools of equilibrium statistical physics. The basic idea, as an efficient Monte Carlo algorithm to determine average values for the Boltzmann distribution in many particle system, however, was developed by a group conducted by Metropolis at Los Alamos in the 1950s. They recognized that instead of evaluating the Boltzmann factor for a randomly chosen microscopic state it is more efficient to use a Markov chain algorithm in which a state will be chosen with a probability defined by the Boltzmann distribution. Nowadays this method is named importance sampling. A similar evolutionary rule was suggested by \citet{glauber_jmp63} who wished to study time-dependent processes for the Ising model whose dynamics is not prescribed by the quantum mechanics. Since that time several other algorithms were suggested and used widely for the investigation of equilibrium and non-equilibrium phenomena in many-particle systems.

Now we describe the above-mentioned rules in historical order by using the terminology of evolutionary game theory. The system is started from an arbitrary (typically random) initial state and during the evolution the following elementary steps are repeated. A randomly chosen player $x$ can modify her strategy ${\bf s}_x$ to ${\bf s}_x^{\prime}$ chosen at random from her options with a probability $w({\bf s}_x \to {\bf s}_x^{\prime})$ dependent on the payoff variation $\Delta u = u_x({\bf s}_x^{\prime},{\bf s}_{-x})-u_x({\bf s}_x,{\bf s}_{-x})=V({\bf s}_x^{\prime},{\bf s}_{-x})-V({\bf s}_x,{\bf s}_{-x})$.

For the Metropolis algorithm the player always adopts the new strategy ${\bf s}_x^{\prime}$ if it is worthy ($\Delta u > 0$). For the opposite cases the player accepts the strategy ${\bf s}_x^{\prime}$ with a probability decreasing exponentially with the possible loss, that is,
\begin{eqnarray}\label{eq:metropolis}
w^{\rm (M)}({\bf s}_x \to {\bf s}_x^{\prime})= \left\{
\begin{array}{l} 1 \;\; \mbox{for} \; \Delta u \ge 0 \;, \\
                 e^{\Delta u/K} \;\; \mbox{for} \; \Delta u <0 \;
\end{array} \right .
\end{eqnarray}
where $K$ denotes the magnitude of noise (or temperature in physical systems). Notice that this rule agrees with those we used in Sec.~\ref{sec:searchNE} if $K \to 0$.

For the Glauber dynamics
\begin{equation}\label{eq:glauber}
w^{\rm (G)}({\bf s} \to {\bf s}_x^{\prime})={1 \over
1+e^{-\Delta u/K}} .
\end{equation}
This evolutionary rule is convenient for the analytical calculation as demonstrated in \cite{szabo_pr07}.

The transition rate for the logit rule \cite{blume_l_geb93, fudenberg_eer98} is defined as
\begin{equation}\label{eq:loglin}
w^{\rm (log)}({\bf s}_x \to {\bf s}_x^{\prime})=\lambda_x { e^{u({\bf s}_x^{\prime},{\bf s}_{-x})/K} \over
\sum_{s_x^{\prime \prime }}e^{u({\bf s}_x^{\prime \prime},{\bf s}_{-x})/K}} \;.
\end{equation}
where the summation runs over all the possible strategies of the player $x$ and $0 < \lambda_x \le 1$. The choice of $\lambda_x=1$ provides the most efficient algorithm in the Monte Carlo simulations.

The most relevant common feature of the above dynamical rules is that they all drive the potential games into the Gibbs state where the probability $p({\bf s})$ of a strategy profile ${\bf s}$ is given by the Boltzmann distribution as
\begin{equation}\label{eq:boltzmann}
p({\bf s})={1 \over Z} e^{U({\bf s})/K}
\end{equation}
where
\begin{equation}\label{eq:boltzmann_Z}
Z=\sum_{\bf s} e^{U({\bf s})/K} .
\end{equation}
The normalization factor $Z$ is called the partition function and plays a crucial role in statistical physics because the partial derivatives of $\ln{(Z)}$ reproduce relevant thermodynamical quantities. For example, the average value of the potential can be given as
\begin{equation}\label{eq:pd_lnZ}
{\cal U} = {\partial \ln{(Z)} \over \partial (1/K)}\;.
\end{equation}
The relevance of these features is enhanced by the exact solutions obtained by \citet{eggarter_prb74, baxter_82, yang_zr_pre94} for several connectivity structures.

In addition, the conditions of detailed balance are also satisfied between all the possible forward-backward transition pairs. As a result, for the Boltzmann distribution the transition ${\bf s}_x \to {\bf s}_x^{\prime}$ and its backward version (${\bf s}_x^{\prime} \to {\bf s}_x$) appear with the same frequency, i.e.,
\begin{equation}\label{eq:detbal}
w({\bf s}_x \to {\bf s}_x^{\prime}, {\bf s}_{-x}) e^{U({\bf s})/K}
= w({\bf s}_x^{\prime} \to {\bf s}_x, {\bf s}_{-x}) e^{U({\bf s}^{\prime})/K}\;
\end{equation}
for $\forall\,x, \forall\, s_x^{}, s_x^{\prime}$, and $\forall\,
s_{-x}$. These conditions are satisfied if
\begin{equation}\label{eq:detbal2}
{ w({\bf s}_x \to {\bf s}_x^{\prime}, {\bf s}_{-x}) \over
w({\bf s}_x^{\prime} \to {\bf s}_x, {\bf s}_{-x})}
  =  e^{[U({\bf s}^{\prime}) - U({\bf s})]/K} .
\end{equation}
Due to this relationship the final stationary state (the Boltzmann distribution) and also the satisfaction of detailed balance remain unchanged if the transition rates $w({\bf s}_x \to {\bf s}_x^{\prime})$ and $w({\bf s}_x^{\prime} \to {\bf s}_x)$ are multiplied by the same coefficient that can be varied from pair to pair of strategy profiles involved [see $\lambda_x$ in Eq.~(\ref{eq:loglin})].

For all the above dynamical rules only one player has modified her strategy in the consecutive elementary steps. For the Kawasaki dynamics \cite{kawasaki_pr66} a randomly chosen player $x$ exchanges strategy with one of her neighbors with a probability dependent on their summarized payoff variation that is not affected by the interaction between them. The payoff variation arising from the rest of interactions can be built up from two consecutive individual strategy changes as before. Consequently, the summarized payoff variation is identical to the potential variation $\Delta U=U(s_x,s_y,s_{-x,y})-U(s_y,s_x,s_{-x,y})$ and the transition rate can be given by the following expression:
\begin{equation}\label{eq:kawasaki}
w^{\rm (K)}({\bf s} \to {\bf s}_x^{\prime})={1 \over
1+e^{-\Delta U/K}}
\end{equation}
used in many solid state applications when studying diffusion or transport in systems where the number of particles (strategies) is conserved. For the two-strategy systems the corresponding dynamical graph is given by the nodes of an $N$-dimensional hypercube and the edges are defined by those diagonals of faces where two neighboring players exchange their different strategies with each other.

\subsection{Statistical physics and thermodynamics}
\label{sec:spt}

For the microscopic description of a multi-agent system we should define the individual strategy for all participants, which requires a huge amount of data. At the same time, for the macroscopic description of the same system we use only a few parameters, {\it e.g.} average portion of a strategy, average payoff and noise level, as it happens in the thermodynamic description of a gas when using the concept of pressure, temperature, and density, without any knowledge about the position and velocity of atoms contained in the gas. Statistical physics gives a general framework to describe the relationships between the macroscopic (thermodynamic) quantities that are influenced by the microscopic interactions.

The fundamental assumption of the {\it equal a priori probability} of the accessible microscopic states serves a basis for the statistical description. According to this approach, those macroscopic states can be observed in the stationary state that are realized by the largest number of microscopic states. Due to the law of large numbers the latter approach defines well the macroscopic state in the limit $N \to \infty$ as detailed in the textbooks of statistical physics [see {\it e.g.} \cite{landau_80,
toda_91}] based on early works by \citet{boltzmann_877}, \citet{gibbs_02} and \citet{szilard_zp29}.

Now we briefly survey the relevant mathematical background following a modern formalism suggested by \citet{jaynes_pr57} and using the terminology of evolutionary game theory \cite{blume_l_geb93, blume_l_geb03, cohen_mh_pa13} as before. Within this framework the system is described by the $p({\bf s})$ probability distribution of the microscopic states ${\bf s}$ and the problem of prediction becomes equivalent to the maximization
of Shannon entropy \cite{shannon_49}
\begin{equation}\label{eq:entropy}
{\cal S}=- \sum_{\bf s} p({\bf s}) \ln {p({\bf s})}  \,
\end{equation}
under several constraints. The crucial problem of extending the maximum entropy principle to non-physical systems lies in the adequate choice of constraints. This approach is used successfully in the analysis of complex \cite{haken_88} and chaotic systems.

In the present case $p({\bf s})$ is normalized, that is,
\begin{equation}\label{eq:psn}
\sum_{\bf s} p({\bf s}) = 1 \,
\end{equation}
and we can fix the average value of the potential as
\begin{equation}\label{eq:avpot}
 \sum_{\bf s} p({\bf s}) U({\bf s}) = {\cal U}  \,.
\end{equation}
Using the standard method the extremum under the above constraints can be evaluated by solving the following set of equations:
\begin{equation}\label{eq:condextr}
{\partial \over \partial p({\bf s})}\left[ S + \lambda_1 \sum_{\bf s} p({\bf s}) + \lambda_2  \sum_{\bf s} p({\bf s}) U({\bf s}) \right]= 0    \,
\end{equation}
where the Lagrange multipliers $\lambda_1$ and $\lambda_2$ are determined by taking into consideration the conditions (\ref{eq:psn}) and (\ref{eq:avpot}). Straightforward calculations yield that the optimal probability distribution becomes equivalent to the Boltzmann distribution for $\lambda_2=1/K$ [see (\ref{eq:boltzmann}) and (\ref{eq:boltzmann_Z})] and for a suitable choice of the average value of the potential (${\cal U}$) depending on $K$ (strictly) monotonously.

In the real utilization of the above extremum principle, it is more convenient to perform a Legendre transform \cite{callen_60, alberty_pac01, zia_ajp09}. Using the analogy of the Helmholtz free energy in thermodynamics, we introduce now a thermodynamic potential
\begin{equation}\label{eq:thermpot}
\Phi= {\cal U} + K {\cal S}  \,
\end{equation}
that has a maximum in the equilibrium state for a fixed $K$ characterizing the temperature in statistical mechanics or noise level in evolutionary potential games in the sense defined by the logit dynamical rules [see {\it e.g.}, (\ref{eq:loglin})].

Notice that for spatial systems with short range interactions both the entropy $S$ and the average value of potential (as well as $\Phi$) are proportional to the system size $N$ and are considered as extensive quantities in the thermodynamical descriptions.

Counterexamples are systems with long-range interactions or systems on some types of small-world networks where we cannot apply the traditional methods directly. In the wide field of statistical physics there are some new directions and approaches to describe the macroscopic behavior of the mentioned systems when considering complex \cite{haken_88}, or other dynamical systems \cite{graham_prl84, graham_prl91}, and those non-equilibrium systems that behave like the superpositions of the Boltzmann distributions \cite{beck_pa03, hanel_pnas11}. Surveying the latter directions goes beyond the scope of the present work, our attention will be focused on the description of phenomena occurring in large spatial systems.

In the following sections we will draw a parallel between the kinetic Ising/Potts models and social systems for a particular evolutionary rule. In fact, for both types of these systems the applicability of thermodynamics is limited to the thermostatics. According to quantum mechanics each microscopic state of the Ising type systems is an eigenstate and remains unchanged forever. The time-dependence in the kinetic Ising model is introduced by assuming a uniform interaction between the spins and a heat reservoir characterized by the temperature $K$. Under these conditions the total energy of the Ising system fluctuates and the first and second laws of thermodynamics become meaningless. The failure of the third law of thermodynamics is related to the high degeneracy of the ground state as discussed later for several cases.

In social and biological systems the application of a similar stochastic evolutionary rule is justified by its mathematical simplicity and the analogies we can exploit in the interpretation of the phenomena. In these systems the value of $K$ can characterize: (i) the fluctuation of payoffs; (ii) the errors made by players during their decision process; and (iii) the magnitude of risk the player accepts in the hope of finding a better (long-term) solution. In social systems the value of $K$ can be even considered as a personal feature and also the subject of the coevolutionary process. The first simulations in such coevolutionary models have indicated the homogenization of $K$ for the coexistence of strategies within the prisoners dilemma region if both the strategy and $K$ adoption are controlled by pairwise imitation rule \cite{szabo_epl09}. Similar results for Glauber type rules would increase the relevance of thermostatics for these systems.

\subsection{Consequences of the extremum principles}
\label{sec:cep}

The principle of the maximum entropy serves as a mathematical background describing the intimate relationship between statistical mechanics/physics and thermodynamics. This principle explains the laws of thermodynamics as well as the relevance of the Boltzmann distribution and connects the meaning of the Lagrange multipliers and the intensive quantities of thermodynamics. The Boltzmann distribution itself implies several
general relationships among the first and second partial derivatives of the thermodynamic potentials \cite{callen_60, tisza_66, callen_85} or the partition function as illustrated by Eq.~(\ref{eq:pd_lnZ}). Examples are the Gibbs-Duhem relations and the Gibbs' phase rule. This knowledge is summarized in equation of states quantifying the relations among the intensive and extensive thermodynamical quantities for macroscopic systems composed of different atoms, ions, and molecules.

The simplicity of the linear response theory and the fluctuation-dissipation theorem are also direct consequences of the Boltzmann distribution \cite{kubo_pr66} where the effect of a small perturbation obeys a simple expression in linear approximation.

In addition to the general thermodynamical relationships, the extremum principles can also be used to evaluate the thermodynamical quantities in the knowledge of the microscopic interactions. We briefly survey now the essence of the cluster variation methods providing a general framework for the traditional mean-field and pair approximations in the approximative description of the lattice systems. For the sake of simplicity our description is focused on systems with interactions between the equivalent neighboring players located on the sites of a lattice. Within this approach the translation invariant microscopic states can be described by configuration probabilities on a small cluster of neighboring sites. For example, $p_1(s_1)$ defines the probability of the strategy $s_1$ at each site of the lattice while $p_2(s_1,s_2)$ describes the probability of finding $s_1$ and $s_2$ strategies on two neighboring sites. For rotational invariant arrangement of strategies ({\it e.g.}, on square lattices where the horizontal or vertical directions are equivalent) we can assume that $p_2(s_1,s_2)$ is independent of both the position and direction of the pair. These quantities are normalized, that is,
\begin{eqnarray}\label{eq:p12norm}
&&\sum_{s_1} p_1(s_1)= 1  \nonumber \,, \\
&&\sum_{s_1,s_2} p_2(s_1,s_2)= 1  \,,
\end{eqnarray}
and satisfy the compatibility conditions:
\begin{equation}\label{eq:compcond}
\sum_{s_2} p_2(s_1,s_2)= \sum_{s_2} p_2(s_2,s_1)= p_1(s_1)  \,.
\end{equation}
Evidently, one can use larger clusters of $n$ sites to describe the stationary state in this system and the corresponding quantities satisfy similar compatibility conditions. The larger the cluster we study, the more accurate is the approximate solution
\cite{morita_jsp90, gratias_p82}.

With the application of Bayes' theorem we can build up the configuration probabilities for a large cluster as a product of configuration probabilities of smaller clusters as detailed in the paper by \citet{gutowitz_pd87}. This approach allows us to derive adequate approximations for some other quantities, {\it e.g.}, correlation function and correlation length \cite{szabo_pr07}. Furthermore, the application of Bayes' theorem plays a fundamental role within the dynamical cluster techniques (developed to study stationary states in non-equilibrium lattice systems) when a set of coupled
differential equations is derived by taking into consideration the contribution of all the elementary steps \cite{dickman_pra90}. In comparison to the dynamical cluster techniques the so-called cluster variation method provides a more convenient way to
evaluate the configuration probabilities. With the use of these quantities one can express both the entropy and the average value of the potential for a given lattice structure. For example, on a $d$-dimensional cubic lattice the value of ${\cal U}$ is expressed as
\begin{equation}\label{eq:Uav}
{\cal U}=N d \sum_{s_1,s_2} V_{s_1 s_2} p_2(s_1,s_2) \,,
\end{equation}
while the entropy can be approximated at the level of pair approximation \cite{bethe_prsa35, kikuchi_pr51} as
\begin{equation}
{\cal S} \simeq {\cal S}^{\rm (2p)} =- Nd \sum_{s_1,s2} p_2(s_1,s_2) \ln {p_2(s_1,s_2)}
+ N(2d-1) \sum_{s_1} p_1(s_1) \ln {p_1(s_1)} \,.
\label{eq:S2p}
\end{equation}
When using larger cluster sizes one can give better approximations for the entropy \cite{kikuchi_jcp67, morita_jmp72}, while expression (\ref{eq:Uav}) remains unchanged for the case of pair interactions. For the cluster variation methods the thermodynamic potential [{\it e.g.}, $\Phi$ as defined by (\ref{eq:thermpot})] is expressed as a function of configuration probabilities and its maximum value is determined by varying these parameters. The standard calculation may be simplified by considering only the
independent parameters of the configuration probabilities. For example, if only two strategies are allowed, $s_1, s_2 = 1, 2$, then the one- and two-site configuration probabilities can be expressed with only two parameters as
\begin{eqnarray}\label{eq:p2param}
&&p_1(1)= c  \nonumber \,, \\
&&p_1(2)= 1-c  \nonumber \,, \\
&&p_2(1,1)= q  \nonumber \,, \\
&&p_2(1,2)= c-q  \,, \\
&&p_2(2,1)= c-q  \nonumber \,, \\
&&p_2(2,2)= 1-2c+q  \nonumber \,
\end{eqnarray}
where $c$ can be interpreted as the portion (or density) of strategy 1 in the whole system and $c-q$ describes the density of domain walls separating homogeneous territories. In that case the equilibrium values of $c$ and $q$ are given by the solution of the following equations:
\begin{equation}\label{eq:equilc}
{\partial \Phi(c,q) \over \partial c}= 0 \,,
\end{equation}
and
\begin{equation}\label{eq:equilq}
{\partial \Phi(c,q) \over \partial q}= 0 \,,
\end{equation}
where $\Phi(c,q)={\cal U}(c,q)+K{\cal S}(c,q)$ is obtained by substituting Eqs.~(\ref{eq:p2param}) into (\ref{eq:S2p}) and (\ref{eq:Uav}). In general, Eqs.~(\ref{eq:equilc}) and (\ref{eq:equilq}) have more than one possible solutions ($0\le p_2(s_2,s_2)\le 1$). In the latter case the real solution is the one where $\Phi(c,q)$ reaches its maximum.

Thus the cluster variation method gives us a unique way to evaluate the equilibrium value of $c$ and $q$ together with all the related quantities as a function of $K$. Besides it, one can derive approximate phase diagrams if the actual model has several solutions with different symmetries, as it happens frequently in solid state systems [examples and further references are given in \cite{de_fontaine_96, udvardi_jpc96}]. In evolutionary games on lattices the average total payoff can be given as
\begin{equation}\label{eq:avpo}
{\cal A}=N d \sum_{s_1,s_2} A_{s_1 s_2} p_2(s_1,s_2) \,,
\end{equation}
in the knowledge of the payoff matrix ${\bf A}$ and the nearest neighbor strategy configuration probabilities. Apart from the symmetric case ${\bf A}={\bf A}^{T}$, the quantity ${\cal A}$ has no analogous concept in solid state physics
because it may contain terms neglected in the evaluation of the pair potential ${\bf V}$. At the same time, this quantity plays a key role in the investigation of social dilemmas.

It is worth mentioning that the cluster variation method at the level of pair approximation reproduces the exact result for the one-dimensional systems with nearest neighbor interactions. For these systems the probability distribution of strategies can be given by the Bayesian formula as
\begin{equation}\label{eq:prob1d}
p({\bf s}) = p_2(s_1,s_2) \prod_{s_x, x=2}^{x=N-1} {p_2(s_x,s_{x+1}) \over p_1(s_x)} \,,
\end{equation}
that satisfies the compatibility conditions, that is, any pair configuration probability can be reproduced by summing over the rest of sites, and the resultant entropy becomes equivalent to (\ref{eq:S2p}) for $d=1$ in the limit $N \to \infty$.

Equations from (\ref{eq:p12norm}) to (\ref{eq:prob1d}) give a mathematical basis for the use of the cluster variation method at the level of two-site approximations. This calculation becomes simpler at the level of one-site approximation (that is equivalent to the traditional mean-field approximation), when we have only one parameter ($c$) to be determined according to (\ref{eq:equilc}). The corresponding expressions for ${\cal U}$ and ${\cal S}$ can be derived with the assumption $p_2(s_1,s_2)=p_1(s_1)p_1(s_2)$ that
simplifies the entropy as
\begin{equation}\label{eq:S1p}
{\cal S} \simeq {\cal S}^{\rm (1p)}=- N  \sum_{s_1} p_1(s_1) \ln {p_1(s_1)} \,.
\end{equation}
It is emphasized that the one-site approximation ignores the role of the topological features of the connectivity structure. More precisely, only the number of co-players is involved in ${\cal U}$. This approach may give an adequate description of phenomena
for those (homogeneous) connectivity structures where the number of neighbors is large enough. For the square lattice the results of the one- and two-site approximations will be contrasted with Onsager's exact result in the following section.

The generalization of the cluster variation method for larger number of strategies and/or for larger size of clusters is straightforward. When increasing these parameters one can study a rich variety of sophisticated phenomena although the calculations become more complex and time-consuming. At the same time we can reduce the number of independent parameters by taking into consideration the additional symmetries. From this point of view the use of the corresponding three- and four-site approximations is advantageous on triangular and square lattices.

Finally we mention that the pair approximation can be applied successfully on Bethe lattices where this approach gives a more accurate prediction as illustrated by
\citet{vukov_pre06} who compared the analytical approximate results with Monte Carlo simulations performed on a locally similar random regular graph for large sizes.

In the above description we have assumed the equivalence between both the players and their location. There exist, however, several spatial structures ({\it e.g.} non-Bravais lattices) or games ({\it e.g.}, matching pennies) where we should distinguish the sites and/or players. The cluster variation method may be extended to these cases by introducing the configuration probabilities on several types of one-, two-, or $n$-site clusters by considering also the sublattice structure.

\section{ISING MODELS}
\label{sec:ising}

The investigation and application of the Ising type models have a
long history as surveyed by \citet{brush_rmp67, niss_ahes05, niss_ahes09, niss_ahes11, sornette_rpp14}. The original idea was first described by \citet{lenz_pz20} as a
simple model to study the ferromagnetism. The one-dimensional version of this lattice model was studied by \citet{ising_zp25}, who was the PhD student of Lenz. Unfortunately, the one-dimensional model is not suitable to describe the ferromagnetic-paramagnetic transition observed in magnetic material when increasing the temperature \cite{bozorth_51,mattis_65} because the domain walls, that can be considered as point defects in the one-dimensional systems, prevent the formation of long range order \cite{landau_80}.

The existence of the phase transition was shown by \citet{peierls_pcps36} using an argument later improved by \citet{griffiths_pr64}. The approximate methods (like mean-field approximation \cite{bragg_prsa34} and pair approximation \cite{bethe_prsa35}) have confirmed the presence of a continuous phase transition. The exact solution in the absence of external magnetic field on the square lattice with nearest neighbor interactions was obtained by \citet{onsager_pr44}. In the following decades this magnetic model was extensively investigated on different lattice structures assuming ferromagnetic or anti-ferromagnetic interactions. Due to its simplicity and the knowledge of the exact solution the model was frequently used to check the accuracy of different approximation techniques \cite{newell_rmp53}. Since that time the Ising models have been fundamental mathematical models in statistical physics representing a robust universality class of critical phase transitions \cite{toda_91,domb_74,kawasaki_72,stanley_71}.

The original Ising model can be directly applied to explain the magnetic behavior only for a few materials. On the other hand, the equivalent lattice gas models are widely used to derive theoretical phase diagrams for alloys \cite{kittel_04}, metal-hydrogen systems \cite{alefeld_78}, solid electrolytes \cite{dieterich_ap80}, intercalation \cite{dresselhaus_ap02} or non-stoichiometric compounds \cite{kosuge_94}, etc. The analogy to the Ising models is due to the fact that in all these models only two states of the lattice sites are distinguished. For example, in the Cu-Au alloys either Cu or Au atoms can be at one site $x$; in the palladium-hydrogen systems an interstitial void $x$ of the metal matrix can be empty or occupied by a H atom; in the superionic conductor AgI, iodide ions form a rigid body-centered cubic lattice and the smaller mobile silver ions can be present or missing in a tetrahedral void represented by the site $x$. The total energy of these systems can be built up from pair interactions between the neighboring sites in the knowledge of the physical properties.

The flexible interpretation of the Ising model implies its applicability to many other systems involving high-energy physics [for a survey see the review by \citet{pelissetto_pr02}] and social models with players located on a network \cite{galam_mjs82, krause_pa13}. The latter situation occurs when the connected individuals can choose between two options, {\it e.g.}, using meter or yard as unit of length; following drive left or drive right rule in traffic; using Windows or Linux operating systems, {\it etc}. All these examples represent the so-called coordination game and are analogous to ferromagnetic systems \cite{galam_mjs82, herz_jtb94, lee_ih_res00, brock_res01, weisbuch_pa07, galam_pa10, grauwin_acs11}.

\subsection{Systems equivalent to Ising models}
\label{sec:mtoI}

In the magnetic Ising model the spin variables $\sigma_x=+1, -1$ refer to upward and downward magnetic moments at site $x$ of a network. The strength of the spin-spin interaction between the neighboring sites $x$ and $y$ is denoted by $J_{xy}$ and we can assume the presence of a site-dependent external magnetic field $h_x$. For any microscopic state ${\bf \sigma}=(\sigma_1, \ldots , \sigma_N)$ the Hamiltonian (or potential energy) function of this multi-spin system is given as
\begin{equation}\label{eq:hising}
H({\bf \sigma})=-\sum_{\langle x,y \rangle} J_{xy} \sigma_x \sigma_y - \sum_{x} h_x \sigma_x \,
\end{equation}
where the summation runs over all nearest neighbor pairs (denoted by $\langle x,y \rangle$ as it is done for the multi-player games). For ferromagnetic interactions the so-called coupling constants are positive ($J_{xy}>0$) and the system achieves one of the ordered ground states with minimal energy in the ferromagnetic phase ($\sigma_x=+1$ at each site if $h_x>0$ or $\sigma_x=-1$ if $h_x<0$. A similar optimal strategy distribution occurs in a multi-agent system with coordination type pair interactions if one of the two strategies is preferred by suitable self-dependent payoffs.

In order to clarify the relationship between the Ising models and the multi-agent, two-strategy, potential games first the site energies are shared among the pair interactions as
\begin{equation}\label{eq:hsharing}
h_x=\sum_{y^{\prime}} h^{\prime}_{x}(y^{\prime}) ,
\end{equation}
where the summation runs over $y^{\prime}$ which are the interacting neighbors of $x$.
In that case the Hamiltonian is composed of pair interactions involving the corresponding contributions of the site energies from both sites in the following way
\begin{equation}\label{eq:hisingpair}
H({\bf \sigma})=\sum_{\langle x,y \rangle} H_{xy}(\sigma_x,\sigma_y)
\end{equation}
where
\begin{equation}\label{eq:Hxy}
H_{xy}(\sigma_x,\sigma_y)= -J_{xy}\sigma_x \sigma_y -
h^{\prime}_x(y) \sigma_x - h^{\prime}_y(x)\sigma_y
\end{equation}
that may also be written as
\begin{equation}
H_{xy}(\sigma_x,\sigma_y)= -J_{xy}\sigma_x \sigma_y -
{h^{\prime}_x(y)+h^{\prime}_y(x) \over 2} (\sigma_x +\sigma_y) -
{h^{\prime}_x(y)-h^{\prime}_y(x) \over 2} (\sigma_x -\sigma_y).
\label{eq:Hxynonsym}
\end{equation}
The latter form of the pair interactions is convenient for the comparison with the potential of a nonsymmetric $2 \times 2$ game given by Eq.~(\ref{eq:pot_bmg_decomp}) because the first term is analogous to the coordination type interaction as the four possible values of the product $\sigma_x \sigma_y$ define a matrix equivalent to ${\bf f}(4)$. Similarly, in matrix notation the second (third) term of (\ref{eq:Hxynonsym}) becomes equivalent to the second (third) term of the expression (\ref{eq:pot_bmg_decomp}) with a suitable choice of the coefficients. Using this analogy a multi-agent, two-strategy, potential game can be mapped onto an Ising model on the same network with an adequate choice of the parameters $J_{xy}$ and $h^{\prime}_x(x^{\prime})$ whereas the local site energy is defined by (\ref{eq:hsharing}). More than one game can be mapped onto the same generalized Ising model by varying the game parameters under the conditions of (\ref{eq:hsharing}).

In a human system it is natural to assume distinct payoffs for each pair, therefore in the corresponding Ising models both the coupling constants $J_{xy}$ and the local magnetic fields $h_x$ can be considered as random parameters. The resulting random Ising models or spin glasses will be discussed briefly in Sec.~\ref{sec:spdsu}.

The traditional Ising model was introduced to study solid state phenomena where the interacting particles are identical and the applied magnetic field is homogeneous, consequently $J_{xy}=J$ and $h_x=h$, and the connectivity network is a translation invariant lattice where each site has $z$ neighbors. Similar conditions can be satisfied for multi-agent evolutionary games with equivalent interactions between the players residing on the sites of the same lattice. In that case the local site energy is shared equally among the $z$ pair interactions becoming
\begin{equation}\label{eq:Hxytrad}
H_{xy}(\sigma_x,\sigma_y)= -J\sigma_x \sigma_y - {h \over z} (\sigma_x + \sigma_y).
\end{equation}
For the quantitative relationship now we compare $-H_{xy}$ with the pair potential (\ref{eq:socdilV}) obtained in the notation (\ref{eq:socdilG}) of social dilemmas.
Drawing a parallel between these two systems, the $\sigma_x=+1$ ($\sigma_x=-1$) spin state corresponds to the strategy $s_x=D$ ($s_x=C$). For this convention one can deduce the following relationship between the parameters of the Ising model and social systems:
\begin{equation}
J={1-T-S \over 4} \quad \mbox{and} \quad {h \over z}= {T-1-S \over 4},
\label{eq:Jh_ST}
\end{equation}
that is, the Ising type spin-spin interaction is analogous to the coordination game whereas the magnetic field plays the role of the driving force validating the risk dominance.

\begin{figure}[ht]
\centerline{\epsfig{file=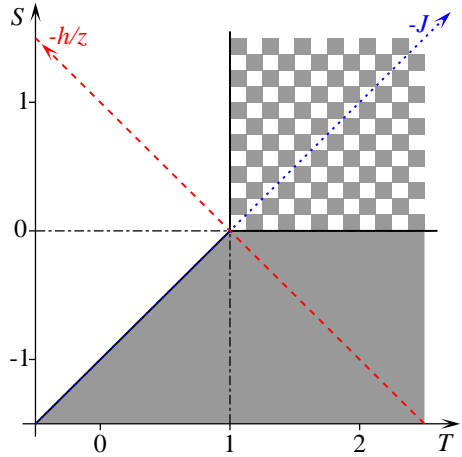,width=6cm}}
\caption{(Color online) Mapping the Ising parameters $J$ and $h$ onto the $S-T$ plane of payoffs in the notation of social dilemmas defined by Eq. (\ref{eq:socdilG}).}
\label{fig:isi_map}
\end{figure}

These linear relations determine the values of $J$ and $h$ as a function of the payoff parameters $T$ and $S$. Figure \ref{fig:isi_map} shows the orthogonal Descartes coordinate axes for the $h/z$ and $J$ (denoted by red dashed and blue dotted lines) on the $S-T$ plane we used in Sect. \ref{sec:proppot} for the classification of games according to the flow graphs and Nash equilibria.
Now the $S-T$ (or $h-J$) plane is divided into three regions with respect to the thermodynamically stable states of the Ising model in the low noise limit. Within the white territory a ferromagnetic, {\it i.e.}, homogeneous spin down ($\sigma_x=-1$), is stabilized that refers to uniform cooperation in games (where $T<1$ and $S>T-1$) involving the harmony games and the upper half of the stag hunt games. The opposite ferromagnetic state ($\sigma_x=+1$ or $s_x=D$) occurs within the dark region ($S<0$ and $T>S+1$) representing the second half of the stag-hunt games and the prisoner's dilemmas.

Within the third region ($T>1$ and $S>0$) an anti-ferromagnetic spin arrangement is realized when the neighboring spins point to opposite directions. On the square lattice the up and down spins form a chessboard like pattern which we used in Fig.~\ref{fig:isi_map} for the illustration of this spatial structure. This anti-ferromagnetic spin arrangement is twofold degenerated as the simultaneous reversal of all spins does not affect the total energy (\ref{eq:hising}) even in the presence of $h$. On the other hand, the application of a magnetic field $h$ influences the local stability of spin states and this is the reason why a ferromagnetic phase is stabilized if $|h|$ exceeds a threshold value when $J<0$, as it is also discussed for games \cite{szabo_jtb12}. These thermodynamically stable (ordered) states are realized by the logit rules in the limit $K \to 0$ and coincide with the preferred pure Nash equilibria determined by the strategy pair $(i,j)$ for which $V_{ij}$ is maximal.

In the present system the highest average (total) income can be achieved by uniform cooperation if $2R>T+S$, otherwise the sublattice ordered arrangement of the $C$ and $D$ strategies provides the highest total payoff. Despite this total payoff optimum the system falls into the state of "tragedy of the community" within the dark territory indicated in Fig.~\ref{fig:isi_map}. In the zero noise limit the latter structure occurs for $h<0$, except the sublattice ordered region where the sufficiently strong repulsive interactions prevent the formation of a homogeneous state. In some sense the negative magnetic field can be interpreted as a force driving the social system into the state of the tragedy of the community.

Notice furthermore that the case of $J=(1-S-T)/4=0$ refers to the absence of interactions between the neighboring spins. In such situations the independent spin reversals are controlled only by the magnetic field $h$ and the system is equivalent to the collection of non-interacting spins that can be described analytically when considering only a single spin. The corresponding games are called ''equal-gains-from-switching'' \cite{nowak_aam90} or ''dummy'' games \cite{facchini_td97} in the literature. The donation game \cite{sigmund_10} represents one of the well investigated examples.

The analytical study of the anti-ferromagnetic spin arrangements requires the division of the lattice into two sublattices ($x \in X$ and $y \in Y$) in a way discussed in Sec.~\ref{sec:mppg}. Additionally, the concept of the Ising model was extended by introducing staggered magnetic fields that are uniform within a sublattice. Thus, $h_x=h+h_s$ and $h_y=h-h_s$ where $h_s$ defines the strength of the staggered magnetic fields. Now the pair interaction
\begin{equation}\label{eq:Hxystag}
H_{xy}(\sigma_x,\sigma_y)= -J\sigma_x \sigma_y - {h \over z} (\sigma_x + \sigma_y)
- {h_s \over z} (\sigma_x - \sigma_y).
\end{equation}
has a third term quantifying the effect of the staggered magnetic field that favors one of the sublattice ordered spin arrangements. The Ising model with staggered magnetic field is equivalent to an evolutionary potential game on the same bipartite network.
Comparison of the matrix notation of Eq.~(\ref{eq:Hxystag}) with the $2 \times 2$ potential matrix (\ref{eq:pot_bmg_decomp}) explains that the staggered  magnetic field
\begin{equation}\label{eq:hstageq}
{h_s \over z}= \alpha^{\prime}(7) ,
\end{equation}
where $z$ denotes the number of neighbors in the translation invariant connectivity structure.

The introduction and application of the staggered magnetic field have no practical importance in the investigation of the magnetic materials. This quantity becomes relevant from theoretical point of view when justifying the equivalence of order-disorder transitions observed for the ferromagnetic and anti-ferromagnetic Ising models when varying the temperature. To be more precise, the effect of $h_s$ on the ferromagnetic state ($J>0$ and $h=0$) is equivalent to the effect of $h$ on the anti-ferromagnetic state.

Finally we mention that the sublattice-dependent local site energy plays an important role in those lattice gas models where two types of sites are distinguished. For example, in the crystals of Pd-H system \cite{alefeld_78}, CaF$_2$ \cite{dieterich_ap80}, and alkali-fullerides \cite{udvardi_jpc96} the mobile atoms/ions can stay both in the tetrahedral and octahedral sites of the face-centered cubic lattice formed by the rigid components.

\subsection{Potts models}
\label{sec:potts}

The two-state Ising model without magnetic field was extended by \citet{potts_mpcps52} who introduced a lattice model with $n$ ($n \ge 3$) equivalent states at each site. Previously \citet{ashkin_pr43} studied a particular four-component version of the Ising model on a square lattice. The lattice model with $n$ states for a common interaction energy between the different local states was also suggested by \citet{kihara_jpsj54}. The name of Potts model was proposed by \citet{domb_74a} and used worldwide since that time. For a comprehensive survey of the early results we can suggest consulting the review by \citet{wu_fy_rmp82}.

In the terminology of game theory the Potts models correspond to multi-agent games with symmetric $n$-strategy pair interactions where the payoff matrix and potential are given by $n \times n$ unit matrices, that is, both the payoffs and potential matrices ${\bf V}$ can be expressed by Kronecker's delta as
\begin{equation}\label{eq:pottsm}
{A}_{ij}={B}_{ij}= {V}_{ij}(x,y)= J \delta_{ij},
\end{equation}
where $i,j=1, \ldots , n$. For $J>0$ this pair interaction is equivalent to a generalized coordination game with $n$ equivalent Nash equilibria when the players choose the same option. This game involves all the ${\bf d}(p,q)$ [$1\le p < q \le n$; see Eq.~(\ref{eq:isingsubgames})] coordination type interactions with the same strength.

In the literature of physics the Potts model is extended by introducing an external magnetic field ($h>0$) or a self-dependent component preferring one of the strategies. The presence of this term suppresses the critical transition in a way as it occurs in the Ising model.

The main features of the corresponding ferromagnetic Potts model are well described in the field of statistical physics. This set of models is used for the classification of the universal behaviors appearing in the order-disorder transitions when the temperature (noise) is increased. Due to its simplicity the investigation of the Potts model played an important role in the exploration of the critical phase transitions and it became a key model when testing different methods and approaches. It turned out that the critical behavior of the Potts model is richer and more general than that of the Ising model.

Originally the Potts model was considered as the simplest mathematical model exhibiting an order-disorder phase transition for $n>2$ while its kinetic versions allow the investigations of the time-dependent ordering processes. In the light of the robustness and universality of critical phase transitions it was later realized that in many substances the phase transitions can be interpreted by the application of various Potts models. For example, three-fold degenerated ordered structures can be formed by atoms adsorbed on single crystal surfaces (for $1/3$ coverage) \cite{alexander_pla75, domany_jap78} or by mobile ions in layered solid electrolytes ({\it e.g.} silver $\beta$-alumina) \cite{gouyet_jpl80}. A large variety of the experimental realizations of the two-dimensional, $n=4$ Potts model was discussed by \citet{domany_prb78}.

In Sect.~\ref{sec:sosl} we will discuss briefly a lattice gas model for illustrating the emergence of equivalent ordered structures on a square lattice. The general and universal properties of the order-disorder phase transitions for the $n$-state Potts model will be detailed in Sec.~\ref{sec:cplat} while the most relevant features of the ordering process are surveyed briefly in Sec.~\ref{sec:ordproc}.

\section{FEATURES OF ISING MODELS}
\label{sec:pfkim}

Despite its simplicity the Ising model can be used to demonstrate a surprisingly wide range of phenomena including different types of ordered structures in the stationary state when varying the range and strength of interactions on lattices or graphs. Besides it, the kinetic Ising model is capable of illustrating a large number of dynamical processes how they evolve towards the final stationary state. Now we briefly survey the most relevant phenomena that can help us understand the behavior of the more general evolutionary games on graphs.

\subsection{Spontaneous symmetry breaking}
\label{sec:ssb}

In the absence of a magnetic field $h$ the ferromagnetic Ising model with nearest neighbor interactions on the square or cubic lattices has two equivalent ordered states that appear with the same probability according to the Boltzmann distribution (\ref{eq:boltzmann}). Similarly, the Boltzmann distribution predicts the presence of the two equivalent anti-ferromagnetic ordered structures with the same probability even in the presence of a magnetic field if its strength does not exceed a threshold value. On the contrary, in the practice we see only one of the ordered structures in sufficiently large systems.

\begin{figure}[ht]
\centerline{\epsfig{file=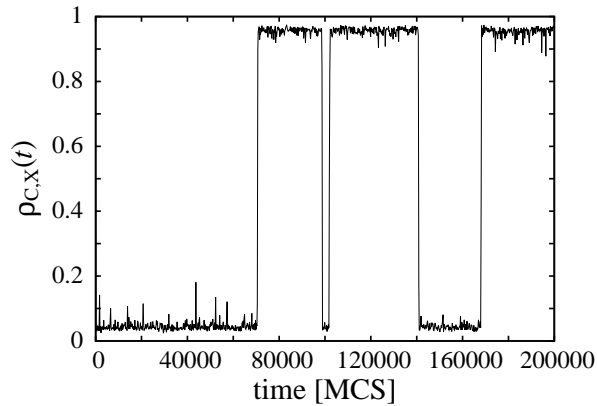,width=7.8cm}}
\caption{Time-dependence of the $C$ strategy frequency in the sublattice $X$ for the hawk-dove game on square lattice with nearest neighbor interactions. The fluctuations are smoothed by averaging over 100 MCS. The Monte Carlo simulation is performed for $T=1.5$, $R=1$, $S=0.5$, $P=0$, $K=0.5$, and $L=10$.}
\label{fig:hd_t}
\end{figure}

\begin{figure}[ht]
\centerline{\epsfig{file=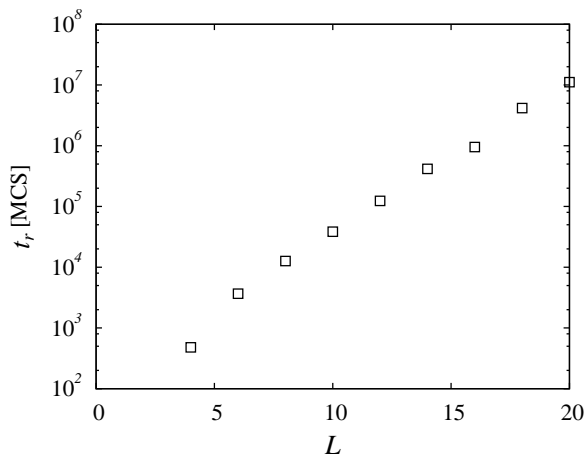,width=7.8cm}}
\caption{MC data for the size dependence of the average reversal time obtained on a square lattice for $T=1.5$, $R=1$, $S=0.5$, $P=0$, and $K=0.5$.}
\label{fig:rt_l}
\end{figure}

To resolve the above discrepancy Fig.~\ref{fig:hd_t} shows a typical time-dependence of the frequency of $C$ strategy as a function of time in one of the two sublattices of a small square lattice for the hawk-dove region at a low noise level when the sublattice ordered strategy arrangements are favored.
Figure~\ref{fig:hd_t} illustrates that the system alternates between two "ordered" states and the transition time is significantly shorter than the average residence time ($t_{r}$) the system stays in the vicinity of one of the ordered states. The value of the average residence time can be estimated by counting the state reversals during a sufficiently long Monte Carlo simulation that gives $t_{av} \simeq 3 \times 10^4$ MCS for the parameters used for results plotted in Fig.~\ref{fig:hd_t}. Figure~\ref{fig:rt_l} shows that $t_{av}$ increases exponentially with the linear size for the values of payoffs and noise level given in the caption of Fig.~\ref{fig:hd_t}. Consequently, for sufficiently large sizes we may find the system staying in one of the macroscopic states during the whole sampling time that may exceed many years or even the age of the Universe. The selection of one of the "stationary" states depends on the initial conditions and the sequence of random numbers used in the simulation.
Although the average residence time depends on other parameters (payoffs and noise) the spontaneous symmetry breaking described above remains valid for other values as well as for many other systems in the limit $N \to \infty$. This feature is contrary to the prediction of the Boltzmann distribution suggesting the presence of both types of ordered microscopic states with equal probability. In other words, the infinite system is nonergodic because the time and ensemble averages would give different results below the critical transition point for $h=0$. An identical behavior occurs for the anti-ferromagnetic lattice models (for $h_s=0$) as well as for the Potts models. This feature has rigorously been studied for a long time in the theory of stochastic phenomena \cite{liggett_85}.

It is emphasized that the above mentioned discrepancy vanishes if the equivalence of the ordered structures is broken by applying $h \ne 0$ for the ferromagnetic Ising models. In that case it is convenient to consider the zero-field magnetization as an average value in the limit $h \to +0$. A similar trick can be applied for the anti-ferromagnetic Ising model ($h_s \to +0$) and also for the Potts models.

\subsection{Mean-field theory}
\label{sec:meanfieldtehory}

In most of the cases the mean-field theory gives an adequate qualitative prediction about the system behavior. \citet{bragg_prsa34} suggested expressing the contribution of interactions via the introduction of an effective magnetic field $h_{eff}$ characteristic to the average magnetization ($m=\langle \sigma_{\bf x} \rangle$) in translation invariant systems. For a $d$-dimensional hyper-cubic lattice
\begin{equation}\label{eq:heff}
h_{eff}=z J m \,
\end{equation}
where $z=2d$ is the number of interacting neighbors.

\begin{figure}[ht]
\centerline{\epsfig{file=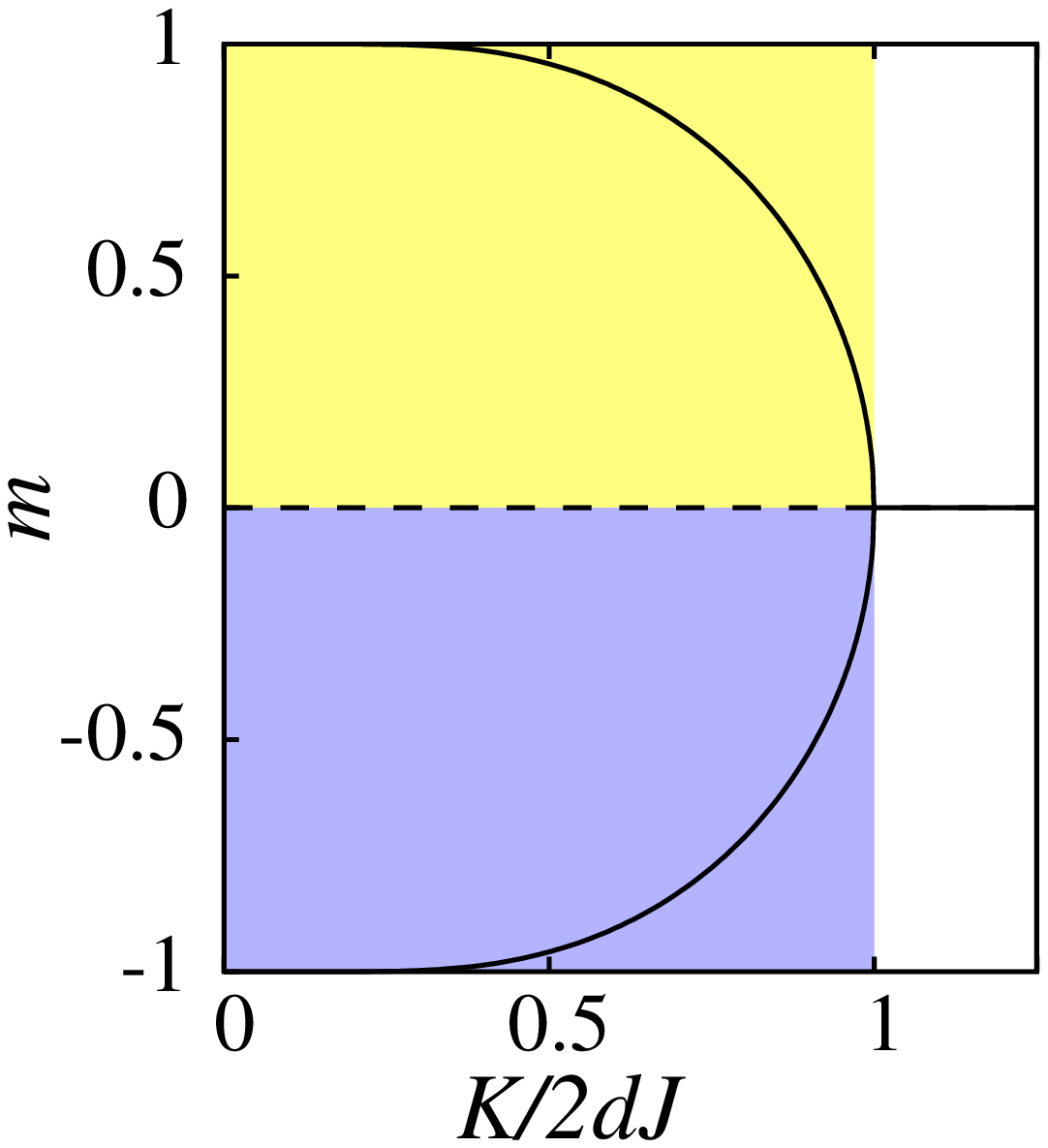,width=4cm}\epsfig{file=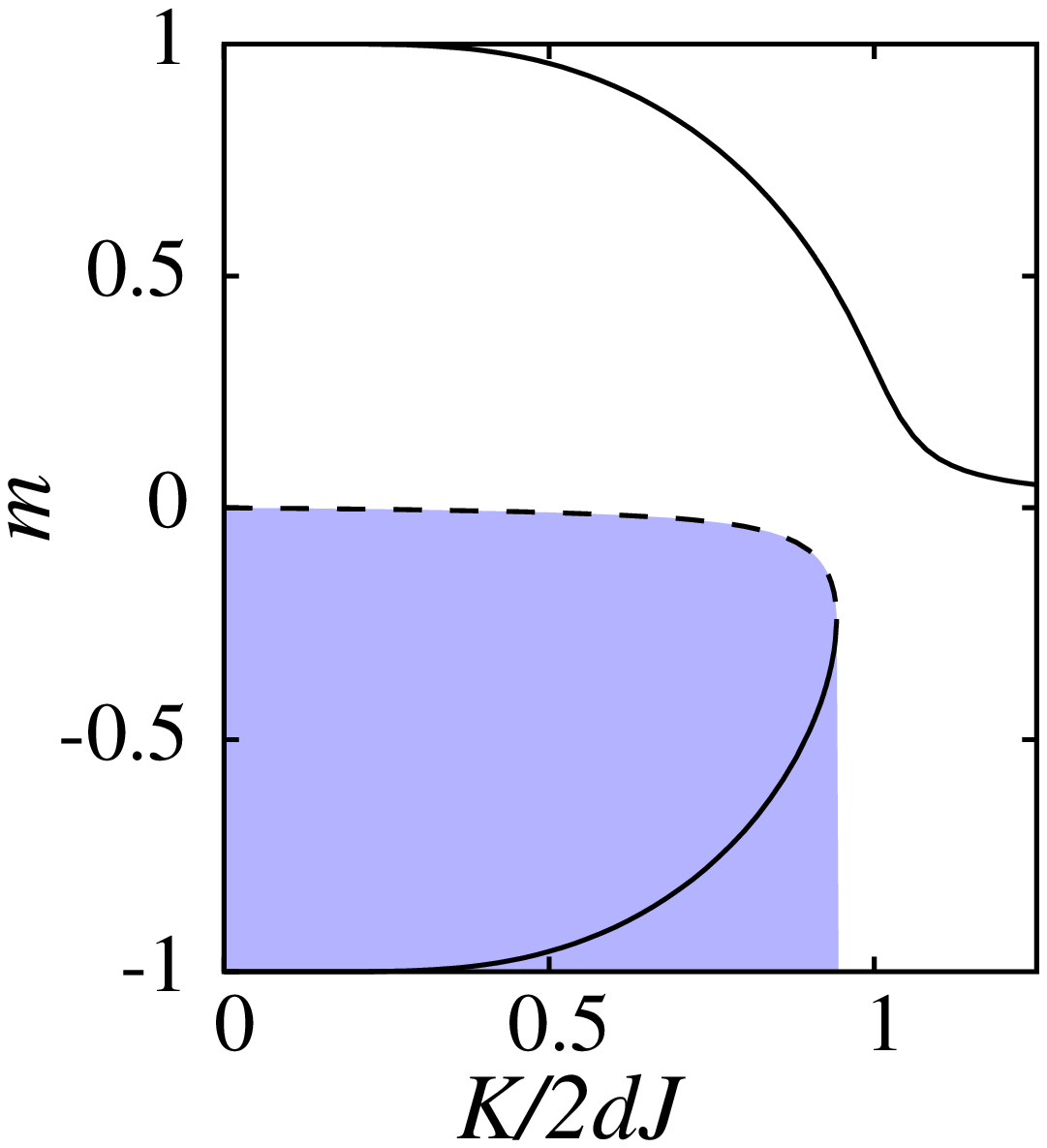,width=4cm}}
\caption{(Color online) Prediction of mean-field theory for the magnetization $m$
{\it vs.} $K/2dJ$ for the Ising model at $h=0$ (left) and $h=0.01$
(right). Solid (dashed) lines denote attractor (repellor)
solutions. Colored regions refer to three basins of attraction.}
\label{fig:ising_mf}
\end{figure}

In the presence of a magnetic field $h+h_{eff}$ the average magnetization for a single spin is determined by the Boltzmann distribution as
\begin{equation}
m=\tanh{[(h+h_{eff})/K]} = \tanh{[(h+2dJm)/K]} \,.
\label{eq:m_mf}
\end{equation}
This implicit equation has a trivial solution $m=0$ if $h=0$. This (paramagnetic) solution is the only solution above a critical noise level ($K>K_c=2dJ$). In the opposite cases ($K<K_c$) two additional equivalent (symmetric) solutions are found as illustrated on the left plot of Fig.~\ref{fig:ising_mf}.
Both solutions tend towards a homogeneous state ($m(K)=\pm 1$) and $m^2(K)$ vanishes linearly when $K \to K_c$ from below. More precisely,
\begin{equation}\label{eq:mcr_mf}
m(K) \simeq {3(K_c-K) \over K_c}^{1/2} \,
\end{equation}
in the close vicinity of $K_c$.

In the presence of an external magnetic field ($h>0$) the typical solutions are illustrated in the right plot of Fig.~\ref{fig:ising_mf}. Notice that Eq.~(\ref{eq:m_mf}) has also three solutions at low values for $K$ while in the opposite limit the solution becomes unique. Contrary to the zero-field case now there is a solution that varies smoothly when $K$ is increased from zero to $\infty$.

It is emphasized that the thermodynamically stable solution is distinguished by the extremum principle at the level of one site approximation. In that case the one-site probabilities are expressed as $p_1(+1)=(1+m)/2$ and $p_1(-1)=(1-m)/2$ and from the resulting thermodynamical potential $\Phi(m)$ Eq.~(\ref{eq:m_mf}) can be deduced by the extremum condition $\partial \Phi (m) / \partial m =0$. Direct comparison of the values of $\Phi(m)$ for the different solutions justifies the preference of the upper (thicker solid) curve in the right plot of Fig.~\ref{fig:ising_mf}.

The dynamical mean-field equation can also be used to determine the dynamical stability of the above solutions. The corresponding equation of motion summarizes the contributions of spin reversals for the Glauber type evolutionary rules, that is,
\begin{equation}
{\partial m \over \partial t} = -{1+m \over 2}{1 \over 1+e^{(4dJm+2h)/K}}+
{1-m \over 2}{1 \over 1+e^{-(4dJm+2h)/K}} .
\label{eq:m_dmf}
\end{equation}
Evidently, in the stationary state this equation becomes equivalent to Eq. (\ref{eq:m_mf}). At the same time Eq.~(\ref{eq:m_dmf}) allows us to determine the basin of attractor for each solution as denoted in Fig.~\ref{fig:ising_mf} by different colors. Namely, if the system is started from a state with a magnetization $m(0)$ then $m(t)$ evolves vertically towards the corresponding attractor.  For high noise levels the system always develops into the only stationary solution. For $h=0$ and below the critical noise level ($K<K_c$), however, the final state depends on the initial state, that is, if $m(t=0)>0$ then the system develops towards the positive $m(K)$ stationary solution. For $K<K_c$ the trivial solution $m=0$ is a separatrix. In the presence of an external field the separatrix is distorted as illustrated in Fig.~\ref{fig:ising_mf}. Consequently, when varying the initial magnetization the dynamical mean-field theory allows the system to remain in both ferromagnetic states contrary to the extremum principle favoring only one showing continuous variation in $m(K)$.

Thermodynamically unstable (or meta-stable) states can be observed very frequently in the nature. For example, after a suitable variation of temperature (or other thermodynamical quantities) materials can evolve into (or remain in) a meta-stable phase like the supercooled liquids and gases. The disordered phase in alloys (cooled down fast from a high temperature) remains present for a long time despite the fact that phase segregation is favored thermodynamically at low temperatures ({\it e.g.}, Cu-Au alloys) \cite{kittel_04}. The magnetic hysteresis represents another example where the reversal of a weak magnetic field is not accompanied directly by the reversal of $m$. Similar phenomena are present in biological and social systems when varying the payoff parameters or dynamical rules in time \cite{wang_wx_pre08, wolpert_pre12, hua_dy_cpb13}.

In the light of the above phenomena the results of the mean-field approach can be interpreted as a message that the mean-field conditions do not support the system to achieve the thermodynamically stable state. Such a situation can occur in a social system where the players select $z$ co-players at random from the whole population and determine their strategies in the spirit of the Glauber dynamics. Later we will show that for short-range interactions the system can evolve into the thermodynamically stable state due to the enhanced role of fluctuations.

In fact, the main shortage of the mean-field theory is related to neglecting the fluctuations. This is not dangerous when there are a large number of neighbors (here for $z=2d \gg 1$) and the law of the large numbers ensures small variance in $h_{eff}$. This is the reason why mean-field type behavior is expected in many spatial multi-player systems if the spatial dimension $d$ exceeds a critical value, that is, if $d>d_c$.

In a low-dimensional system the fluctuations affect significantly the system behavior. Due to this effect the prediction of mean-field theory is wrong in the one-dimensional systems (with short-range interaction) that do not have long-range ordered arrangements at finite temperatures. Otherwise the mean-field theory also gives a qualitatively good picture about the phase transitions in more complicated systems.

\subsection{Series expansions and duality}
\label{sec:duality}

The prediction of mean-field theory can be improved by the application of low and high noise series expansions allowing us to extract additional properties of the Ising model. Now we study the traditional ferromagnetic Ising model without magnetic field ($h_x=0$) on a square lattice with uniform interactions $J_{xy}=J$ between the nearest neighbors in order to illustrate the duality that is an inherent symmetry of this system.

In the low noise limit ($K \to 0$) the partition function [see Eq.~(\ref{eq:boltzmann_Z})] can be approximated by summing the contribution of those states that are present with the highest probabilities for a finite $N$ as
\begin{equation}
Z=e^{2NJ/K} [1 + N e^{-4 \cdot 2 J/K}
+ 2N e^{-6 \cdot 2 J/K} + \cdots ] \,.
\label{eq:lnserexp}
\end{equation}
Here the first term gives the contribution of the ordered 'up spins' state ($\sigma_x=+1$); the second one comes from states where only a single spin is reversed inside the ordered arrangement, and the third term summarizes the contribution of states where only two nearest-neighboring spins are flipped. The argument of the corresponding exponential functions reflects the interfacial energy of the island of spins reversed.

Similar expressions can be evaluated for other lattices even in the presence of a homogeneous magnetic field. The resulting expression can be used to derive analytical approximations for thermodynamical quantities ({\it e.g.} magnetization, energy, specific heat) in the low noise/temperature limit.

\begin{figure}[ht]
\centerline{\epsfig{file=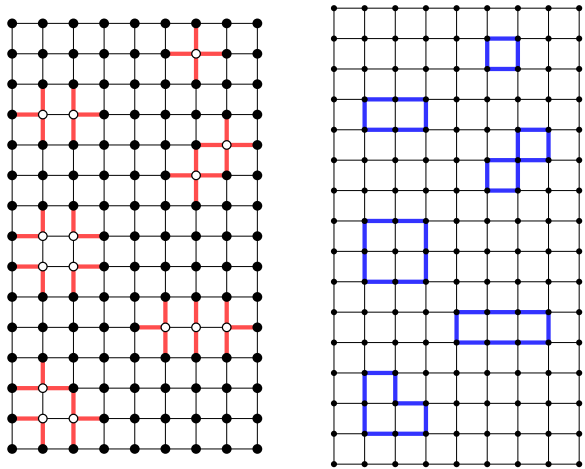,width=8.0cm}}
\caption{(Color online) Leading pairs of contributions to the partition function for the low and high-noise series expansions on the square lattice. For the low noise case (left) islands of down spins are indicated by open bullets in the sea of up spins (closed bullets) while thick (red) edges denote spin-pairs contributing to the interfacial energy. In the high noise limit (right) thick (blue) edges denote loops of neighboring spin-pairs giving non-zero contribution when summing over the spins.}
\label{fig:dualconf}
\end{figure}

The high noise series expansion is based on the following transformation suggested by \citet{waerden_zp41}. Accordingly, the contribution of a spin-pair to the Boltzmann factor is divided into two parts as
\begin{equation}\label{eq:vanderwaerden}
e^{{J \over K} \sigma_x \sigma_y} =  \cosh{(J/K)}[1 + \sigma_x \sigma_y \tanh {(J/K)}] \,
\end{equation}
where the second term goes to zero if $K \to \infty$. Using this expression the partition function obeys the following form:
\begin{equation}
Z = \sum_{{\bf \sigma}} \prod_{\langle x,y \rangle } e^{{J \over K} \sigma_x \sigma_y} = [\cosh{(J/K)}]^{2N} \sum_{\bf \sigma} \prod_{\langle x,y \rangle } [1 + \sigma_x \sigma_y \tanh {(J/K)}]
\label{eq:Zhighnoise}
\end{equation}
where the product runs over all possible nearest neighbors. The product can be expanded out into $2^{2N}$ terms, where most of them [{\it e.g.} $\sigma_x \sigma_y \tanh {(J/K)}$ or $(\sigma_x \sigma_y) (\sigma_{x^{\prime}}\sigma_{y^{\prime}}) \tanh^2{(J/K)}$] have vanishing contribution to the partition function after the sum is over the spin configurations. Exceptions are those products of spin pairs where each $\sigma_x$ has an exponent of 2 or 4 when their value is one for all the $2^N$ spin configurations. The latter constellations of different pairs have a general geometrical feature, namely, the connected $\sigma_x \sigma_y$ nearest neighbors form a closed loop (or collection of loops) as illustrated in Fig.~\ref{fig:dualconf}.

As a result, the leading contribution to the partition function can be written as
\begin{equation}
Z=2^N[\cosh{(J/K)}]^{2N}[1 + N [\tanh{(J/K)}]^{4} + 2N [\tanh{(J/K)}]^{6} + \cdots ]\, .
\label{eq:hnse}
\end{equation}
Comparing the expressions between the rectangular brackets in Eqs.~(\ref{eq:lnserexp}) and (\ref{eq:hnse}) indicates their similarity that remains valid for arbitrary high order of the approximations as justified by combinatorial methods detailed in the reviews by \citet{wannier_rmp45}, \citet{newell_rmp53}, and \citet{domb_ap60}. This feature serves as a basis for the so-called duality transformation mapping the low-noise behavior at $J/K$ onto a high noise Ising system with $J/K^{\prime}$ on the square lattice if
\begin{equation}\label{eq:duality}
e^{-2J/K}=\tanh{(J/K^{\prime})} \, .
\end{equation}
This formula reflects that the square lattice Ising model is self-dual, that is $e^{-2J/K^{\prime}}=\tanh{(J/K)}$. In short, the dual of the dual is the original system. This symmetry was exploited by \citet{kramers_pr41} in the first exact evaluation of the critical temperature.

Before going to the discussion of the critical transition we briefly mention several properties related to the high-noise series expansions or duality. First we emphasize that the concept of duality remains valid for other two-dimensional lattices. That means, for example, that the low-noise behavior on a triangular lattice can be mapped into the high noise behavior on the honeycomb lattice as detailed in the above mentioned reviews. Notice furthermore that both the high- and low-noise series expansions can be performed when the coupling constant depends on $x$ therefore it involves the possibility that the low noise behavior of an Ising model with inhomogeneous coupling constant can be mapped onto another inhomogeneous Ising model at high noises (disregarding the irrelevant prefactors in Eqs. (\ref{eq:lnserexp}) and (\ref{eq:hnse})) \cite{wegner_jmp71}.

Notice furthermore, that the high-noise series expansion reflects the relevance of loops
formed by interacting spin pair throughout the connectivity structure. So this approach indicates directly the absence of corrections for the loop-free networks, namely, on the one-dimensional lattice with open boundary or on the tree-like structures including Cayley trees. As a result, the partition function of the Ising model on loop-free networks can be given as
\begin{equation}\label{eq:Zontree}
Z=2^N[\cosh{(J/K)}]^{N_l} \,
\end{equation}
where $N_l$ is the number of links (connected spin pairs) in the given structure. For example, $N_l=N-1$ for the one-dimensional lattice where the vanishing correction ($[2\sinh{J/K}]^N$) of the periodic boundary condition can also be evaluated as it comes from the only loop containing all the sites/edges. Notice that $N_l=N-1$ for all the tree-like structure (independent of the degree distribution).

\subsection{Critical phase transitions on lattices}
\label{sec:cplat}

In the quantitative analysis of the critical phase transition(s) from the disordered state to the ordered one, several quantities play fundamental roles \cite{fisher_pr67, stanley_71}. For example, in a ferromagnetic system the average magnetization $m$ is defined as
\begin{equation}\label{eq:op}
m={1 \over N}  \sum_x \langle \sigma_{\bf x} \rangle_{h \to +0}
\end{equation}
where the sum runs over all sites ${\bf x}$ of the lattice. In general $m$ is considered to be an order parameter where $\langle \cdots \rangle_{h \to +0}$ denotes long-time average in the stationary state in the presence of an external magnetic field with strength $h \to +0$. For any finite value of $h$ the probability of the states of disfavored magnetization is suppressed in the Boltzmann distribution in the limit $N \to \infty$. Using this approach we can avoid the difficulties related to the nonergodicity mentioned above.

In the equivalent evolutionary potential games the above order parameter can also be used for the quantitative analysis of the ordering by substituting $\sigma_x=+1$ ($\sigma_x=-1$) if $s_x=D$ ($s_x=C$).

For the analysis of an anti-ferromagnetic system the lattice sites are divided into two disjoint sublattices (${\bf x} \in X$ and ${\bf y} \in Y$) as discussed in Sec.~\ref{sec:mppg}. This anti-ferromagnetic system can be transformed into a ferromagnetic model by substituting $J \to -J$ and $\sigma_y \to -\sigma_y$ $\forall {\bf y} \in Y$. This transformation is accompanied by a transfer of the homogeneous magnetic field $h$ into a staggered magnetic field $h_s$ and {\it vice versa}. In the absence of magnetic fields this transformation explains the equivalence of the corresponding order-disorder transitions.

For these sublattice ordered spatial structures the anti-ferromagnetic order parameter is usually defined as
\begin{equation}\label{eq:opaf}
m={2 \over N} \left[ \sum_{{\bf x} \in X} \langle \sigma_{\bf x} \rangle_{h_s \to +0} -
\sum_{{\bf y} \in Y} \langle \sigma_{\bf y} \rangle_{h_s \to +0} \right] \,
\end{equation}
where the averages are evaluated in the limit $h_s \to +0$. Henceforth we will not denote these assumptions.

\begin{figure}[ht]
\centerline{\epsfig{file=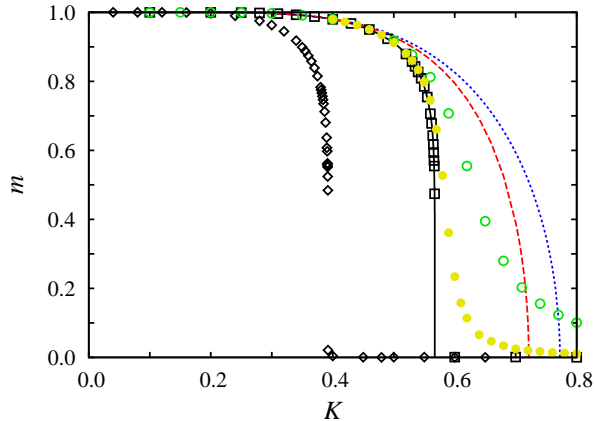,width=7.8cm}}
\caption{(Color online) Monte Carlo results for the order parameter $m$ as a function of noise for evolutionary potential games with Glauber dynamics on square lattice at several values of payoff parameters: in the notation of social dilemmas:  $T=1.5$, $S=0.5$ (boxes); $T=1.4$, $S=0.3$ (diamonds); $T=0.5$, $S=-0.49$ (open circles); and $T=0.5$, $S=-0.499$ (closed circles). The exact solution for the corresponding Ising model is denoted by a solid line. Dotted (blue) and dashed (red) lines illustrate the prediction of the cluster variation method for the levels of one- and two-site approximations.}
\label{fig:op_cp}
\end{figure}

In Fig.~\ref{fig:op_cp} Monte Carlo data of the order parameter are indicated by symbols when varying the noise level (temperature) for an evolutionary social dilemma game with nearest neighbor interactions on the square lattice. Data obtained for Glauber dynamics at parameters $T=1.5$ and $S=0.5$ are identical to the case of an anti-ferromagnetic Ising model without the external magnetic fields ($h=h_s=0$) for which the exact solution was obtained by \citet{onsager_pr44} as
\begin{equation}\label{eq:onsager}
m=\left[1-[\sinh{(\ln (1+\sqrt{2}){K_c \over K})} ]^{-4}
\right]^{1 \over 8} \,,
\end{equation}
where $K_c=2J/\ln{(1+\sqrt{2})}$. The results illustrate clearly how the system undergoes an order-disorder transition at a critical noise level. As mentioned above, this value of $K_c$ can also be evaluated from the duality relation by substituting $K=K^{\prime}=K_c$ into (\ref{eq:duality}). For low noises ($K<K_c$) the order parameter varies from 1 monotonously and continuously to zero when approaching the critical point and remains $m=0$ if $K>K_c$. Similar transition can be observed for other evolutionary games (where $T-1=S$) corresponding to the zero-field Ising model. More precisely, the $m(K/K_c)$ functions coincide and $K_c$ is proportional to $|T-1+S|$.

It is well known that for the anti-ferromagnetic Ising model the main features of the critical transition remain unchanged in the presence of an external magnetic field $h$ whereas the critical noise level decreases as illustrated by data obtained for $T=1.4$ and $S=0.3$. This behavior is related to the fact that $h$ does not influence the total energy of the two equivalent ordered spin arrangements. Similar effects can be observed for the ferromagnetic Ising model when a staggered magnetic field ($h_s$) is switched on for $h=0$ due to the above-mentioned intimate relationship between the ferromagnetic and anti-ferromagnetic Ising models. On the contrary, the application of $h$ to the ferromagnetic systems smoothed the abrupt variation of the order parameter as illustrated by open and closed circles in Fig.~\ref{fig:op_cp}. These Monte Carlo data show how the magnetization tends towards the exact solution if the corresponding magnetic field goes to zero. Notice, furthermore, that $m$ remains positive for arbitrary noise levels while $m \to 0$ if $K \to \infty$. In addition, the symmetries prescribe that $m(-h)=-m(h)$. Evidently, similar behavior occurs for the anti-ferromagnetic system in the presence of a staggered field.

\begin{figure}[ht]
\centerline{\epsfig{file=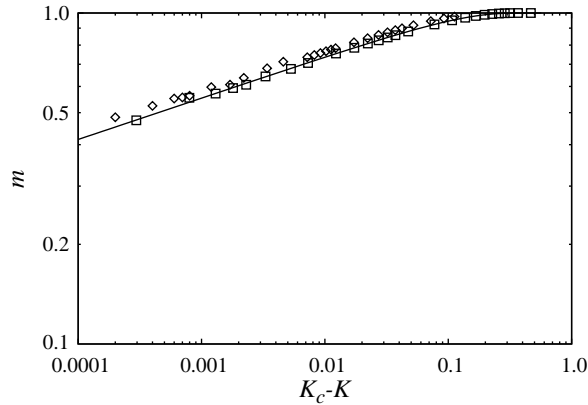,width=7.8cm}}
\caption{Log-log plot of the order parameter as a function of $K_c-K$ for results plotted in Fig.~\ref{fig:op_cp}.}
\label{fig:beta}
\end{figure}

Notice the excellent coincidence between the analytical and numerical results we have obtained for $T=1.5$ and $S=0.5$ while Monte Carlo data (with $m <0.5$) are missing in the close vicinity of the critical point. The latter deficiency is a direct consequence of the technical difficulties related to the general features of this type of critical transitions. Namely, the average value of the order parameter vanishes algebraically, more precisely,
\begin{equation}\label{eq:beta}
m \propto (K_c-K)^{\beta} \,,
\end{equation}
with $\beta = 1/8$ if $K \to K_c$ from below for all the cases ($T>1$ and $S>0$) when the chessboard-like ordered strategy arrangement transforms into a disordered ($m=0$) state as illustrated in Fig.~\ref{fig:beta}.
In the absence of the exact two-dimensional solution for $h \ne 0$ the latter universal feature was suggested by \citet{griffiths_prl70} and justified by \citet{rapaport_jpc71} who used approximation methods in the investigation of the transition from the disordered (paramagnetic) state into the anti-ferromagnetic structure in the presence of a magnetic field. The exploration of this universal behavior in the critical phase transitions was also motivated by the experiments. In fact, the first critical exponent (or divergency) was observed when measuring the specific heat as a function of temperature in the absence of magnetic field. Within the context of thermodynamics the different versions of the specific heat are described as the second partial derivative of the suitable thermodynamic potential and denoted as
\begin{equation}\label{eq:alpha}
c_h \propto |K_c-K|^{\alpha} \,,
\end{equation}
where the index $h$ refers to fixed magnetic field for the anti-ferromagnetic systems. For the ferromagnetic system $h=0$ is required otherwise the external field $h$ suppresses the divergence of $c_h$ in parallel with the elimination of the power law behavior in the magnetization as shown in Fig.~\ref{fig:op_cp}. Accordingly, instead of the power law divergency, one can observe a peak in the $K$-dependence of the specific heat at the vicinity of $K_c$ and the height of this peak decreases if $h$ is increased. Evidently, the internal symmetries of the Ising model imply that the application of a staggered magnetic field ($h_s$) results in similar consequences for the transitions from anti-ferromagnetic to paramagnetic phase.

In agreement with the features mentioned above and with the fluctuation dissipation theorem [see \cite{marconi_pr08} with further references therein] the magnetization depends sensitively on the external magnetic field at the critical point ($K=K_c$), namely,
\begin{equation}\label{eq:m_h}
m \propto |h|^{1/\delta} \,.
\end{equation}

In the understanding of the universal features of the critical behaviors, the analysis of the correlation functions played crucial roles. The general versions of correlation functions are defined in the translation invariant stationary states as
\begin{equation}
g^{(i,j)}({\bf x},t)=\langle n_{y}^{(i)}(t^{\prime}) n_{x+y}^{(j)}(t+t^{\prime}) \rangle - \langle n_{y}^{(i)}(\tau)\rangle \langle n_{x+y}^{(j)}(t+t^{\prime}) \rangle
\label{eq:cfs}
\end{equation}
where $i$ and $j$ denote strategy labels ($i,j=1, \dots , n$), $\langle \cdots \rangle$ refers to averaging over all sites ${\bf y}$ and time $t^{\prime}$. Notice that here we use the vector notation of sites in the argument of the correlation functions although in many numerical analyses ${\bf x}$ refers to horizontal (or vertical) spatial distances. $n_{x}^{(i)}$ is the extended definition of the occupation variable at time $t$ and site ${\bf x}$, namely,
\begin{eqnarray}
n_{x}^{(i)}(t)= \left\{
\begin{array}{l} 1 \;\; \mbox{for} \; s_x = i \;, \\
                 0 \;\; \mbox{otherwise} \;
\end{array} \right . \; .
\label{eq:occvar}
\end{eqnarray}
For sublattice ordered structures it is convenient to prescribe that ${\bf x} \in X$  and ${\bf y} \in Y$. For short range interactions the one-time correlation function decreases exponentially with the horizontal or vertical distance as
\begin{equation}\label{eq:cfs1ta}
g^{(i,j)}({\bf x},0) \simeq e^{-|{\bf x}| \over \xi_{ij}}
\end{equation}
if $|{\bf x}| \to \infty$. The one-site correlation function shows similar behavior, namely,
\begin{equation}\label{eq:cfs1sa}
g^{(i,j)}(0,t) \simeq e^{-t \over \tau_{ij}}
\end{equation}
if $t\gg 1$. Both the correlation time $\tau_{ij}$ and the correlation length $\xi_{ij}$ depend on the noise level $K$ as well as on the potential parameters. In the Ising model for $h=0$ we can omit the indices ($i$ and $j$) and the corresponding quantities diverge as
\begin{equation}\label{eq:nuperp}
\xi \propto |(K_c-K)|^{-\nu_{\perp}}
\end{equation}
and
\begin{equation}\label{eq:nupar}
\tau \propto |(K_c-K)|^{-\nu_{\parallel}}
\end{equation}
with $\nu_{\perp} =\nu_{\parallel}=1$ when approaching the critical point. We note that $\nu_{\parallel}$ characterizes the so-called critical slowing down in the time-dependence of correlations. Beside it, the intensity of the fluctuation of order parameter given as
\begin{equation}\label{eq:chim}
\chi_m=N \Bigl{\langle} \left[ \langle \sum_x \sigma_{\bf x} \rangle
- \sum_x \sigma_{\bf x} \right]^2 \Bigr{\rangle}
\end{equation}
also diverges. More precisely,
\begin{equation}\label{eq:chi}
\chi_m \propto |(K_c-K)|^{-\gamma} .
\end{equation}
In the stationary state at the critical point the correlation function decreases algebraically, that is,
\begin{equation}\label{eq:cfs1tcp}
g({\bf x},0) \propto |{\bf x}|^{2-d-\eta_{\perp}} .
\end{equation}
All these features jointly cause serious technical difficulties in Monte Carlo simulations if one wishes to determine the order parameter with a reliable accuracy in the close vicinity of the critical point, because both the system size and sampling time should be chosen to be significantly larger than the correlation length and time.

In fact the exponents introduced above are not independent. One of the main results of the statistical physics in the 20th century is the explanation of the universal behaviors in the critical phase transitions. Using the scaling hypothesis and the renormalization-group techniques, introduced by \citet{kadanoff_rmp67} and \citet{wilson_kg_rmp83}, relations have been explored between the above defined critical exponents. One form of this idea assumes that the singular part of the thermodynamic potential $\Phi(K,h)$ in the close vicinity of the critical point $K_c$ is a generalized homogeneous function of $\varepsilon=|K-K_c|/K_c$ and $h$, {\it e.g.},
\begin{equation}\label{eq:scalehyp}
\Phi_s(\lambda^{a_h}h,\lambda^{a_{\varepsilon}} \varepsilon) =
\lambda \Phi_s(h,\varepsilon) \,
\end{equation}
as surveyed briefly by \citet{stanley_rmp99}. This concept implies that two parameters ($a_h$ and $a_{\varepsilon}$) determine the values of critical exponents. The validity of the scaling hypothesis is confirmed by numerous experiments and theoretical calculations by observing data collapse when a thermodynamic quantity of two variables ({\it e.g.} $m(h,K-K_c)$) forms a single curve, if we use suitable scales in a two-dimensional plot. The latter feature reflects a universal behavior and is valid in the close vicinity of the critical point for the systems belonging to the Ising universality class.

The concepts of renormalization group techniques helped us understand the phenomena and relevant conditions resulting in the universal behaviors (for a brief survey of the essence and history of this approach see the review by \citet{fisher_rmp98}). This approach distinguishes relevant ({\it e.g.}, spatial dimension $d$, number $n$ of possible states, and additional symmetries) and irrelevant (lattice structure) quantities and gives an adequate description and explanation of the universal behavior at the critical point (and also in its close vicinity). These investigations have justified that the diverging quantities tend asymptotically towards a power law behavior characterized by the same exponent on both sides of the critical point. For example, when approaching the critical point $\xi \simeq C_1 (K_c-K)^{\gamma}$ if $K<K_c$ and  $\xi \simeq C_2 (K-K_c)^{\gamma}$ if $K>K_c$ ($C_1 \ne C_2$). This is the reason why we used the same notation for the exponents below and above the transition point.

The renormalization group techniques have confirmed quantitatively the universal scaling hypothesis between an upper and lower spatial dimension $d$. The predictions of the scaling hypothesis are simple consequences of the properties (\ref{eq:scalehyp}) preserved by the Legendre transform of $\Phi(K,h)$ and its derivatives. Since the critical exponents are directly related to $a_h$ and $a_{\varepsilon}$, one can derive {\it scaling laws} by eliminating these variables. Using this method three scaling laws can be derived:
\begin{eqnarray}
\alpha + 2 \beta + \gamma &=& 2 \,, \label{eq:exprel1} \\
\alpha + \beta (\delta +1)&=& 2 \,, \label{eq:exprel2} \\
(2 - \eta_{\perp}) \nu_{\perp} &=& \gamma  \,. \label{eq:exprel3}
\end{eqnarray}
which are independent of the spatial dimension $d$, whereas the fourth, so-called {\it hyper scaling law}, indicates directly the relevance of the spatial dimension as
\begin{equation}\label{eq:exprel4}
2 - \alpha = d \nu_{\perp} .
\end{equation}

Due to these relations only two of the six static exponents are independent. For the practical identification of this class of universal behavior in any models one needs to check the values of only two static exponents.

For the systems belonging to the Ising universality class, the values of most relevant exponents as a function of the spatial dimension $d$ are summarized in Table \ref{table:isingexp}. The reader can find a comprehensive list of the theoretical and experimental results in the review by \citet{pelissetto_pr02} and in the book by \citet{odor_08}. The one-dimensional results are missing here due to the absence of the critical transition. For $d \ge 4$ the critical behavior and exponents are well described by mean-field theory. Furthermore, the two-dimensional results are extracted from the exact results and the value of $\alpha =0$ refers to logarithmic divergence ($c_h \propto -\log{\varepsilon}$) in the specific heat.

\begin{table}{}
\begin{center}
\begin{tabular}{|c|l|l|l|}
\hline
exponent&$d$=2&$d$=3&$d$=4(MF)\\
\hline
$\alpha$&0(log)&0.110(1)&0\\
\hline
$\beta$&1/8&0.34265(3)&1/2\\
\hline
$\gamma$&7/4&1.2372(5)&1\\
\hline
$\delta$&15&4.789(2)&3\\
\hline
$\eta$&1/4&0.0364(5)&0\\
\hline
$\nu_{\perp}$&1&0.6301(4)&1/2\\
\hline
\end{tabular}
\end{center}
\caption{Critical exponents of the $d$-dimensional Ising model.}
\label{table:isingexp}
\end{table}

Finally we emphasize that a similar universal critical behavior can even occur in other non-equilibrium models where the probability distribution of the microscopic states differs from the Boltzmann distribution. For example, \citet{perez_pd02} have reported this type of universal critical transition in an Ising model, where the spins are reversed simultaneously as in the stochastic cellular automata surveyed by \citet{wolfram_pd84}.

The critical phenomena have also been explored for the Potts models on different lattices and graphs. Most of our knowledge comes from a wide scale of analytical (approximation) methods or numerical techniques and are justified by experiments. As it is known, the mean-field description gives a qualitatively correct picture of the phase transition in the Ising model. \citet{kihara_jpsj54} found that the mean-field approach predicts first-order phase transition for all $n > 2$. Subsequently, the more sophisticated methods have clarified that the order-disorder transitions are of first-order for $n > 4$ at $d=2$; $n \ge 3$ at $d=3$; and $n>2$ at $d=4$ in homogeneous spatial systems \cite{wu_fy_rmp82}.

Contrary to the mean-field prediction, the order-disorder transitions are continuous for all the two-dimensional lattices if $n=2$, 3, or 4. These critical transitions exhibit power law behavior in several quantities in the close vicinity of the transition point as detailed above. The corresponding critical exponents are listed in Table~\ref{table:potts2dgexp} where the values in column $n=2$ are equivalent to the critical exponents of Ising model given also in Table~\ref{table:isingexp}. Evidently, the listed values of the critical exponents satisfy the relations expressed by (\ref{eq:exprel1})-(\ref{eq:exprel4}) as these are derived from general scaling laws.

\begin{table}{}
\begin{center}
\begin{tabular}{|c|l|l|l|}
\hline
exponent&$n=2$&$n=3$&$n=4$ \\
\hline
$\alpha$&0(log)&1/3&2/3\\
\hline
$\beta$&1/8&1/9&1/12\\
\hline
$\gamma$&7/4&13/9&7/6\\
\hline
$\delta$&15&14&15\\
\hline
$\eta$&1/4&4/15& 1/2\\
\hline
$\nu_{\perp}$&1&5/6&2/3\\
\hline
\end{tabular}
\end{center}
\caption{Critical exponents {\it vs.} $n$ for the two-dimensional Potts models.} \label{table:potts2dgexp}
\end{table}

It is emphasized that the above robust behavior occurs in homogeneous spatial systems where $n$ individual strategies or degenerated ground states can be distinguished. Deviations from these universal behaviors appear when the spatial systems become inhomogeneous or the connections between the players are described by different graphs \cite{dorogovtsev_rmp08}.

\subsection{Ordering in other relatives of the Ising model}
\label{sec:oorim}

\begin{figure}[ht]
\centerline{\epsfig{file=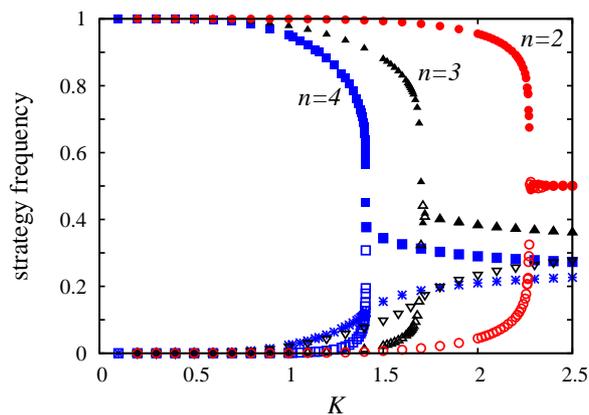,width=7.8cm}}
\caption{(Color online) Strategy frequencies as a function of $K$ for spatial evolutionary games if the interaction is defined by ${\bf d}(1,2)$ [see Eq.~(\ref{eq:isingsubgames})]. The Monte Carlo data of the 1st (2nd) strategies are indicated by closed (open) symbols for $n=2$ (red $\circ$), $n=3$ ($\bigtriangleup$), and $n=4$ (blue boxes). The frequencies of the third and fourth strategies for ($n=4$) are indicated by (blue) pluses and crosses while the symbol $\bigtriangledown$ shows the frequency of the third strategy for $n=3$.}
\label{fig:pn234}
\end{figure}

The decomposition of payoff matrices for $n$ strategies has highlighted the relevance of those basis games, denoted as ${\bf d}(p,q)$, that possess Ising type interactions between the $p$th and $q$th strategies. These models have not yet been investigated systematically. The Monte Carlo simulations on a square lattice with logit dynamics indicate an Ising type ordering process when increasing the noise level $K$ as it is illustrated in Fig.~\ref{fig:pn234} for three values of $n$. For all the three interactions ${\bf d}(1,2)$ the system prefers the homogeneous $s_x=1$ or $s_x=2$ state in the low noise limit and $(\rho_1 - \rho_2) \to 0$ if $K$ approaches $K_c$ from below. The preferences of the strategies 1 and 2 can be observed for $K>K_c$ because of the formation of their cluster when $\rho_1=\rho_2>\rho_n$ (for $n>2$). However, in the limit $K \to \infty$ the fluctuations suppress these correlations and all the $n$ strategies are present with the same probability ($1/n$).

The first numerical investigations \cite{szabo_pre14b, szabo_pre15b, vukov_pre15} support the theoretical expectations predicting Ising type critical transitions at $K_c(n)$ for $n=3$ and 4. The preliminary Monte Carlo results show that $K_c(n)$ decreases if $n$ is increased and the phase transition becomes a first order one if $n$ exceeds a threshold value.

\begin{figure}[ht]
\centerline{\epsfig{file=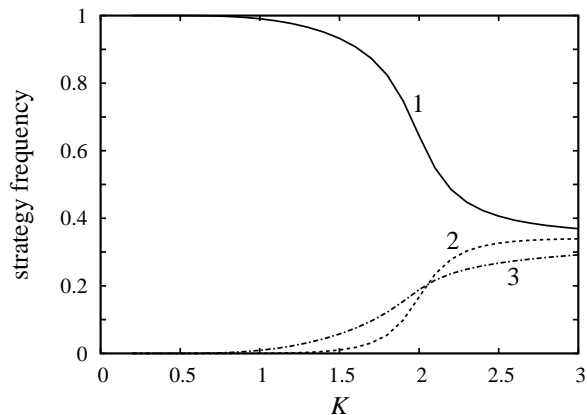,width=7.8cm}}
\caption{Strategy frequencies {\it vs.} noise level $K$ if the interaction is given by ${\bf A}={\bf d}(1,2)+0.1 \cdot {\bf d}(1,3)$ in the three-strategy multi-agent evolutionary potential game.}
\label{fig:d12e13}
\end{figure}

In the above-mentioned systems the critical phase transitions are smoothed if the interaction is perturbed by switching on a self-dependent component or an additional coordination type interaction that can prefer the first (or the second) strategy. The $\rho_i(K)$ functions in Fig.~\ref{fig:d12e13} are obtained for an interaction that results in the dominance of strategy 1 in the low noise limit. In this plot we were able to illustrate the Monte Carlo data by lines due to the absence of the fluctuation enhancement in the region of smooth transition.

As mentioned previously, the symmetric potential matrix can possess two equivalent preferred Nash equilibria [$(p,q)$ or $(q,p)$ if $\max (V_{ij})= V_{pq}=V_{qp}$] even for $n>2$. On the square lattice these interactions yield two equivalent sublattice ordered strategy arrangements when the players select strategies $p$ and $q$ in the sublattices $X$ and $Y$ or conversely if $K \to 0$. When increasing the noise level $K$ these systems exhibit an order-disorder phase transition at $K_c$. The strategy frequencies in the sublattices show a similar $K$-dependence plotted in Fig.~\ref{fig:pn234}. Contrary to the case of equivalent homogeneous ordered states, the equivalence between the sublattice ordered strategy distribution is not destroyed if the interaction is perturbed weakly  because their effects are identical on both sublattices. Consequently, the Ising type sublattice ordering takes place in a wide region of the payoff parameters. The latter phenomenon is resembling the effect of the external magnetic field $h$ on the anti-ferromagnetic ordering if its value does not exceed a threshold value (see Figs. \ref{fig:isi_map} and \ref{fig:op_cp}).

\subsection{Critical phase transitions on networks}
\label{sec:cponw}

The investigation of the Ising model on different networks also has a long history as detailed in the review by \citet{dorogovtsev_rmp08}. First we discuss the simplest cases where the players are distributed on a Cayley tree or Bethe lattice that is considered as an infinite Cayley tree without its periphery. Despite the strong relationship between these connectivity structures, the Ising model exhibits fundamentally different behaviors on them as it was emphasized in early works of \citet{eggarter_prb74, mullerhartmann_prl74, wang_yk_jpa76}.

The first important difference is due to the fact that a relevant portion of the nodes of the Cayley tree belongs to the periphery (the leaf nodes in graph theory terminology) where the players have only one neighbor and the presence of this boundary affects the macroscopic behavior significantly, due to the large number of these sites. On the other hand, this feature is utilized in the determination of the exact solution \cite{eggarter_prb74, baxter_82, ostilli_pa12}. The absence of the ordered states on the Cayley tree for any finite values of $K$ is related to the fact that when removing a single edge this structure is divided into two independent parts where opposite ordered structure can be formed. Due to this feature, arbitrary deviation in the magnetization can be achieved by generating a single point defect that is always present at finite $K$ in the sufficiently large systems. A similar reason explains the absence of ordered strategy arrangements (and phase transition) in the one-dimensional lattice where one of the long-range ordered phases occurs only for $K=0$.

\citet{mullerhartmann_prl74} have shown that the thermodynamic potential (free energy) on the Cayley tree becomes a nonanalytic function of the magnetic field $h$ below a critical temperature $K_{BP}$ given by the pair approximation applied by \citet{bethe_prsa35} and \citet{peierls_pcps36}. More precisely, the leading part of the nonanalytic behavior is proportional to $h^{\kappa}$ where the exponent $\kappa$ increases monotonously from 1 to $\infty$ as the temperature goes from $0$ to $K_{BP}$. This phenomenon is accompanied by a diverging magnetic susceptibility below the transition point. \citet{melin_jpa96} have shown that this unusual behavior may be related to the large number of metastable states that are stable with respect to single-spin flips.

\begin{figure}[ht]
\centerline{\epsfig{file=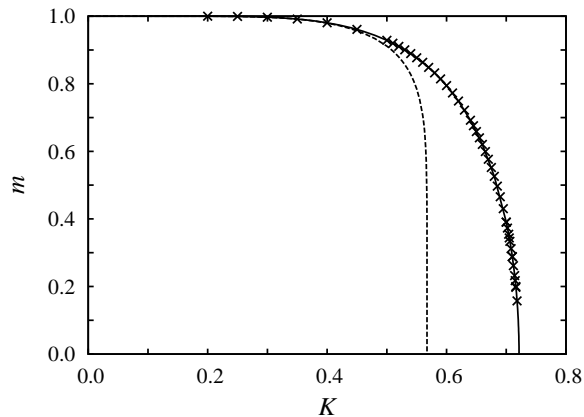,width=7.8cm}}
\caption{Order parameter $m$ as a function of noise $K$ for evolutionary hawk-dove game with Glauber dynamics at payoff parameters $T=1.5$, $S=0.5$ if the players are located on a random regular graph of degree 4. Xs indicate MC data for $N=2 \times 10^6$. The solid line denotes the prediction of the two-site cluster variation method for the Bethe lattice and the Onsager results (on square lattice) are illustrated by dotted line.}
\label{fig:ising_bethe}
\end{figure}

The use of Bethe lattice eliminates the boundary layer and assumes translation invariance ({\it i.e.} the equivalence of sites). Using this assumption we can derive analytical results that predict mean-field type order-disorder phase transition if $K$ is increased. Due to the absence of loops in this structure the two-site cluster variation method (called also Bethe-Peierls or pair approximation) reproduces the exact result as indicated in Fig.~\ref{fig:ising_bethe} where the MC data are obtained on a random regular graph for large sizes as the MC simulations cannot be performed on an infinitely large Bethe lattice.
In network analysis \cite{albert_rmp02, newman_siamr03} and graph theory \cite{bollobas_98} it is well known that the average length of the shortest loops for the random regular graphs increases with $\ln{N}$. Consequently, this approach can give an adequate description of the system behavior for those cases where the sparse and long loops of the connectivity structure do not modify the behavior relevantly. This approach is capable of explaining more striking differences in those phenomena where a random regular graph is substituted for the square lattice in a three-strategy evolutionary game \cite{szabo_pre02b}.

In the last decades the statistical analysis of different networks initiated the reinvestigation of traditional lattice models on a wide scale of complex networks. Most of these networks are claimed to provide a better description of the connections that have emerged in social interactions, neural systems, communication, transportation, etc. Some of these network models are capable of describing a continuous transition from a spatial lattice to random or random regular graphs, while others create novel features not involved in the traditional concept of lattices. Now we briefly outline what happens to the Ising and Potts models on several types of networks.

The so-called small-world effect takes place in networks where the average shortest distance between two nodes is proportional to $\ln{N}$. The Bethe lattices, Cayley trees, and random regular graphs possess this feature. \citet{watts_dj_n98} proposed a way of building the small-world effect into spatial lattices. In the first model they considered a one-dimensional chain (ring) with first- and second-neighbor interactions. The small world feature is achieved by removing $qN$ connections from the lattice and connecting one of their end points to another site chosen at random ($0 < q \ll 1$). This method does not modify the average number of neighbors though it destroys the regularity. At the same time, it can be applied for any $d$-dimensional lattices. The number of neighbors is conserved at each site by the method proposed by \citet{newman_pla99} where connections are interchanged between pairs of connected sites selected randomly. Soon after the network models had been published, different phenomena were investigated on the resultant structures.

Studying the one-dimensional Ising model with the above rewired lattices or with a lattice obtained by adding a portion $q$ of new links, \citet{barrat_epjb00} and \citet{gitterman_jpa00} have observed the appearance of ordered structure at finite noise levels ($K < K_c(q)$) in sufficiently large systems. It is found that $K_c$ increases monotonously with $q$ and $K_c \propto -z / \log{(q)}$ for low values of $q$ where $z$ ($z \ge 4$) is the number of neighbors in the starting one-dimensional chain. They also reported a mean-field type phase transition at $K_c$.

These types of small-world structures were reinvestigated by \citet{herrero_pre02} and \citet{chatterjee_pre06} when the initial structure was a square ($d=2$) or cubic ($d=3$) lattice. Their analyses have confirmed the presence of the ordered state at low temperatures in agreement with the expectation. The increase of $K_c$ was similar to those found for the one-dimensional cases, that is, $\delta K_c \propto 1/\log {(q)}$ and the phase transitions were of mean-field type.

Up to now we have studied the Ising models on regular or quasi-homogeneous networks where the degrees of nodes were close to their average value. Fundamentally different behaviors were reported by \citet{dorogovtsev_pre02} and \citet{leone_epjb02} who studied the Ising model on random graphs with a degree distribution having a fat tail. Most of these investigations are performed on networks having a power-law degree distribution $p(z) \propto z^{-\upsilon}$ for large number ($z$) of neighbors. Using different approaches these authors studied the general features of the ferromagnetic phase transitions for different values of $\upsilon$. The analyses showed mean-field type phase transitions when $\upsilon >5$. These approaches indicated non-trivial critical exponents for $3 \le \upsilon \le 5$ as listed in Table~\ref{table:isingexp_sfg}. Note that behavior of the fluctuation is not affected by the value of $\upsilon$ within this region, while $\alpha$, $\beta$, and $\gamma$ depend on $\upsilon$ in a way that breaks  the scaling relations (\ref{eq:exprel1}) and (\ref{eq:exprel2}).

\begin{table}{}
\begin{center}
\begin{tabular}{|c|l|l|l|}
\hline
exponent&$\upsilon >5$&$\upsilon=5$&$3< \upsilon <5$ \\
\hline
$\alpha$&1st order&0(log)&$(5-\upsilon)/(\upsilon -3)$\\
\hline
$\beta$&1/2&1/2$^{\star}$&$1/(\upsilon -3)$ \\
\hline
$\gamma$&1&1&1\\
\hline
$\delta$&3&3$^{\star}$&$\upsilon -2$ \\
\hline
\end{tabular}
\end{center}
\caption{Critical exponents {\it vs.} $\upsilon$ for the Ising models on scale-free graphs. $^{\star}$ refers to logarithmic corrections described in \cite{dorogovtsev_pre02, leone_epjb02}.}
\label{table:isingexp_sfg}
\end{table}

For $\upsilon \le 3 $ the role of the high degree nodes (hubs) becomes relevant in the limit $N \to \infty$ and the system stays in a ferromagnetic state for arbitrary noise level $K$. Within this range of $\upsilon$, however, a size effect can be observed because the value of $N$ limits the maximum of $z$, too. As a result, for most of these finite systems the Ising model exhibits an order-disorder transition at $K_c \propto \ln{(N)}$ according to the theoretical investigations mentioned above, in agreement with the results of Monte Carlo simulations \cite{aleksiejuk_pa02}. Many realistic networks and network models belong to this class, therefore the latter feature governs the behavior in both biological and social systems.

In the above small-world networks the additional random links can be interpreted as a way of introducing long-range interactions. Similarly to the L{\'e}vy flights, the strength of long-range interactions can be weakened by creating the additional links with a probability $p(l)\propto l^{-\Delta}$ where $l$ is the Euclidean distance between the sites to be connected and $\Delta > 0$. The behavior of the resulting system is similar to those obtained by introducing long-range interactions, as discussed in the following section.

Fundamentally different behaviors were reported by \citet{gefen_prl80, gefen_prl83} who studied the Ising and Potts models on several fractal lattices discussed by \citet{mandelbrot_77}. In the 80s the analysis was extended by \citet{bhanot_prl84, dauriac_jpa86, bonnier_prb88, monceau_prb98}. Some of the fractal lattices were generated by an algorithm producing a Sierpinski carpet. For example, a large hypercube of a $d$-dimensional lattice with $\tilde{n}^{\tilde{k}d}$ sites is divided into $\tilde{n}^d$ equal hypercubes and $(\tilde{n}^d-\tilde{m})$ of them are removed. The same process is repeated for the rest of the smaller hypercubes. After the $\tilde{k}$th segmentation steps we get a fractal lattice characterized by a fractal dimension $d_f=\ln{(\tilde{m})} / \ln{(\tilde{n})}$ that can be tuned from 0 to $d$. The resultant self-similar structures made these types of models attractive for the application of renormalization group techniques and other exact methods. A series of investigations have clarified that the behavior of the Ising model on these lattices depends not only on the fractal dimensions, but is also affected significantly by other topological features like the order of ramification and lacunarity \cite{gefen_prl80, carmona_prb98}.

The order of ramification ${\cal R}$ measures the number of links to be cut in order to isolate an arbitrarily large portion of the network. If ${\cal R}$ is finite then the ferromagnetic ordering is missing at finite noise levels. Notice that this criterion is the generalized version of those we used above for explaining the absence of ordering in the one-dimensional lattice and Cayley trees. Among the fractal lattices the quasi-linear structures (generated similarly to the Koch-curve) \cite{gefen_jpa83} and some versions of the Sierpinski gasket \cite{gefen_jpa84} represent networks on which the Ising model has no long-range order (and phase transition) at finite noise level $K$.

On the contrary, phase transition occurs if ${\cal R}= \infty$ in the limit $N \to \infty$. The first papers \cite{gefen_jpa84} indicated clearly that some critical exponents depend also on the lacunarity $\Lambda$ \cite{mandelbrot_83}, which quantifies the deviation from the translation invariance. For the above mentioned Sierpinski carpets the lacunarity $\Lambda$ measures how the removed small boxes are distributed. At high lacunarity the removed boxes form a large hole, while at low lacunarities these boxes are distributed ''homogeneously''.

The critical exponents of the resultant phase transition are studied quantitatively only on a few fractal lattices. For example, \citet{monceau_prb98} and \citet{carmona_prb98} studied the critical behavior on the above mentioned Sierpinski carpet for $d=2$, $\tilde{n}=3$, $\tilde{m}=8$ where $d_f=\ln{(8)} / \ln{(3)}=1.8927\ldots$. Despite the technical difficulties their results are consistent with the suggestion of \citet{wu_yk_pra87} stating the existence of a weaker universality class. Accordingly, the static exponents may vary with the geometrical parameters of the fractal ({\it e.g.}, $d_f$ and lacunarity) and the relations between the critical exponents (\ref{eq:exprel1})-(\ref{eq:exprel4}) are valid if $d_f$ is substituted for $d$ in (\ref{eq:exprel4}). In a subsequent paper \citet{monceau_pa04} studied the critical behavior on fractals sharing the same fractal dimension but having different lacunarities. It was found that the long-range order at the critical point decays faster when the lacunarity is increased for a given $d_f$.

At the end of this section it is worth mentioning that most of the above discussions are based dominantly on approximate results supported by numerical simulations. Very recently, however, \citet{dembo_aap10}, \citet{montanari_ptrf12}, and \citet{dembo_ap13} have performed a more rigorous mathematical analysis of the Ising and Potts models on locally tree-like structures and their results support the picture sketched above.

\subsection{Long-range interactions on lattices}
\label{sec:longrange_ising}

In ionic compounds the Coulomb interaction plays a crucial role in the formation of the microscopic arrangements of ions even in the cases when screening complicates the analysis in the presence of opposite charges \cite{dieterich_ap80}. For the ferromagnetic materials the dipole-dipole interaction is responsible for the emergence of a proper magnetic domain structure that results in an almost zero remanence magnetization in the soft magnets \cite{bozorth_51}. In solid solutions the interaction between two large interstitial atoms is mediated by the lattice distortion and it can be well approximated by a suitable long-range interaction. This interaction drives the segregation process. In social and biological systems the diffusion of opinion or chemical products can also mediate a similar long-range interaction. For the investigation of these type of interactions, the formalism of the Ising model provides a good mathematical background, although huge technical difficulties arise from the large number of interactions to be considered in the real phenomena. In short, the analysis of the Ising and Potts models with long-range interactions on lattices is recently at a beginning stage.

There are, however, several results that are worthy of a brief outline. In many cases the coupling constant $J_{xy}$ between the sites ${\bf x}$ and ${\bf y}$ is expressed by an algebraic function $J_{xy}=1/|{\bf x}-{\bf y}|^{d+\Delta}$ where $|{\bf x}-{\bf y}|$ is the Euclidean distance on the $d$-dimensional lattice and $\Delta$ quantifies the type of long-range interactions. Assuming a ferromagnetic interaction, the total effect (sum) of the coupling constants $J_{xy}$ over the whole system helps preserve the ferromagnetic state. This quantity as well as the total energy, however, diverges if $\Delta \le 0$ as
\begin{equation}
\sum_{y, y \ne x} J_{xy}=\sum_{y, y \ne x} {1 \over |{\bf x}-{\bf y}|^{d +\Delta}}
\simeq a(d) \int_{1}^{\infty} {1 \over r^{1+\Delta}} dr
\label{eq:heff_mflr}
\end{equation}
where $a(d) r^{d-1}$ refers to the surface of a $d$-dimensional sphere of radius $r$. Consequently, the ferromagnetic state is stabilized if $\Delta \le 0$. Since the early work of \citet{ruelle_cmp68} the systematic analysis of the Ising and Potts models has clarified what happens for $\Delta >0$.

\citet{ruelle_cmp68} proved rigorously the absence of long-range order in the one-dimensional Ising ferromagnetic model if $\Delta > 1$. On the other hand, \citet{dyson_cmp69} proved the existence of a phase transition in these systems if $0<\Delta <1$. For $\Delta=1$ the existence of phase transition was justified later by \citet{dyson_cmp71} and by \citet{imbrie_cmp88} who described an intermediate phase where the two-point correlation function exhibits power law decay with a $K$-dependent exponent below $K_c$. Using the renormalization group approach \citet{fisher_prl72} have shown that the one-dimensional Ising model exhibits mean-field-like behavior for $0<\Delta < 0.5$ while non-trivial ($\Delta$-dependent) exponents are found for $0.5< \Delta < 1$.
For higher dimensions the analyses were focused on two regions of $\Delta$ where mean-field or $\Delta$-dependent exponents are predicted by the renormalization group techniques.

\subsection{Sublattice ordered structures on lattices}
\label{sec:sosl}

If the range of interactions exceeds the nearest neighbors on a lattice then we may face a wide variety of sublattice ordered states in the low noise limit. The extensive investigations of these systems were unavoidable in solid state physics where the effects of the second- and third-neighbor interactions with different strengths are studied in many materials. For example, the adequate description of the oxygen ordering in the Cu-O layer of the super-conducting YBa$_2$Cu$_3$O$_{7-\delta}$ compounds required to take into consideration four terms of interactions \cite{wille_prb89}.

The introduction of the additional second- and third-neighbor interactions (denoted as $J_2$ and $J_3$) does not cause qualitatively different behaviors if all the coupling constants favor the ferromagnetic structure, that is, if $J_1, J_2, J_3 >0$. The presence of the additional terms increases the value of $K_c$. Fundamentally different consequences occur, however, if the additional interactions do not support the formation of the ordered arrangement dictated by the first-neighbor interactions. In the latter cases numerous types of ordered structure may emerge. For the illustration of these phenomena we study now a simple model investigated in detail first by \citet{binder_prb80} and recently by \citet{yin_j_pre09} and \citet{de_queiroz_pre11}.

\begin{figure}[ht]
\centerline{\epsfig{file=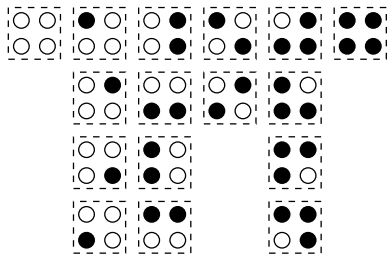,width=5.2cm}}
\caption{All the possible spin arrangements within a $2 \times 2$ cluster of sites that are used for generating four-sublattice ordered arrangements on a square lattice.}
\label{fig:2x2patt}
\end{figure}

Consider an Ising model on the square lattice with equivalent anti-ferromagnetic interactions ($J_1=J_2=-1$) between the first and second-neighbors for the presence of a magnetic field $h$. One can assume that all the possible ordered structures can be built up by tiling, that is, by repeating one of the $2 \times 2$ patterns shown in Fig.~\ref{fig:2x2patt}.
The resulting long-range ordered arrangements can be well described by a four-sublattice
description defining the average magnetization for each sublattice. Notice that patterns within the second, third and fifth column of Fig.~\ref{fig:2x2patt} can be transformed into each other by a rotation of 90$^{\circ}$ and the corresponding ordered structures have the same potential (or energy) in the macroscopic system. All these three types of ordered structures are fourfold degenerated and involve the possibility that the corresponding order-disorder transitions belong to the Potts model universality class for $n=4$ and $d=2$.

\begin{figure}[ht]
\centerline{\epsfig{file=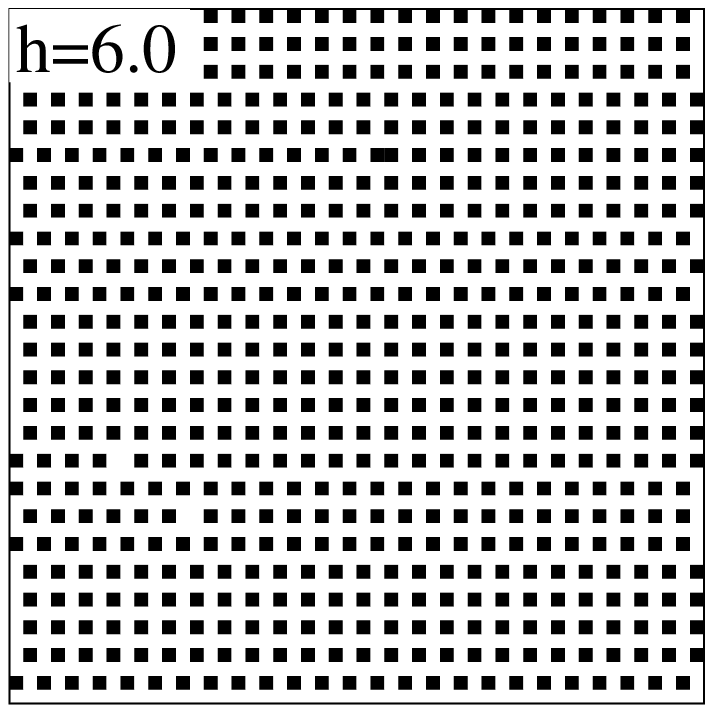,width=4cm}
\epsfig{file=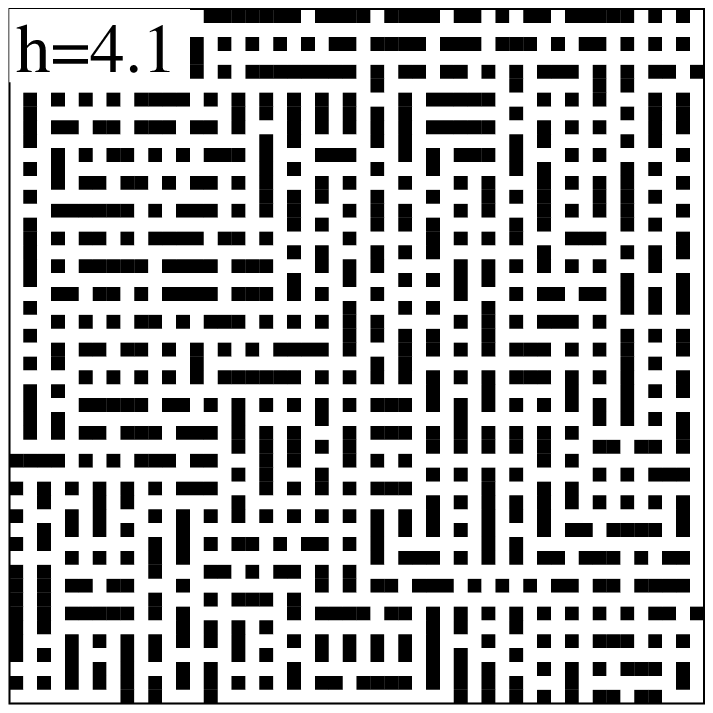,width=4cm}} \vspace{0.1cm}
\centerline{\epsfig{file=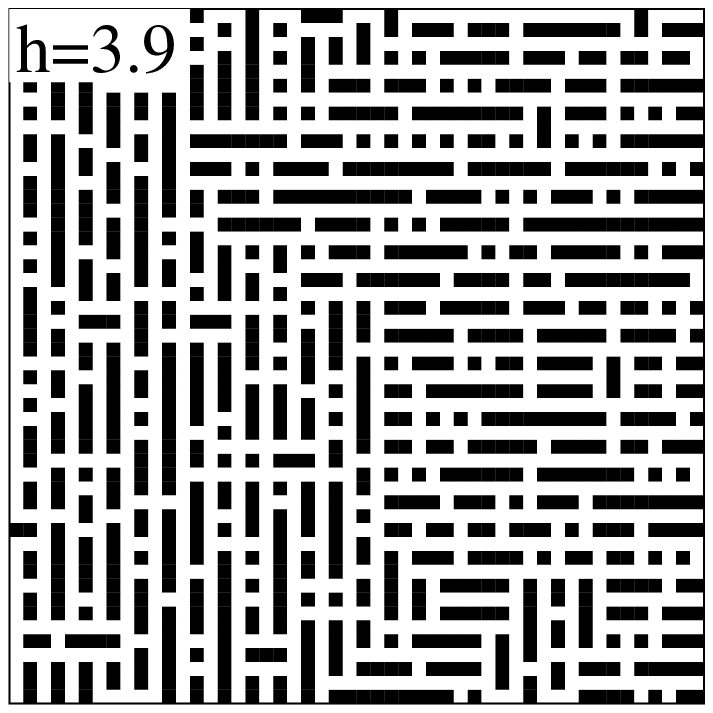,width=4cm}
\epsfig{file=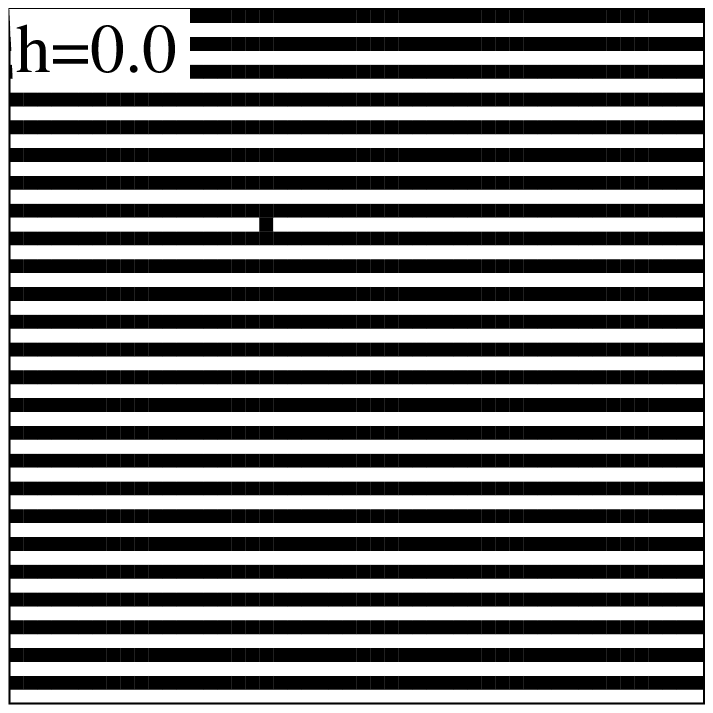,width=4cm}}
\caption{Ordered and partially ordered spin arrangements in the kinetic Ising model on the square lattice when the first- and second-neighbor anti-ferromagnetic coupling constants are equivalent and $K=0.3 \cdot |J|$ for different external magnetic fields ($h$) indicated by figures inside the ($50 \times 50$) snapshots.}
\label{fig:isi8}
\end{figure}

Contrary to the naive expectations, simulations have indicated a more complex behavior both in the spatial patterns and in the phase transitions. Figure \ref{fig:isi8} illustrates four snapshots when increasing $h$. All these simulations are performed at a low noise level in order to keep the strip like anti-ferromagnetic arrangement (see the bottom right plot in Fig.~\ref{fig:isi8}) almost free of point defects. The applied magnetic fields are selected according to the phase diagram reported by \citet{yin_j_pre09} and represent different phases. Besides these four phases there exists a disordered (paramagnetic) spin arrangements stabilized for sufficiently high noise level ($K>K_c(h)$) or high fields $|h|>8$.

The fourfold degenerated strip-line anti-ferromagnetic structure undergoes a phase transition to the paramagnetic phase at $K_c(h=0)=2.08$. This curious phase transition will be discussed later. In the zero noise limit this ordered structure is stable as long as $|h|<4$. For $4<|h|<6$ a partially ordered structure, called row-shifted ($2 \times 2$) phase, can be observed in the low noise limit. The corresponding spin arrangements can be constructed from the completely ordered pattern built up from one of the $(2 \times 2)$ blocks of the second columns in Fig.~\ref{fig:2x2patt} that consist of ferromagnetic and anti-ferromagnetic rows (columns) alternately. For the present interactions the total energy (or potential) is not changed if the anti-ferromagnetic rows are shifted horizontally at random. In the resultant partially ordered phase the entropy remains finite in the limit $K \to 0$. Due to the symmetries the equivalent column-shifted states can be constructed in a similar way. Thus within a region of $h$ dependent on the temperature this system has two equivalent sets of partially ordered structures. Within this region one can observe a spontaneous symmetry breaking that is analogous to the formation of ferromagnetic or anti-ferromagnetic structures if the system is started from an ordered structure at a low noise level. This process is driven by the increase of entropy while the coexistence of the domains of row- and column-shifted phases is prevented by the positive interfacial energy.

Between the strip-line and row-shifted structures, it has been possible to distinguish two additional phases by \cite{yin_j_pre09} in the phase diagram at low noises. The upper-right snapshot of Fig.~\ref{fig:2x2patt} illustrates the spin arrangement in the paramagnetic phase with short-range correlations resembling all the above mentioned ordered phases. The appearance of the disordered (paramagnetic) phase is common at low noises in the phase diagrams between basically different ordered phases. In the present model, however, another type of structure (see the left-bottom snapshot in Fig.~\ref{fig:2x2patt}) is observed. The latter phase can be considered as a poly-domain pattern of the strip-line phases where the presence of interfaces is stabilized as it occurs in water-oil emulsions or other self-assembling systems \cite{gompper_94, henriksen_pre00}. Remarkably, the distinction of this phase required sophisticated techniques.

Recently different methods have been applied by \citet{yin_j_pre09} and \citet{de_queiroz_pre11} for the classification of the phase transitions in the present model. According to these investigations, the phase transitions from the ordered or partially ordered structure into the disordered one are continuous but not universal. Both transitions exhibit power law behavior with exponents dependent on the magnetic field. In other words, these transitions do not belong to the above mentioned universality classes of the Ising and Potts models.

The above features are not unique. Similar phenomena can be observed for the present model when varying the ratio $J_2/J_1$ or when additional interactions are switched on \cite{yin_j_pre10}. In the paper by \citet{dublenych_pre11} the reader can find dozens of patterns that can be used for building up long-range ordered structures of the Ising model on triangular lattices, if first- and second-neighbor interactions are taken into consideration. Evidently, analogous phenomena are expected in lattices for higher dimensions if the range of interactions is increased and also in evolutionary potential games when the number of strategies exceeds 2.

\subsection{Frustration}
\label{sec:frust}

\begin{figure}[ht]
\centerline{\epsfig{file=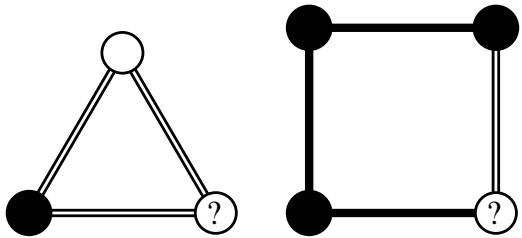,width=7cm}}
\caption{Two typical topological conditions yielding frustration. The left panel shows three players with two strategies (closed and empty bullets) located on a triangle for pairwise anti-coordination games denoted by double lines. The question mark refers to the equivalence between the two strategies for the third player. The right panel shows four players on a square network who play pairwise coordination (simple edge) or anti-coordination (double edge) game with each other as denoted. For the given three-player strategy profile the fourth player has two equivalent strategies.} \label{fig:frustrat}
\end{figure}

In the previous section we have faced situations when the ground state was infinitely degenerated (see the upper-left snapshot of Fig.~\ref{fig:2x2patt}). Such a situation becomes natural on lattices where three spins (players) interact with each other via equivalent anti-ferromagnetic pair interactions (anti-coordination games) as denoted on the left panel of Fig.~\ref{fig:frustrat} \cite{diep_04}. This phenomenon was first described by \citet{wannier_pr50} who studied the anti-ferromagnetic Ising model on a triangular lattice. In fact, for the anti-ferromagnetic interactions, the basic reason is related to the presence of loops with an odd number of edges. Such topological situation occurs on many other lattices with a finite clustering coefficient ({\it e.g.}, square lattice with the first- and second-neighbor interactions), in some curious lattice structures [{\it e.g.}, the Cairo pentagonal lattice studied by \citet{rojas_pre12}] that includes five-edge loops.

On the contrary in social networks the length of loops may be either odd or even, consequently, the frustration is unavoidable on these structures for the anti-coordination games.

\begin{figure}[ht]
\centerline{\epsfig{file=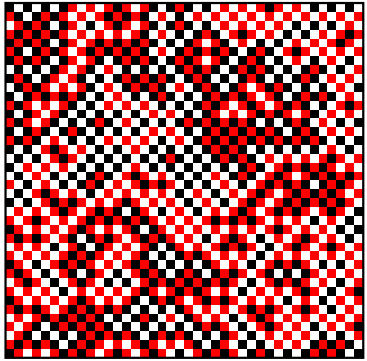,width=5cm}}
\caption{(Color online) Typical snapshot for the repulsive three-state Potts model on the square lattice at low noise levels.}
\label{fig:repotts}
\end{figure}

The other source of frustration is related those models where the coupling constants are chosen to be $+J$ or $-J$ at random as it is illustrated in the right panel of Fig.~\ref{fig:frustrat}. Some consequences of the frustration are well illustrated in Fig.~\ref{fig:repotts} where one can observe six different domains representing the chessboard and anti-chessboard arrangements of two of three strategies on the square lattice for games equivalent to the repulsive three-state Potts model.
\citet{berker_jpa80} have shown that the typical size of these domains (or the correlation length) increases algebraically when the noise level $K$ is decreased. Similar features were reported previously by \citet{wannier_pr50} for the Ising model on a triangular lattice.

A direct consequence of the frustration can be the high level of degeneracy of states having maximal potential. If the number of these degenerate states increases with $N$ exponentially then the specific entropy remains finite in the zero noise limit, contrary to the third law of thermodynamics. This discrepancy vanishes if the games are not equivalent, as it occurs in social systems. On the contrary, if we study a system with many edges of small potential differences in the dynamical graph, then the systems may reach the thermodynamically stable state faster.

\subsection{Effects of randomness}
\label{sec:spdsu}

For many multi-agent systems we cannot assume equivalent games representing the interactions among the participants. In Sec.~\ref{sec:mtoI} it is shown that if the interactions are limited to symmetric $2 \times 2$ games then the potential of this multi-agent system can be mapped onto an Ising model where the values of $J_{xy}$ and $h_x$ vary within a wide range. The effects of these types of randomness have been studied for four decades and the results are documented in the literature of spin glasses. The experimental investigations of these systems were first motivated by the unusual magnetic behavior of some metal alloys  ({\it e.g.} AuFe and CuMn) \cite{cannella_prb72}. From a theoretical point of view these phenomena demanded the development of new concepts and approaches. Comprehensive descriptions of the models, methods, phenomena, and perspectives are given in the books by \citet{mezard_87} and by \citet{stein_13}. For most of the investigations the effects of randomness (occurring in $J_{xy}$ and $h_x$) are considered separately. Exceptions are represented by the one-dimensional random Ising models \cite{normand_jpa85}. Recent trends in the research of evolutionary game theory offer further possibilities for the introduction of randomness via the consideration of personal features and the wide scale of connectivity structures. Here we survey only a few phenomena that seem to be important for information processing \cite{nishimori_01}, decision theory \cite{galluccio_pa98}, social \cite{stein_13} and biological \cite{stein_92} systems.

Spin glasses are systems combining quenched randomness and frustration. For the Ising type spin glasses the parameters $J_{xy}$ and $h_x$ are random variables and are characterized by some probability distribution functions. The accurate mathematical form of the probability distribution of the $J_{xy}$ values is not important in general. For technical reasons the Gaussian distribution is chosen with a mean value of zero (denoted as $\langle \langle J_{xy} \rangle \rangle_{J}=0$ and $\langle \langle J_{xy}^{2} \rangle \rangle_{J}=1$ where $\langle \langle \cdots \rangle \rangle_{J}$ refers to averaging over all pairs). An alternative approach was suggested by \citet{domany_jpc79} for the so-called $\pm J$ model where $J_{xy}=J$ ($J_{xy}=-J$) is chosen at random with a probability $p$ ($1-p$). Here we remind the reader that $\langle \langle J_{xy} \rangle \rangle_{J}>0$ does not guarantee the ferromagnetic order in the low noise limit in the absence of magnetic field ($h_x=0$). When investigating the $\pm J$ model on a square lattice \citet{domany_jpc79} has explained the existence of the ferromagnetic order in the low noise limit if $p > p_c$ ($p_c \simeq 0.83$ on the square lattice). It was found that the percolation of unfrustrated squares was responsible for the ferromagnetic state and the critical point ($p_c$) is associated with the adequate percolation threshold value.

The spin glass model was introduced and studied by \citet{edwards_jpf75} for a uniform Gaussian distribution ($\langle \langle J_{xy} \rangle \rangle_{J}=0$ and $\langle \langle J_{xy}^{2} \rangle \rangle_{J}=1$). In agreement with the experimental results this model has indicated the presence of a "spin glass" phase at low noise levels ($K<K_{sg}$). The main characteristic of the spin glass phase is the existence of a large number of free-energy valleys that are inaccessible from each other in the limit $N \to \infty$. Within the spin-glass phase the average magnetization is zero and there are ferromagnetic domains with different sizes and orientations. During the evolution the system remains in the vicinity of one of the given free-energy valleys similarly to the ferromagnetic phase where only two (ordered) microscopic states are distinguished. For the quantification of the remanence of these spin arrangements \citet{edwards_jpf75} have introduced an order parameter which is similar to the one-site two-time correlation function characterizing the coincidence of the "average" spin directions $\langle \sigma_x^{\rm (1)}\rangle$ and $\langle \sigma_x^{\rm (2)}\rangle$ in the vicinity of times $t_1$ and $t_2$ which are far from each other. If the system remains in the same free-energy valley then $\langle \sigma_x^{\rm (1)}\rangle = \langle \sigma_x^{\rm (2)}\rangle = \langle \sigma_x \rangle$ and the spin glass phase can be characterized by the so-called Edwards-Anderson order parameter:
\begin{equation}\label{eq:qEA}
q_{EA}={1 \over N} \sum_{x} \bigl{\langle} \langle \langle \sigma_x \rangle^{2} \rangle \bigr{\rangle}_{J} .
\end{equation}

At high temperatures $q_{EA}=0$ because each spin can flip into the opposite state within a short time. $q_{EA}=1$ refers to frozen patterns at $K=0$. Deviations from $q_{EA}=1$ arise from thermal fluctuations and frustration.

A more detailed insight into the nature of spin glasses is provided by considering an exactly solvable model introduced by \citet{sherrington_prl75}. It is an Ising model on a complete network (interactions exist between all spin pairs), therefore this model could be well investigated by the mean-field theory using the replica trick. For the analytical analysis of this model, \cite{sherrington_prl75} \cite{thouless_pm77} have justified the phase transition at $K_c=1$ from the paramagnetic state to the spin glass phase when decreasing the temperature and explained the magnetic behavior observed in experiments.

In order to study the inherent relationship between two glassy states \citet{parisi_prl79} has suggested using a more complicated order parameter:
\begin{equation}\label{eq:qP}
q_{\alpha \beta}={1 \over N} \sum_{x} \langle \sigma_x \rangle_{\alpha} \langle \sigma_x \rangle_{\beta} .
\end{equation}
that quantifies the average overlapping between two frozen spin patterns $\alpha$ and $\beta$ for a given randomness in $J_{xy}$. Assuming that the state $\alpha$ takes place with a probability $W^{\alpha}$ for a given randomness, \citet{parisi_prl79, parisi_jpa80} introduced a probability distribution of overlap values as
\begin{equation}\label{eq:Pod}
P(q)=\sum_{\alpha , \beta} W^{\alpha} W^{\beta} \delta(q-q_{\alpha \beta}) .
\end{equation}
It is generally assumed that $q_{\alpha \alpha}$ is independent of $\alpha$ in the limit $N \to \infty$. In that case, $q_{\alpha \alpha}=q_{EA}$ and $P(q)$ exhibits two Dirac-delta peaks at $\pm q_{EA}$, referring to the global spin reversal symmetry of the Ising model. For the Sherrington-Kirkpatrick model, \citet{parisi_jpa80, parisi_prl83} has shown the hierarchical (tree-like) structure in the overlapping $q_{\alpha \beta}$ for a given randomness in $J_{xy}$ and discussed the general features of $P(q)$ when averaging over the randomness in $J_{xy}$. It was found that $\langle \langle P(q) \rangle \rangle_J > 0$ if $-q_{EA}<q<q_{EA}$.

For more realistic spin glass models with short range interactions, basically different results have been reported by \citet{newman_pre98, middleton_prb13, billoire_prb14} who studied the averaged overlap probability distribution. The differences are related to the number of energy valleys, the consequences of spatial structure and the absence/presence of frustration. At the same time these investigations raised many new questions. For a brief survey of the differences and recent achievements we suggest reading the recent reviews by \citet{newman_ijmpb10} and by \citet{read_pre14}.

The effect of random-field $h_x$ on the thermodynamical behaviors of the $d$-dimensional ferromagnetic Ising model was investigated by \citet{imry_prl75, villain_jpl82, imbrie_prl84}, using different approaches. In these investigations $J_{xy}=J$, $\langle h_x \rangle =0$, and $\langle h_x^2 \rangle =h_R^2$ are chosen where the mean variance $h_R$ measures the randomness. The complexity of questions related to the existence of ferromagnetic phase in the low noise limit is well reflected by the facts that the initial studies lead to controversial results. The systematic Monte Carlo calculations \cite{gawlinski_prl84, grest_prb86, mackenzie_jpc86, frontera_pre99} have indicated the ferromagnetic phase in the presence of a weak randomness at low temperatures if $d\ge 2$, otherwise there is no average magnetization in the presence of a random field. In other words, the paramagnetic to ferromagnetic critical phase transition can occur if $h_R$ does not exceed a threshold value dependent on the lattice structures. Very recently \citet{shrivastav_pre14} have investigated the morphological properties of the spin arrangement in the two- and three-dimensional systems in the low noise limit and reported power law behaviors in the correlation functions. In that case  the paramagnetic phase possesses spin glass order $q_{EA}>0$, that increases when decreasing the temperature, although the system has only one paramagnetic state.

Finishing this section we emphasize that the evolutionary potential games describing biological and social systems demand the systematic investigation of the effects of additional randomness that can occur in the connectivity structure, the local dynamical rules, and even in the number of strategies. On the other hand, the characteristic features of the spin glasses may be suppressed when slow variation is allowed in the parameters characterizing the randomness.

\section{ORDERING PROCESSES}
\label{sec:ordproc}

A typical two-dimensional ordering process in the Ising model from the random initial distribution towards the equilibrium state was illustrated previously in Fig.~\ref{fig:myothd}. This process can be observed for a fixed noise level below its critical value $K_c$ and it has some universal features. Namely, two equivalent types of ordered strategy arrangements form a domain structure that is topologically similar to the cases where islands are in lakes located within larger islands located in larger lakes {\it etc.} for the infinitely large systems. This pattern can be characterized by the average distance $l(t)$ between two neighboring domain walls (along the horizontal or vertical cross-sections) that give a time dependent contribution to the equilibrium average potential. Namely, after a relaxation time, $U_{eq} - U(t)= \alpha / l(t)$ where $\alpha$ depends on the pair potential and noise level $K$, and $l(t) \propto \sqrt{t}$ as long as $l(t)<L$.

\begin{figure}[ht]
\centerline{\epsfig{file=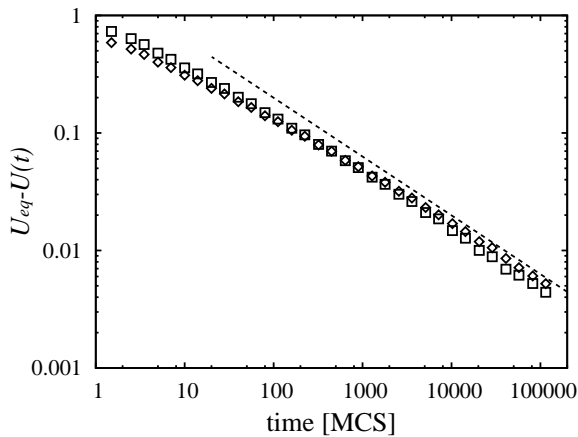,width=7.8cm}}
\caption{Log-log plot of averaged $U_{eq} - U(t)$ {\it vs}. time for $K=0.1$ (squares) and $K=0.34$ (diamonds) at $T=1.5$ and $S=0.2$. The dashed line shows the slope (-1/2) characterizing the theoretical prediction.}
\label{fig:domgrowt}
\end{figure}

Figure \ref{fig:domgrowt} shows two typical Monte Carlo results obtained when the system with hawk-dove parameters is started from a random initial state on a square lattice and the time-dependence of the average potential [$U(t)$] is recorded during the evolution for two noise levels.
Despite the large system size ($L=4000$) the functions $U(t)$ are decorated by fluctuations with an amplitude increasing with time. These undesired fluctuations are suppressed by averaging $U(t)$ (at $t=t_i \simeq 2^{i/2}$) over time intervals $0.8 t_i < t < 1.2 t_i$ for the clear illustration of the asymptotic behavior mentioned.
Notice, that it takes a longer time to achieve the asymptotic behavior when $K$ approaches the critical point due to the critical slowing down as indicated by diamonds in Fig.~\ref{fig:domgrowt}.

Different approaches have been developed to describe the above-mentioned time-dependent processes that remain valid also for the homogeneous three-dimensional systems. All the relevant features are well investigated by continuum description in terms of coarse-grained order parameter field [for detailed discussion we suggest consulting the reviews by \citet{hohenberg_rmp77}, \citet{langer_92}, and  \citet{bray_ap94}]. For this approach the systems are described by a scalar order parameter $\Phi$ dependent on time and continuous space that can be considered as average magnetization (or strategy density) over a small region.

Additionally, the motion of interfaces can be described by geometrical approaches for both the two- and three-dimensional systems as it is detailed by \citet{brakke_78}, \citet{brower_pra84}, and \citet{goldstein_prl91} (with further references therein). For these differential geometric descriptions the two-dimensional system is characterized by a set of closed curves representing the interfaces in the corresponding two-color map. The evolution of these curves is determined by a differential equation taking into consideration the average velocity of the interface depending on its local curvature, direction, and additional symmetries coming from the microscopic dynamical rules. For example, the reduction of the length of interfaces via the so-called curvature driven interface evolution (for a very recent geometrical survey see the work by \citet{garcke_jdmv13}) can explain different domain growing phenomena ($l \propto t^{\alpha}$ with $\alpha = 1/2$, $1/3$, or $1/4$). Additionally, these approaches can describe a wide variety of interfacial instabilities occurring in solid state systems. On the other hand, the applicability of this method is limited to systems possessing one type of interfaces.

\begin{figure}[ht]
\centerline{\epsfig{file=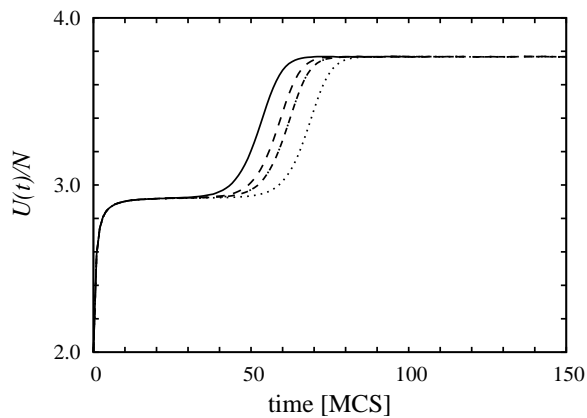,width=7.8cm}}
\caption{Average potential as a function of time during the sublattice ordering process on bipartite random regular graph for a hawk-dove game at $T=1.5$, $S=0.5$, $K=0.5$, and $N=4*10^6$. MC data of four independent runs are illustrated with different types of lines.}
\label{fig:hdrrgut}
\end{figure}

In comparison to the spatial systems the final ordered strategy arrangement is formed significantly faster on small-world connectivity structures due to the absence of large distances. For the demonstration of this phenomenon we have performed MC simulations for an evolutionary hawk-dove game on a random bipartite regular graph with a degree of 4. The system started from a random initial state and the order parameter, the average payoff and potential were recorded after each MC steps performed at a noise level below its critical value. During the first steps the sharp increase of potential (see Fig.~\ref{fig:hdrrgut}) refers to the formation of ordered microdomains.
Within this period the sublattice order parameter remains zero because of the large
number of microdomains of two types. The number of domains decreases gradually and after some time the size of the largest domains becomes comparable to the diameter of the given graph ($\propto \ln{N}$). Thereafter, the largest domain conquers the system. The parallel curves in Fig.~\ref{fig:hdrrgut} refer to similar scenarios and the randomness influences dominantly the time when ruling domain emerges. Evidently, on smaller systems the above behavior is disturbed by the stochastic events occurring both in the generation of the random networks and in the evolution of strategy distribution.

\subsection{Evolution in the limit $K \to 0$}
\label{sec:evolK0}

The pattern evolution in the Ising model on different lattices and graphs has been studied for several years in the zero noise limit \cite{biswas_pre11}. This particular case exhibits some curiosity.

For example, in the one-dimensional Ising model the long-range order can also be formed in the system if it is started from a random initial state while the evolution is controlled by the Glauber dynamics in the limit $K \to 0$ for $h = 0$. In that case the system becomes equivalent to the one-dimensional voter model \cite{clifford_bm73, liggett_85} and the interfaces (separating the opposite domains) move randomly. If two interfaces collide then both are annihilated. Finally the system evolves into one of the homogeneous states with a probability dependent on the initial magnetization. The average domain size increases in time as $l(t) \propto t^{1/2}$ \cite{bramson_ap80}.

In general, the $l(t) \propto t^{1/2}$ scaling law is valid for the two- and three-dimensional lattices, too. However, in several trials the system evolves into a frozen state \cite{lipowski_pa99, spirin_pre01} and these events modify the long-time behavior. The bottom snapshots in Fig.~\ref{fig:tendtoNE} show an example when rectangular boxes of the preferred phases are frozen into the opposite phase if the system is started form a random state being close to the ordered opposite structure. The application of the periodic boundary conditions may also result in frozen patterns in the finite systems. \citet{spirin_pre01} have reported that for about one third of the trials in the two-dimensional lattice the system evolves into a poly-domain state where all the interfaces are horizontal (or vertical). For most of the cases only two strips are formed. It is reported furthermore that the number of frozen states is larger for $d=3$.

Recently \citet{biswas_pre11} have studied the Ising system on a random network created from the one-dimensional lattice by adding new links into the connectivity structure. As a result the lattice sites have different degrees and for some constellations these irregularities were capable of blocking the domain growing processes, independently of the details of the generation of random networks. On the other hand, the freezing disappears in the limit $N \to \infty$ for densely connected networks \cite{das_epjb05}.

The appearance of frozen patterns is expected for the $n$-strategy games (or Potts models) and also for systems where the number of neighbors is increased.

\subsection{Interfacial phenomena and rearrangement through nucleation}
\label{sec:interfacial_phenomena}

\begin{figure}[ht]
\centerline{\epsfig{file=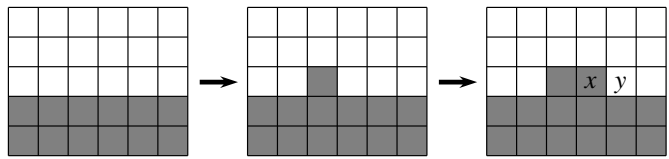,width=9cm}}
\caption{Consecutive steps in the evolution of a horizontal interface separating two ordered strategy arrangements on a square lattice.}
\label{fig:nucl1d}
\end{figure}

Many relevant phenomena for the ordering or reordering processes can be well interpreted by considering microscopically the evolution of an interface separating two ordered phases for the two-dimensional ferromagnetic Ising model. Figure \ref{fig:nucl1d} shows a horizontal interface between the up- and down-spin ordered phases that remains stable at $K=0$ even in the presence of a weak magnetic field. Due to the rare stochastic events at low noise levels, one of the spins may reverse along the interface as denoted by the middle panel in Fig.~\ref{fig:nucl1d}. The ferromagnetic nearest neighbor interactions enforce this spin to flip back, which represents the typical behavior. Sometimes, however, before the reconstruction of the smooth interface one of the neighboring spins may also reverse thus forming a two-spin cluster.

The appearance of a two-spin cluster along this interface results in a new situation when the spins at sites $x$ and $y$ in Fig.~\ref{fig:nucl1d} have similar environments, namely, there are two up-spins and two down-spins in their neighborhood and the preferred spin reversal is determined by the magnetic field. As a result this two-spin cluster can be considered as a nucleon from which the expansion (or growing) of the preferred state can start. The direction of the motion of steps and also its average velocity can be quantified by comparing the probabilities of spin reversals at sites $x$ and $y$ in the third panel of Fig.~\ref{fig:nucl1d}. The quantitative analysis justifies that the average vertical velocity is proportional to the magnetic field (if $|h| \ll 1$).
Throughout these consecutive elementary steps the system will evolve into the thermodynamically stable ferromagnetic state if $0 <K <K_c$.

The average vertical velocity of the above horizontal interface depends on two factors: the average horizontal velocity of the steps and the frequency of these steps along the interface. The second quantity depends on time and also on the nucleation rate characterizing the probability of the appearance of a sufficiently large nucleon. If the nucleation rate is low, then the interfaces are composed of large horizontal and vertical segments, otherwise the domain pattern is almost isotropic.

Similar phenomena can be observed along the interfaces separating the equivalent anti-ferromagnetic domains. In that case the presence of the homogeneous external field $h$ modifies the interface as it is illustrated in the snapshots of Figs.~\ref{fig:NE-sHD}, whereas the the average velocity of the interface is zero. At the same time, the application of a staggered magnetic field prefers one of the ordered spin arrangements to the opposite one and results in an average velocity proportional to $h_s$ (if $h_s \ll 1$).

In the above process the appearance and expansion of the one-dimensional preferred nucleons play the crucial role in the evolution towards the final stationary state. Similar mechanism can be observed in $d$-dimensional lattices when the appearance of sufficiently large islands of the preferred ordered phase is necessary to initiate the transition from an ordered phase to the thermodynamically stable final spin arrangement. That happens when we wish to reverse the direction of magnetization by the application of an external magnetic field. In thermodynamical systems, the formation of the sufficiently large nucleon of the preferred phase is supported by a suitable series of stochastic elementary steps that may occur rarely. Sooner or later, however, some sufficiently large nucleons appear and catalyze the transition towards the stable phase via a domain growing process. The resident time in the meta-stable state(s) may be extremely long, especially at low noise levels. These phenomena are well investigated in a wide scale of physical systems and exploited in many products of high technology.

In two-strategy evolutionary potential games the growth of the $C$ domains on a square lattice can be characterized by the average velocity $v$ of the moving step shown in Fig.~\ref{fig:nucl1d}. If logit rule controls the evolution then
\begin{equation}\label{eq:stepvel}
v={e^{u_x(C)/K} \over e^{u_x(C)/K}+e^{u_x(D)/K}} -
  {e^{u_y(D)/K} \over e^{u_y(C)/K}+e^{u_y(D)/K}} .
\end{equation}
Due to the similar strategy arrangements in the neighborhood the denominators are equal and
\begin{equation}\label{eq:stepvsd}
v={e^{2(1+S)/K} - e^{2T)/K} \over e^{2(1+S)/K} + e^{2T)/K}}
\end{equation}
in the social dilemma notation. As a result, within the region of stag hunt game the condition $v=0$ defines a straight line ($S=T-1$) on the $T-S$ plane separating homogeneous cooperation (${\bf s}_x=C$) and defection (${\bf s}_x=D$) regions in the low noise limit (in agreement with the phase diagram plotted in Fig.~\ref{fig:isi_map}).

\subsection{Interfacial phenomena in three- and $n$-state systems}
\label{sec:interface_nstate}

\begin{figure}[ht]
\centerline{\epsfig{file=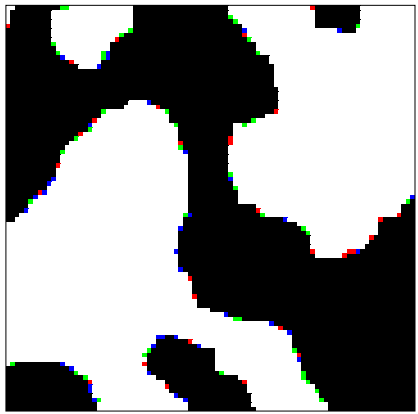,width=5.5cm}}
\caption{(Color online) Two homogeneous domains of strategies 1 (white) and 2 (black) are growing in the two-dimensional system if the interaction is defined by ${\bf A}={\bf d}(1,2)$ for $n=5$. During the evolutionary process the strategies $i=3, 4,$ and 5 (denoted by red, blue and green boxes) occur dominantly at the steps of interfaces separating the white and black territories.}
\label{fig:d12n5dg}
\end{figure}

During evolutionary processes the interfacial phenomena can play crucial roles in the $n$-strategy ($n>2$) systems, too. First we show what happens during the domain growth if the coordination type interaction between strategies 1 and 2 is extended by additional (neutral) strategies. Figure \ref{fig:d12n5dg} illustrates the formation of two homogeneous domains (with strategy 1 and 2) if ${\bf A}={\bf d}(1,2)$ [see the definition (\ref{eq:isingsubgames}) at $n=5$] if the two dimensional system is started from a random initial for logit rule at a noise level ($K=0.4$).
For this low noise level the strategies 3, 4, and 5 can occur very rarely inside the homogeneous domains. At the same time the mentioned strategies will be selected dominantly by the players located along the interfaces where the opposite effects of neighbors (with strategies 1 and 2) are balanced (see the sites $x$ and $y$ indicated in the right plot of Fig.~\ref{fig:nucl1d}). In the present system the appearance of the additional strategies along the interfaces does not modify the general features of domain growing process.

If the interactions are dominated by ${\bf d}(1,2)$ then the equivalence between the strategies 1 and 2 is evidently broken by the presence of self-dependent components and also by the additional coordination type interactions. More precisely, the preference of strategy 1 (or 2) depends on the value of $\gamma_1-\gamma_2$ and also on the strengths of the components ${\bf d}(1,3)$ and ${\bf d}(2,3)$. The competition between these
components may result in different phase transitions as it is discussed by \citet{vukov_pre15} in a three-strategy evolutionary game.

For another example we mention the attractive $n$-state Potts model evolves towards one of the homogeneous ordered states throughout a domain growing process at low noise levels, if the system is started from a random initial state.

\begin{figure}[ht]
\centerline{\epsfig{file=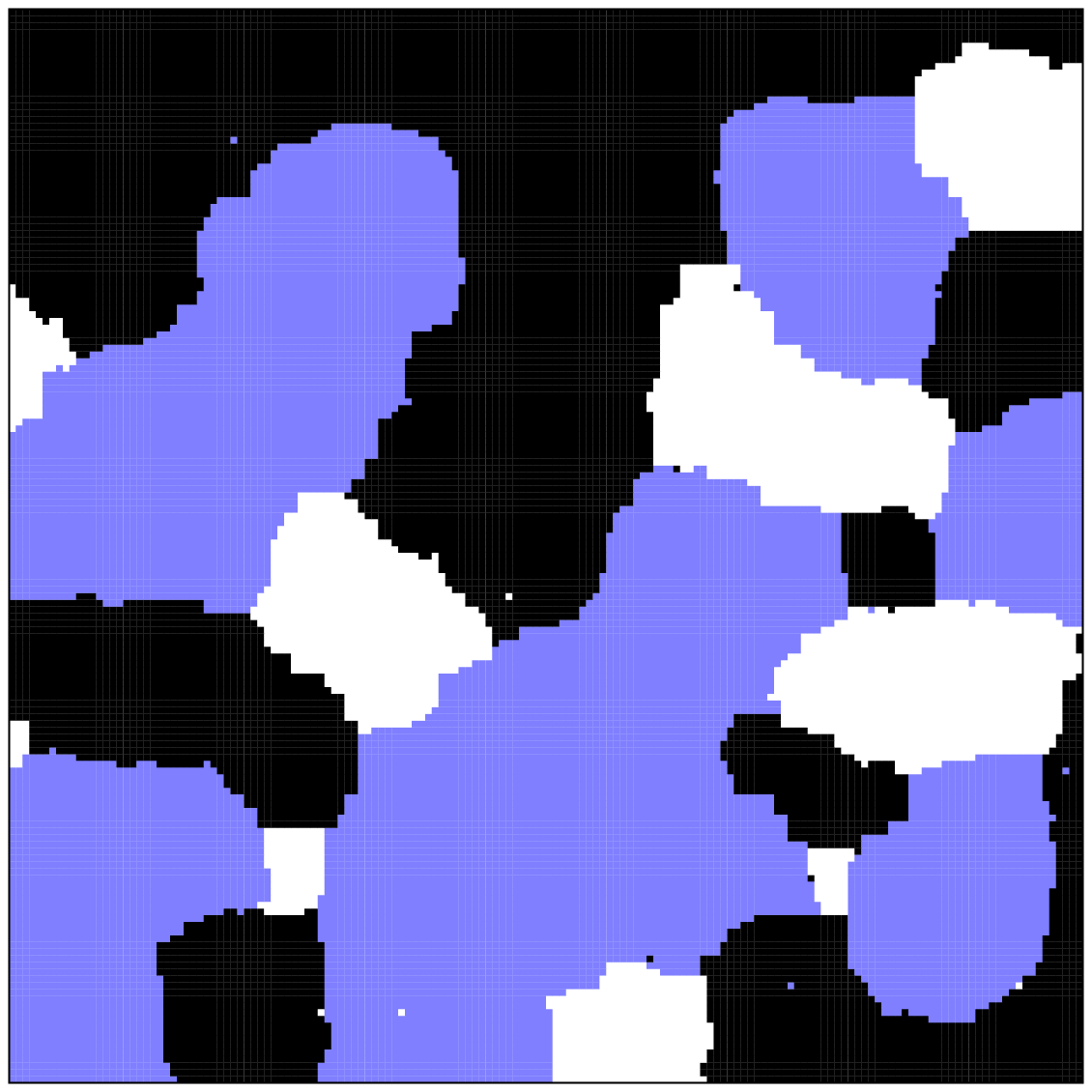,width=5cm} \epsfig{file=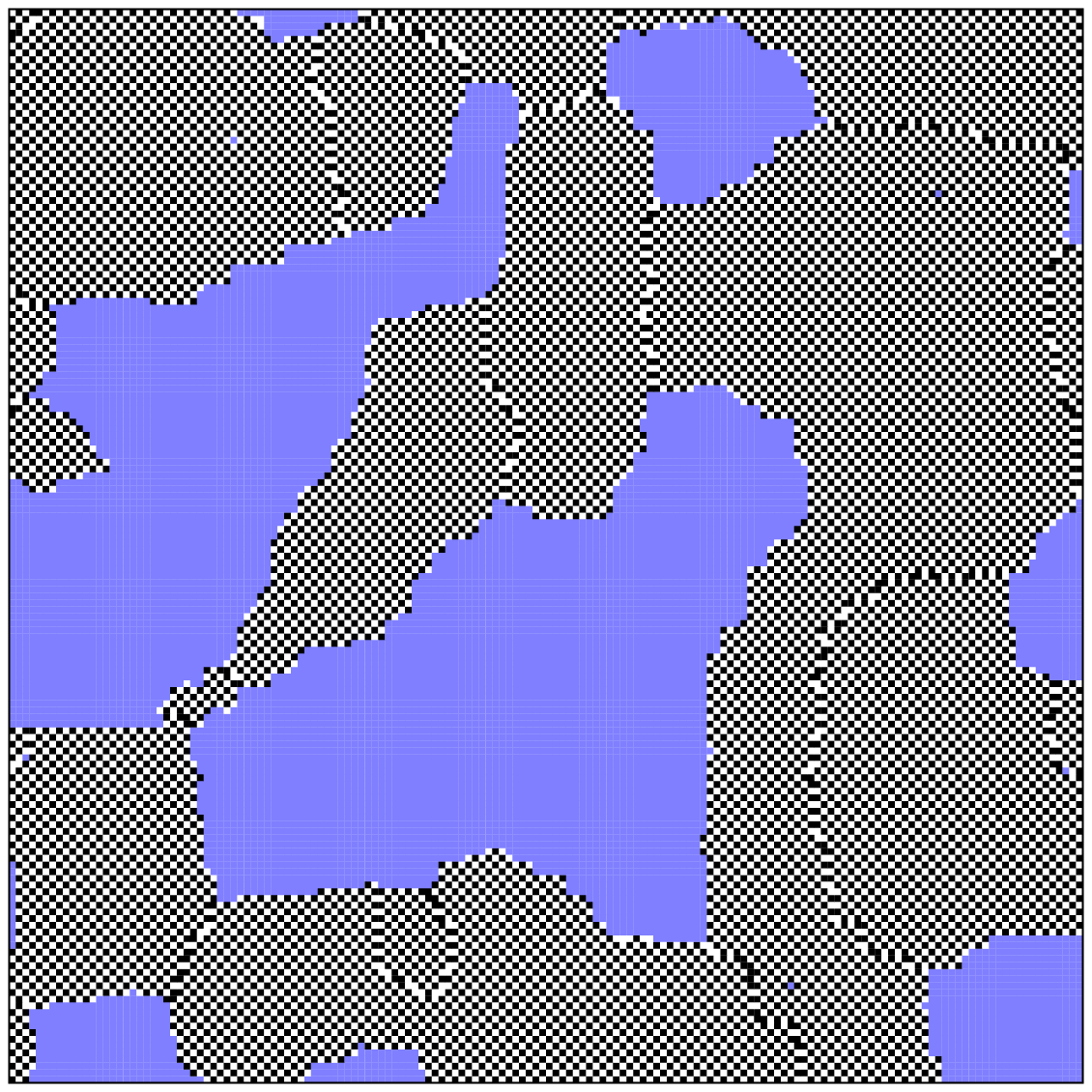,width=5cm}}
\caption{(Color online) Typical strategy distributions on a square lattice for two types of the three-strategy coordination games during the domain growing process after $500$ MCS if the system is started from a random initial state and the evolution is controlled by logit rule at low noise levels ($K \simeq 0.5 K_c$).}
\label{fig:g8_g6}
\end{figure}

A similar phenomenon can also be observed for some spatial coordination games \cite{szabo_pre14b}. The left snapshot of Fig.~\ref{fig:g8_g6} illustrates the spatial distribution of three strategies during the domain growing process in a system where ${\bf A}={\bf d}(1,2)+{\bf d}(2,3)+{\bf d}(1,3)$ for $n=3$].
The right hand snapshot of Fig.~\ref{fig:g8_g6} shows similar domain growing process in a system where the payoff matrix ${\bf A}^{\prime}$ is obtained from ${\bf A}$ by exchanging its first and second rows. As discussed in Sec.~\ref{sec:decomposition_smg} and \cite{szabo_pre14b} there are two other equivalent systems that can be transformed into each other by exchanging two strategy labels in one of the sublattices. A similar symmetry is behind the equivalence of the ferromagnetic and anti-ferromagnetic Ising model on bipartite graphs.

In two-dimensional spatial systems the geometrical features of these domain patterns differ significantly from those occurring in the two-state systems, where the interfaces form closed loops typically in the infinitely large systems (as mentioned in Sect.~\ref{sec:ordproc}). For $n=3$ one can distinguish three types of interfaces that can form closed loops or represent a planar network with three-edge vertices.

In spite of the striking geometrical differences the average domain size increases with the square root of time ($l(t) \propto t^{1/2}$) as reported by \citet{grest_prb88} and the geometrical features of these patterns become similar on the scale of $l(t)$ as it is observed for the Ising model and other field theoretical models surveyed by \citet{bray_ap94}.

In the present system, the existence of the third strategy does not influence the behavior of interfaces if the point defects are distributed sparsely. If the attractive Potts model includes a magnetic field favoring one of the homogeneous states then one can observe the expansion of the favored domains along their interfaces with an average velocity proportional to the magnetic field.

A similar behavior is expected for three-strategy potential games if the payoff matrix ${\bf g}(8)$ is weakly disturbed by additional self-dependent components (\ref{eq:self3x3}). In those cases the difference between the corresponding $\gamma_i$ values will determine the direction (and also the average velocity) of invasions between two "homogeneous" domains. For example, this mechanism results in the prevalence of strategy 1 in the final stationary state if $\gamma_1 > \gamma_2, \gamma_3$.

The right hand snapshot of Fig.~\ref{fig:g8_g6} illustrates a system that is not yet studied systematically within the framework of Potts models. In the latter evolutionary game one can observe a homogeneous domain formed by strategy 3 and two equivalent chessboard like ordered strategy arrangements of the strategies 1 and 2. The most striking feature of this game is related to an inherent symmetry that we have discussed previously when justifying the equivalence of the ferromagnetic and anti-ferromagnetic Ising models on bipartite networks (see Sec.~\ref{sec:cplat}). According to the generalization of the mentioned method, game ${\bf g}(8)$ on a bipartite network becomes equivalent to those where the pair interactions are defined by ${\bf g}(6)$ if the players in sublattice $Y$ exchange the labels of their first and second strategies ($1 \leftrightarrow 2$). The latter transformation is equivalent to the exchange of the first and second row of the payoff matrix. Due to the mentioned relation the three domains are equivalent on the square lattice and after a domain growing process one of these ordered structures will prevail in the finite systems. When increasing the noise level $K$ this system undergoes an order-disorder critical phase transition belonging to the universality class of the three-state Potts model.

One can generate two additional potential games with payoff matrices obtained from ${\bf g}(8)$ by exchanging the labels $1 \leftrightarrow 3$ or $2 \leftrightarrow 3$ for the players staying in sublattice $Y$. The resulting games are similar to ${\bf g}(6)$. These relatives of the Potts model can be constructed as suitable linear combinations of three elementary games (${\bf g}(6)$, ${\bf g}(7)$, and ${\bf g}(8)$ \cite{szabo_pre14b}. It turned out that the corresponding three-dimensional subset of games can be considered as a generalization of the Potts model. Here we have to emphasize that the additional self-dependent components are not capable of preferring one of the sublattice ordered two-strategy structures to its anti-pattern.

The exploration of the three-strategy symmetric potential games is not complete. Even more complex behavior is expected for $n>3$ when the number of interfaces as well as the types of vertices increases with $n$. The preliminary results have indicated the dominance of three-edge vertices that follows a complicated transition/annihilation rule \cite{loureiro_pre12}. The consideration of nonsymmetric games will allow us to study the effect of those types of self-dependent components that distinguish the chessboard and anti-chessboard arrangements of two strategies, as it is done when applying a staggered magnetic field to the anti-ferromagnetic Ising model.

In the above-discussed models the average motion of the invasion fronts is driven by the increase of individual payoffs that is quantified by the increase of ${\cal U}$ in the thermodynamic potential $\Phi$ (\ref{eq:thermpot}) if $K \to 0$. In the maximization of $\Phi$, however, the high entropy of the disordered phase, especially for a large number of strategies, can become the leading term at sufficiently high values of $K$.

\begin{figure}[ht]
\centerline{\epsfig{file=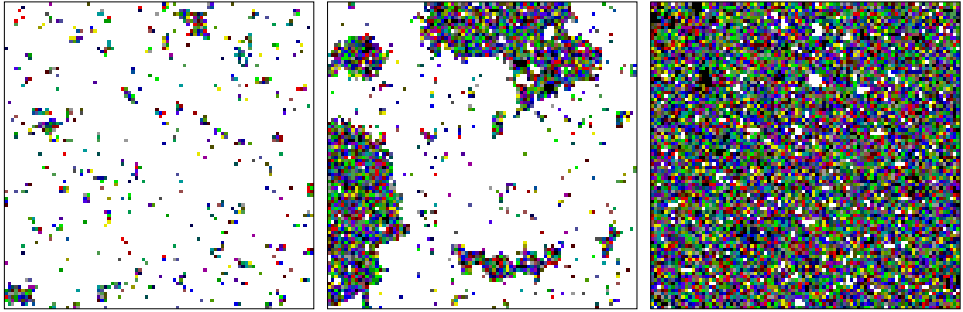,width=13cm}}
\caption{(Color online) Three consecutive snapshots at times $t=200$, 1000, and 3000 MCSs (from left to right) show a domain growing process when the (white) homogeneous spatial strategy distribution (${\bf s}_x=1$) transforms into the disordered phase composed of $n=50$ strategies (distinguished by different gray scales).}
\label{fig:d12disord}
\end{figure}

Figure \ref{fig:d12disord} illustrates a domain growing process on a square lattice from one of the ordered phases (here ${\bf s}_x=1$) to the disordered strategy arrangement if the interaction is described by $d(1,2)$ for $n=50$ at a noise level $K=0.52 > K_c(50)=0.512(2)$. The process begins with a nucleation procedure that is followed by the expansion of the disordered territories where all the strategies are present with approximately equivalent probabilities.

Here the evolution is also controlled by the logit rule, therefore we can estimate the average velocity $v$ of a step along the interfaces in the same way as it is described in Sec.~\ref{sec:interfacial_phenomena}. Neglecting the appearance of strategy 2 the value of $v$ can be approximated as
\begin{equation}\label{eq:stepv50}
v \simeq {e^{2/K} \over e^{2/K}+n-2} -
  {n-2 \over e^{2/K}+n-2}\, .
\end{equation}
Accordingly, the disordered phase expands at the expense of ordered phase if $v<0$, that occurs if $K> 2 / \ln{(n-2)}$ for sufficiently large values of $n$ when the $K$-dependence of strategy frequencies exhibits a first-order phase transition. The criterion $v=0$ gives us an estimation for the critical point, $K_c(n)=2/\ln{(n-2)}$, that agrees very well with the Monte Carlo result given above for $n=50$.

The latter result implies that the high-entropy phases can occur for any noise level if $n$ is sufficiently large. Here it is worth mentioning that similar arguments justify the stability of high-entropy alloys at room temperatures. Recently the high-entropy alloys are studied progressively and considered to be a promising family of materials with a wide scale of applications \cite{senkov_im10, yeh_jom13, carroll_sr15}.

Anyway, the average velocity of an interface can be determined numerically if the Monte Carlo simulations are started from artificial initial state where the competing phases are represented by two domains with equal sizes. Using this method one can evaluate the phase boundaries in the phase diagrams more accurately \cite{szolnoki_pre11, szolnoki_pre11b}, particularly if the first order transition is accompanied with a hysteresis \cite{hintze_pb15} or sensitivity to the initial state \cite{shigaki_pone13}.

\subsection{Slow relaxation in random systems}
\label{sec:evolinrs}

Up to now we have mainly discussed the stationary states of the systems (in the limit $N \to \infty$). In general, homogeneous systems evolve towards the stationary states exponentially if the state is weakly perturbed. We have mentioned two exceptions when the homogeneous system reaches the final state more slowly. In the first case the system evolves form a random initial state into one of the ordered arrangements through a domain growing process and the average domain size (or correlation length) increases with $\sqrt{t}$ as detailed above. The second case occurs at the critical point where system behavior is dominantly controlled by the fluctuations and results in a power law decay in most of the quantities. Now we briefly discuss the slow relaxation processes observed in the Griffiths phase of the random Ising systems.

The slow (nonanalytic) relaxation of the magnetization in the paramagnetic phase of a random ferromagnetic Ising model at $h=0$ was reported by \citet{griffiths_prl69} who studied a diluted Ising model on a lattice where a portion of the lattice sites are not occupied by Ising spins \cite{griffiths_jmp68}. Subsequent analyses have indicated the presence of Griffiths phase in many other random Ising models between the paramagnetic and ferromagnetic (or spin-glass) phases. According to the investigation of different models the one-site two-time correlation function [defined in Sec.~\ref{sec:ordproc}] is found to have a "stretched-exponential" form, $g(0,t) \simeq \exp[- ( t /\tau)^{\kappa}]$ with $0 < \kappa < 1$, depending on the spatial dimension and other details of the system \cite{palmer_prl84, dedominicis_jpl85, randeria_prl85, bray_prl87}.

The related spatial patterns assume the existence of sparse and large ordered domains in the random environment. The thermalization (spontaneous reversal) of these sparse domains is very slow because it requires a long sequence of coherent flipping over a large volume. According to this picture the time-dependence of magnetization (or any order parameter) can be approximated as
\begin{equation}\label{eq:m_t}
m(t)=\int_{0}^{\infty} w(\tau) \exp [-t/\tau]\, d\tau
\end{equation}
with a suitable choice of the weight function $w(\tau)$. \citet{noest_prb88, noest_prl86} has shown that if the relaxation time increases exponentially with the size $n$ of a compact cluster ($\tau \sim \exp[a n]$) and the probability of such clusters decreases exponentially with $n$, then the leading term of the asymptotic behavior of $m(t)$ can be estimated as
\begin{equation}\label{eq:m_powlaw}
m(t) \sim t^{-\theta}
\end{equation}
where $\theta$ ($\theta >0$) depends on $a$ and on other parameters within the Griffiths phase. It is emphasized that similar behavior is reported for several other systems, {\it e.g.}, stochastic cellular automata \cite{noest_prl86} and contact processes with quenched disorder in the environment \cite{moreira_pre96}. Today the contact process \cite{harris_ap74} is considered as the paradigm of systems where the extinction of a species/strategy exhibits a critical transition that belongs to the directed percolation universality class \cite{kinzel_zpb85}. For a survey of the main features of this critical transition we suggest consulting the review by \citet{hinrichsen_ap00}. In the latter system the quenched randomness modifies also the system behavior at the critical point \cite{dickman_pre98}. The presence of Griffiths phase and its consequences are described by \citet{munoz_prl10} for the contact process on complex networks.

The occurrence of the Griffiths phase in the contact process has implied algebraic extinction of a strategy in many evolutionary games where the evolutionary rule is based on imitation in spatial systems with quenched disorder \cite{droz_epjb09}. In most of the evolutionary games with quenched randomness, however, the appearance of Griffiths phase is not investigated although it can cause incorrect numerical data in the vicinity of the critical point(s).

The above theoretical picture supposes that the relaxations of the domains are independent of each other ({\it i.e.}, the rare events are not organized hierarchically). This feature simplifies the numerical analysis of these systems as the Monte Carlo simulations can be performed simultaneously on many "small" systems. In the opposite cases, when the models involve hierarchically constrained dynamical processes \cite{palmer_prl84}, more complex finite-size analyses are required.

At the end of this section we underline the relevance of the Griffiths phase in the evolutionary games modeling biological or social systems where the quenched randomness is assumed naturally \cite{stein_13}. For the numerical analysis of these systems in the Griffiths phase we have no chance to achieve the final stationary state due the the slow algebraic relaxation processes. It means, on the one hand, that in the Griffiths phase the final stationary quantities should be determined by extrapolation of the asymptotic behavior. On the other hand, for the interpretation of the experimental and numerical data we should consider the fact that the system has not achieved its equilibrium state.

The Griffiths phase represents technical difficulties in numerical simulations that can be avoided by introducing slow variation in the randomness that is also a natural ingredient of biological and social systems. The relevant differences between the quenched and temporal randomness are detailed by \cite{hinrichsen_ap00} for the contact process.

\section{DEVIATIONS FROM THERMODYNAMICAL EQUILIBRIUM}
\label{sec:dfepg}

In the absence of potential the existence of Boltzmann distribution becomes meaningless and we cannot apply the results of equilibrium statistical physics and thermodynamics. Additionally, the validity of thermodynamics is dropped when applying an evolutionary rule that breaks the detailed balance and drives the system far away from the Boltzmann distribution, as happens, for example, in the imitation-based dynamics used frequently in many previous investigations. For some types of coevolutionary games the absence of the fixed connectivity structure, the possible changes in personal features and spatial location raise many additional questions whose analyses go beyond the scope of the present work.

In the following sections we discuss briefly some effects of the matching pennies and rock-paper-scissors games that can be studied in multi-agent evolutionary games even for the application of logit rules. The discussion of these games is unavoidable because of their important role in the subgames of any $n \times n$ matrix games, which can affect the system's behavior significantly.

\subsection{Effects of matching pennies}
\label{sec:emp2x2}

The matching pennies game, defined by ${\bf f}^{\prime}(8)$ in its bimatrix form (\ref{eq:newbasis2a}), represents the simplest cyclic interaction. For a two-player evolutionary game with logit rule the unsatisfied player reverses her strategy with a high probability dependent on $K$, while the opposite transitions become rare. Thus this interaction breaks the detailed balance and can be considered as a microscopic force inducing cyclic variations in the systems. The effect of this interaction can be quantified by the probability current(s), measuring the difference in the frequency of forward and backward transitions along the directed edges of the flow graph (see Fig. \ref{fig:matpen}). This current is uniform along the four edges in the stationary state of this stochastic process.

\begin{figure}[ht]
\centerline{\epsfig{file=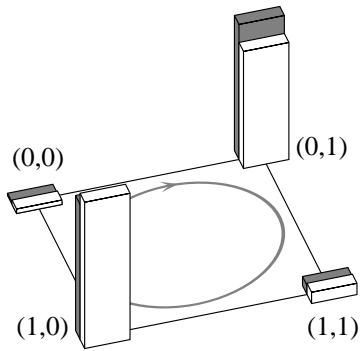,width=5cm}}
\caption{Probabilities of the four strategy profiles are proportional to the height of the dark ($\varepsilon = 0$) and white ($\varepsilon = 0.1$) columns for a two-person hawk-dove game if it is extended by a matching pennies component with a strength of $\varepsilon$. The arrowed gray circle denotes the probability current loop induced by the matching pennies game.}
\label{fig:hdmp2x2c}
\end{figure}

The induced circular probability current creates observable variations in the probability of the strategy profiles. For the illustration of this phenomenon we discuss a system where a weak matching pennies component is added to the payoff of a hawk-dove game. The variations in the probabilities are illustrated in Fig.~\ref{fig:hdmp2x2c}.
In this figure the height of the dark columns denotes the probability of each strategy profile in the stationary state when ${\bf G}={\bf G}^{\rm (sd)}$ as defined by (\ref{eq:socdilG}). For the given parameters ($T=1.4$, $S=0.3$, and $K=0.3$) $p(1,1) < p(2,2)$ whereas $p(1,2)=p(2,1)$. If the latter game is modified by adding a weak matching pennies component to the payoffs (quantitatively ${\bf G}={\bf G}^{\rm (sd)}+\varepsilon {\bf G}^{\rm (mp)}$ with $\varepsilon = 0.1$) then the appearance of probability current is accompanied with a striking variation in the stationary state. The largest increase of $p(2,1)$ can be interpreted as the consequence of a congestion phenomenon. Accordingly, for the maintenance of the circular probability current, the lowest value of p(1,1) plays the role of a narrower bottleneck that creates an increase in $p(2,1)$ to ensure uniform probability current through the four-edge loop. At the same time the value of $p(1,2)$ is decreased, that is, the presence of the matching pennies component destroys the equivalence of the two Nash equilibria.

In the multi-agent version of this evolutionary game on the square lattice, the above microscopic effect occurs for each interacting neighbor and affects the macroscopic behavior. This extension of the models can be performed for those lattices that can be divided into two sublattices ($X$ and $Y$, where the two types of players are located separately). As a result of this breaking of the original symmetry, one of the sublattice ordered strategy arrangements can be preferred in the low noise limit. Evidently, the preference is reversed with the sign of $\varepsilon$. Furthermore, the preference is also reversed if we consider the upper half part of the hawk-dove game, where $S>(T-1)$ and $p(2,2) < p(1,1)$, because here the strategy pair $(2,2)$ plays the role of the narrower bottleneck in the circulation. In fact, the effect of the matching pennies component is similar to the application of a suitable staggered magnetic field in the anti-ferromagnetic Ising model.

Along the line $S=(T-1)$ in the parameter space, the above effect does not work because here $p(1,1)=p(2,2)$. In these systems the presence of the matching pennies component does not destroy the universal features of Ising type critical transition \cite{szabo_pre14}, whereas the value of $K_c$ is reduced (proportionally to $\varepsilon$) if its strength does not exceed a threshold value. In the latter case the pair interaction belongs to the classes of ordinal potential games, because weak contribution of the matching pennies is not enough to change the edge directions in the flow graph.

In non-equilibrium systems the breaking of detailed balance can be well quantified by evaluating the entropy production $I$ (for a survey see \cite{schnakenberg_rmp76}). This quantity is constructed from the frequencies of the forward and backward transitions between states ${\bf s}$ and ${\bf s}^{\prime}$ if these transitions are allowed by an elementary step in both directions. Now the random consecutive elementary steps consist of single site strategy changes from $({\bf s}_x, {\bf s}_{-x})$ to $({\bf s}_x^{\prime}, {\bf s}_{-x})$ appearing with a frequency $W({\bf s}_x \to {\bf s}_x^{\prime})$ in the stationary state. The entropy production summarizes the contributions of the forward and backward transitions for each possible transition pair in the following way:
\begin{equation}
\label{eq:ep}
I={1 \over 2}\sum_{s_x,s_x^{\prime} \atop s_{-x}}
 [W({\bf s}_x \to {\bf s}_x^{\prime}) - W({\bf s}_x^{\prime} \to {\bf s}_x)]
\ln { W({\bf s}_x \to {\bf s}_x^{\prime}) \over W({\bf s}_x^{\prime} \to {\bf s}_x)} .
\end{equation}
Notice that this quantity is always positive ($I>0$) except the case $I=0$ when the conditions of detailed balance are satisfied, {\it i.e.}, if $W({\bf s}_x \to {\bf s}_x^{\prime}) = W({\bf s}_x^{\prime} \to {\bf s}_x)$ $\forall {\bf s}_x, {\bf s}_x^{\prime}, {\bf s}_{-x}$. Despite the large number of transitions in lattice systems the entropy production can be well estimated by recognizing that the transition frequency $W({\bf s}_x \to {\bf s}_x^{\prime})$ depends dominantly on the close neighborhood of player $x$, who modifies her strategy, if the dynamics is controlled by short range interactions. Besides it we can exploit the translation invariance of the lattice system. As a result the specific entropy production ($I/N$) can be estimated by considering the transition frequencies at any site ${\bf x}$ for all possible strategy configurations in its close neighborhood.

For example, during the Monte Carlo simulations on a square lattice one can determine the transition frequencies ${\bf s}_x \to {\bf s}_x^{\prime}$ for all the $2^4=16$ possible strategy configurations of the four nearest neighbor sites and also for the cases when $2^8=256$ configurations are distinguished on the first- and second neighbor sites. In this way we can deduce two approximate results for the specific entropy production ($I/N$) and comparison of them indicates the relevance of the second neighbors although they do not influence directly the transition in the present models. Evidently, the larger the neighborhood, the more accurate is the present approach (for a more detailed description of this approach see \cite{szabo_pre10b}).

\begin{figure}[ht]
\centerline{\epsfig{file=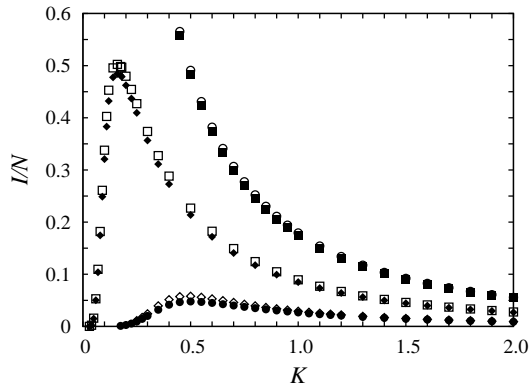,width=7cm}}
\caption{Specific entropy production as a function of $K$ for evolutionary game with ${\bf G}={\bf G}^{\rm (sd)}+\epsilon {\bf G}^{\rm (mp)}$ pair interactions on the square lattice if $T=1.4$, $S=0.3$, and $\epsilon=0.1$ (open diamonds and bullets), 0.175 (open boxes and closed diamonds), 0.25 (open circles and closed boxes). The open symbols denote data obtained for the larger neighborhood.}
\label{fig:hdmpep_k}
\end{figure}

Figure \ref{fig:hdmpep_k} shows the Monte Carlo results for the specific entropy production when varying the noise level for different strengths of the matching pennies component. Here the closed symbols represent data obtained when only the first neighbors are taken into consideration in the identification of ${\bf s}_{-x}$. Notice that the latter approximate data are close to those we obtained when a larger neighborhood (the first and second neighbors of player $x$) is  used for the characterization of ${\bf s}_{-x}$.
The small differences between the two sets of data justify the reliability of this approach. Notice furthermore another general feature: $I/N$ vanishes in the limit $K \to \infty$ when only the randomness controls the players' decisions.

Figure \ref{fig:hdmpep_k} illustrates that the force of ordering (strength of hawk-dove component) blocks the strategy reversals and also the breaking of detailed balance in the low noise limit if $\varepsilon$ is less than a threshold value, more quantitatively, $|\epsilon| < |\epsilon_{th}|$ where $|\epsilon_{th}| = 1/2 \min (|T-1|, |S|)$. In the opposite limit, the specific entropy production diverges if $K \to 0$ and we can observe a remarkably different behavior.

The divergency of $I/N$ in the low noise limit is characteristic to systems where cyclic dominance controls the system behavior and prevents the formation of ordered strategy arrangements. A similar divergency occurs in systems where a finite portion of the transition pairs becomes unidirectional. If the evolution is defined by only the matching pennies component in a spatial evolutionary game then the visual observation indicates random strategy distribution on a square lattice. The quantitative analysis, however, has clearly indicated weak correlations between the second and third neighbors \cite{szabo_pre14} that evidently vanish in the high noise limit.

\subsection{Effects of rock-paper-scissors game}
\label{sec:ditpi}

If the multi-agent evolutionary games are composed of equivalent players with three strategies, then the rock-paper-scissors component is responsible for the deviation from the potential games. The rock-paper-scissors game itself creates a weakly correlated random distribution of the three strategies on a square lattice if a logit rule controls the evolution. The three strategies are present with the same probability ($1/3$) and the numerical investigations indicate a weak spatial correlation that is similar to those found for the matching pennies game. The latter analogy implies that the alternation of the three strategies on each site reflects relevant breaking of the detailed balance, particularly in the limit $K \to 0$ when $I/N$ diverges.

\begin{figure}[ht]
\centerline{\epsfig{file=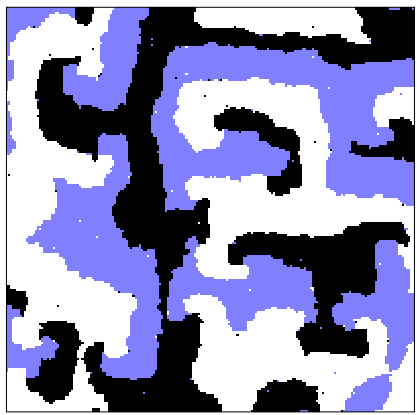,width=5.5cm}}
\caption{(Color online) Rotating spiral arms in the snapshots characteristic to the self-organizing pattern on the square lattice where the pattern evolution is controlled by attractive (ferromagnetic) three-state Potts type interaction and a rock-paper-scissors component with a strength of $\epsilon=0.1$ for logit dynamics at a low noise level.}
\label{fig:rps_p}
\end{figure}

Contrary to the two-strategy games with a matching pennies component, the presence of a weak rock-paper-scissors component can cause more relevant changes in the macroscopic behavior of the multi-agent spatial three-strategy evolutionary games. To demonstrate this, we consider the three-state Potts model on a square lattice if the uniform attractive pair interactions are modified by introducing a weak cyclic dominance, {\it i.e.}, ${\bf G} = {\bf G}^{\rm (Potts)}+\epsilon {\bf G}^{\rm (rsp)}$ where ${\bf G}^{\rm (Potts)}= {\bf d}(1,2)+{\bf d}(2,3)+{\bf d}(1,3)$. As mentioned before, the $n=3$ Potts model evolves into one of the three ordered states at low noises. In the presence of a weak cyclic dominance, however, the domain growth is blocked by the formation of rotating spiral arms as illustrated in Fig.~\ref{fig:rps_p}. The spiral form of the rotating edges is a direct consequence of the fact that the component $\epsilon {\bf G}^{\rm (rsp)}$ induces an average invasion velocity independent of the direction of the interface and of the distance measured along the interface from the center of the given three-edge vortex (vortex means rotating vertex). Due to the cyclic symmetry, all the three edges of a vortex (anti-vortex) rotate clockwise (anti-clockwise). Sometimes the moving interfaces meet and may annihilate each other or create a new vortex-anti-vortex pair. The latter processes are accompanied with a rearrangement of the connections (represented by the interfaces) between the vortices and anti-vortices. Some geometrical features, namely, the average curvature of interfaces, the average distance of vortices, the average length of interfaces between a vortex and anti-vortex pair, were already investigated by \citet{szolnoki_pre04, szolnoki_pre05} who found that the critical transition is suppressed in the presence of cyclic dominance and the correlation length diverges as $\xi_{ij} \propto 1/\epsilon$ if $\epsilon \to 0$ at sufficiently low noise levels.

If the previous model is modified by additional potential game components that favor one of the homogeneous states, then a sufficiently weak cyclic component cannot prevent the formation of an ordered state at low noises. In these cases the interaction itself might be considered as an ordinal potential game. A further increase in the cyclic component, however, can maintain a self-organizing spatio-temporal pattern (see Fig.~\ref{fig:rps_p}) in which the portions of the three strategies are different. In the practice of evolutionary game theory this mechanism is exploited when the undesired effect of social dilemma is reduced by introducing a third strategy representing voluntarism \cite{hauert_s02}, ''tit for tat'' \cite{nowak_jtb89}, or punishment \cite{fehr_n02}.

It is emphasized that similar rotating spirals are observed in many other systems including Belousov-Zhabotinsky reaction \cite{field_jcp74, showalter_c15}, excitable media ({\it e.g.} cardiac muscle \cite{wiener_aic46}, neural systems \cite{hempel_prl99}), epidemiological models \cite{kermack_prsa27}, and biological/ecological models \cite{durrett_jtb97, kerr_n02, frey_10}. The robustness of similar spatio-temporal patterns is also demonstrated by numerous three-strategy spatial evolutionary games (for a survey see \cite{szabo_pr07}). It is already well-known that the rock-paper-scissors type cyclic dominance helps the maintenance of all the participating strategies/species \cite{may_siam75} even for inhomogeneous invasion rates generating a non-trivial reaction in the populations \cite{tainaka_pla93, tainaka_pla95}. The cyclic dominance can mediate positive or negative feedback throughout the cyclic process that depends on the parity of the number of strategies within the cyclic game \cite{tainaka_jtb99, sato_k_amc02, szabo_pre07}. The consequences of this parity effect can affect the behavior in many systems.

The presence of cyclic interactions is responsible for the survival of a large number of strategies and the bio-diversity in biological systems \cite{may_n91, szabo_pr07}. The investigations of predator-prey models with a large number of species and with a complex structure of cyclic dominance indicate an extremely wide scale of behaviors. In some systems the cyclic components can maintain different subsets of strategies, called strategy associations, that are stabilized against the external invaders by suitable spatio-temporal patterns. These strategy associations can survive simultaneously in a large spatial systems by forming large domains \cite{szabo_jpa05, szabo_jtb07, szabo_pre08b, rulands_jsm11}. In these complex systems the cyclic dominance among the strategy associations can be quantified by determining the average velocity of interfaces separating them  \cite{vukov_pre13, dobrinevski_pre14}.

\begin{figure}[ht]
\centerline{\epsfig{file=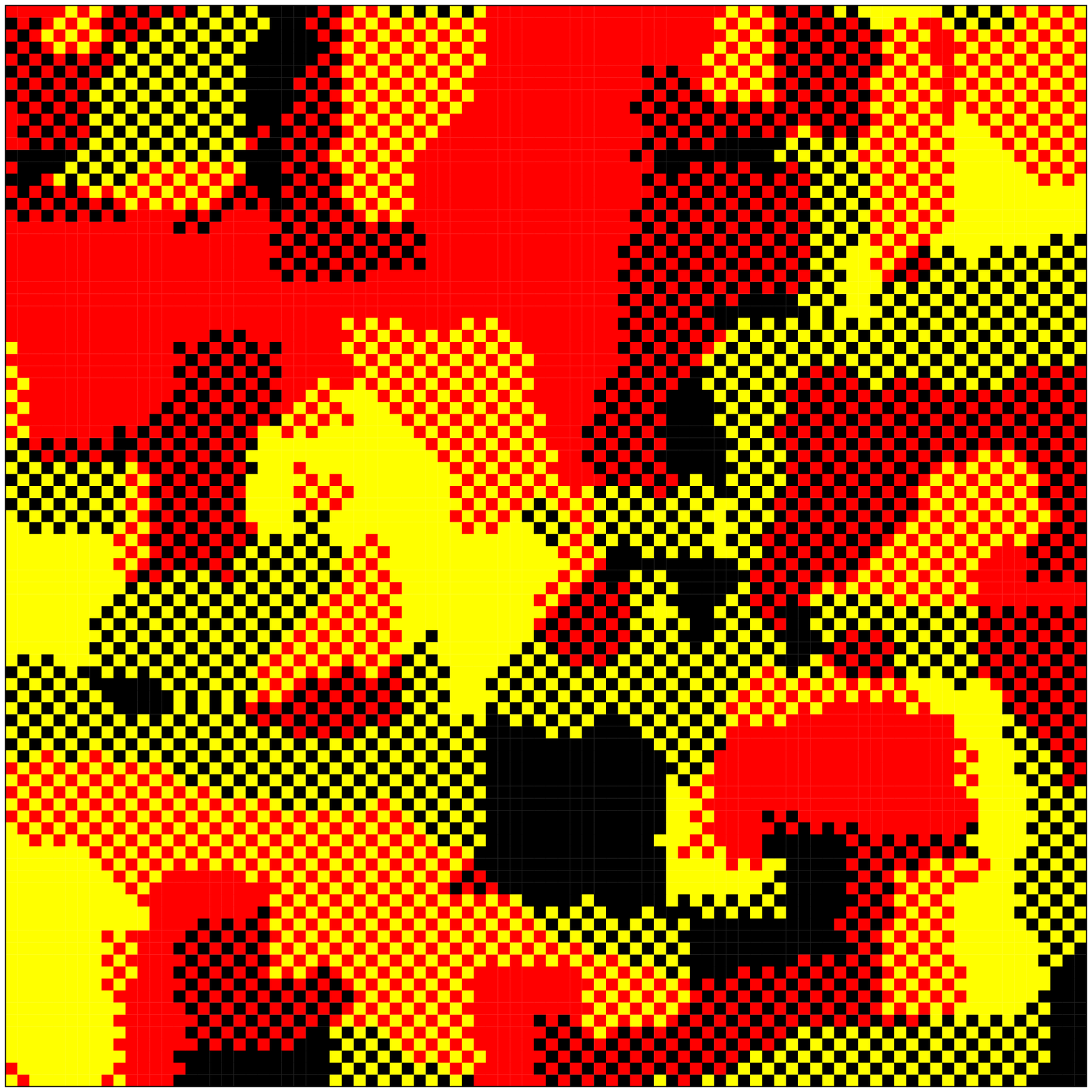,width=5cm}\hspace{1cm} \epsfig{file=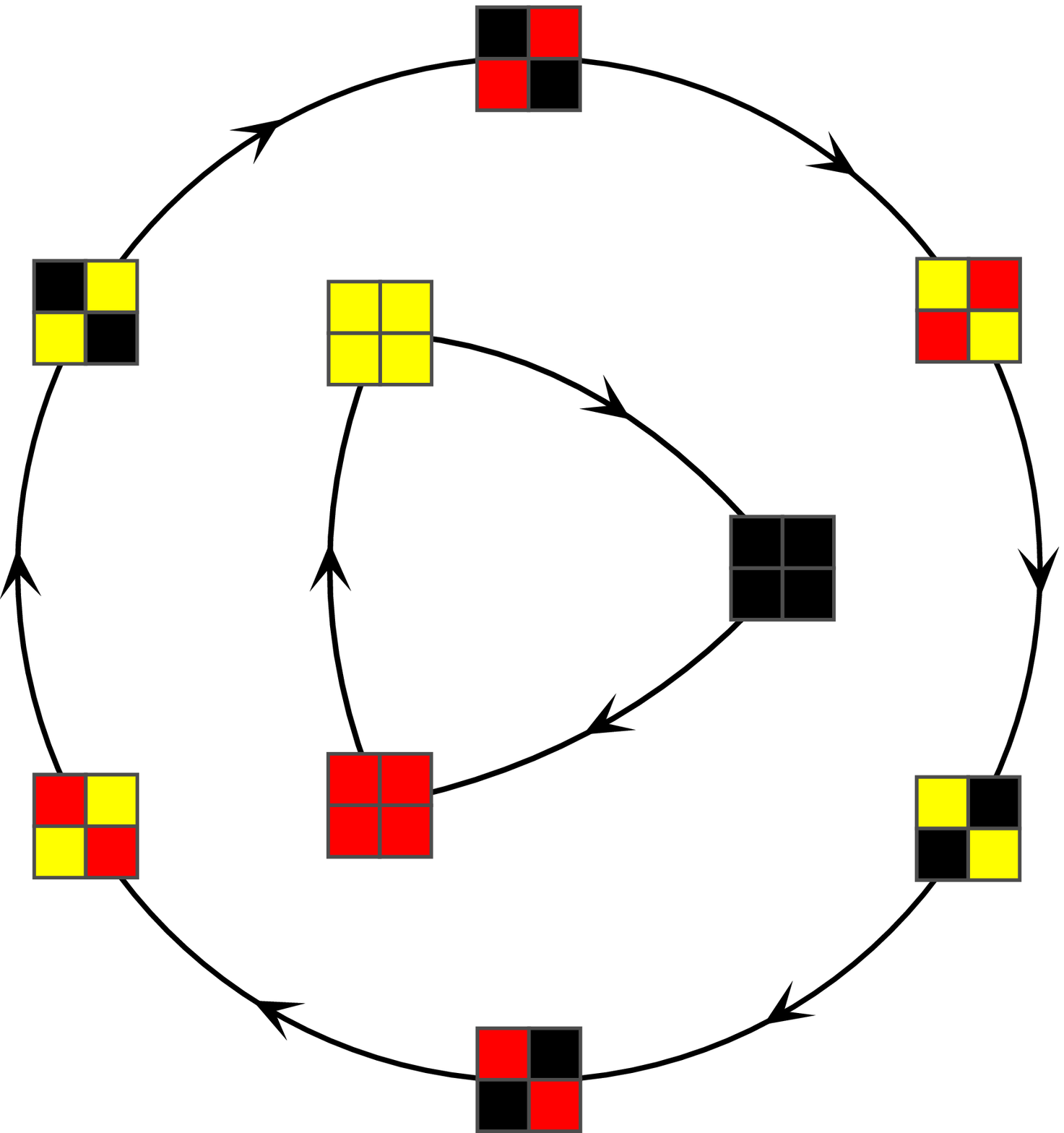,width=5cm}}
\caption{(Color online) Snapshot (left) of a chimera state on a square lattice where the evolution of a rock-paper-scissors game is controlled by a low noise logit rule applied synchronously. The right hand plot illustrates how the spatial patterns change cyclically.}
\label{fig:rps_ca}
\end{figure}

The analysis of the competition between strategy associations is generally based on models with random sequential imitation type evolutionary rule that may even be applied simultaneously for several systems  \cite{wiltermuth_ps09} when the models become similar to stochastic cellular automata \cite{wolfram_rmp83, langton_pd86}. The spatial rock-paper-scissor game with a synchronized stochastic logit update \cite{varga_pre14} has demonstrated the appearance of chimera states which have been intensively studied in the literature of coupled spatial oscillators \cite{abrams_prl04, dudkowski_pre14, santos_pla15, laing_pre15, xie_jb_pre15}. For the repeated two-player rock-paper-scissors game the synchronized logit rule at low noises results in cyclic choices [{\it e.g.} $(1,1) \to (2,2) \to (3,3) \to (1,1)$] until the first mistake. Afterwards the $(1,2) \to (3,2) \to (3,1) \to (2,1) \to (2,3) \to (1,3) \to (1,2)$ cycle is repeated in the absence of mistakes. Such cycles can also occur in the spatial system as it is illustrated in Fig.~\ref{fig:rps_ca}.
The numerical simulations indicate the formation of large domains in which the cyclic choices are stabilized by the neighborhoods suppressing the effect of individual mistakes. In the snapshots one can distinguish nine types of domains representing different phases of these cycles. Additionally the visualization of the pattern evolution indicates clearly the presence of rotating spiral arms due to the cyclic dominance between the oscillating associations.

Finally we mention that the interfaces separating two strategy associations can serve as a location for the emergence of a new strategy association with a proper spatio-temporal structure. Such phenomena were reported in spatial evolutionary games with cyclic dominance between the strategies/species (for $n\le 5$) where the evolutionary rule is based on imitation and site exchange of neighboring players \cite{szabo_jtb07} \cite{vukov_pre13}. The interfaces in Fig.~\ref{fig:d12n5dg} can also exemplify the appearance of a new phase that may play a crucial role in the formation of multicellular living materials.

\section{CONCLUSIONS AND OUTLOOK}
\label{sec:co}

We have reviewed our recent understanding of potential games representing an intimate relationship between physical systems and models applicable to study relevant phenomena in biological and social/economical systems. The analogy becomes particularly
important for those social and biological multi-agent systems where the pair interactions can be well described by symmetric $n \times n$ potential games with logit rules when the systems are driven into the Boltzmann distribution and the general laws of thermodynamics are valid. The application of concepts and methods developed in statistical physics proved to be trivially beneficial for the partnership games where the equivalent players share their income equally. It turned out, however, that the analogies can be directly extended to the potential games representing a wider scale of games.

The evaluation of the potential is demonstrated if it exists. The largest component of the potential matrix identifies the preferred Nash equilibrium (playing the role of ground state in physical systems) even for multi-agent systems composed of uniform pair interactions. This feature implies a simple method for determining the phase diagram at low noises. If the largest component of the potential matrix is located on the main diagonal of the potential matrix then all the players prefer to choose the corresponding strategy independent of the connectivity structure. In general, the latter systems show a thermodynamical behavior represented by the Ising model in the presence of an external magnetic field, even for $n>2$. As Ising type models have already been investigated under a wide scale of different conditions (including symmetries, randomness, networks, ordering processes) therefore many results of statistical physics can be directly adapted to explain the phenomena in evolutionary games, as well. If the largest pair of the potential matrix components occurs outside the main diagonal then the systems have two equivalent preferred Nash equilibria and become similar to the anti-ferromagnetic Ising models. On bipartite graphs these systems exhibit an Ising type order-disorder phase transition when the noise level is increased, otherwise the sublattice ordering can be suppressed by frustration and/or randomness that may result in extremely slow relaxations towards the final stationary state as discussed in the literature of spin glasses and Griffiths phases.

The classification of games into four classes of interactions has helped us conclude general features characterizing the corresponding subset of games. Accordingly, the players are not interested in favoring one of their strategies for games with cross-dependent payoffs, thus they choose strategies at random. For the self-dependent payoffs the players' behaviors can be considered separately from each other. Consequently, all these multi-agent systems with equivalent players and interactions are well described by considering only one player. It is found, furthermore, that the real pair interactions of potential games can be built up as a linear combination of coordination type games between the possible strategy pairs.

It is shown that the presence of the elementary games with cyclic dominance prevents the existence of potential. The consequences of the latter deviations are discussed for several examples challenging the application of methods developed in the field of non-equilibrium statistical physics. In the light of these results one can conclude that these terms of interactions result in self-organizations characterizing living systems.

We have shown that some general questions of traditional game theory become particularly transparent when using the tools of graph theory. Here we used graphs for three purposes. The dynamical graphs visualize the strategy profiles (microscopic states) and the possible transitions between them if only unilateral strategy changes are allowed in the system. Due to the simple structure of the dynamical graph (for the symmetric games) we could determine the number of independent and relevant loops, along which the sum of payoff variations should be zero for the potential games. The flow graph illustrates the preferred strategy changes and simplifies the identification of the pure Nash equilibria (that always exist in potential games) as nodes without outgoing edges. The dominance graph denotes graphically the payoff differences quantified by the antisymmetric part of the payoff matrix. Using these concepts one can distinguish cyclic and hierarchical dominance. In potential games the hierarchical dominance can be related to emergence of social dilemmas occurring even for $n>2$. It is hoped that further graph theoretical investigations can throw light on additional relationships.

When writing this survey we faced challenging questions and interesting phenomena week by week. Some of these problems have already been clarified during the preparation of this work while others remained in the state of "challenging questions". Examples for the former problems are the decomposition of matrix games, the identification of different classes of interactions, the relevance of cyclic dominance, the inherent symmetries involved in the matrices, and a series of interesting phenomena. The list of the latter examples is much longer and contains the identification of ordinal potential games, the elucidation of inherent symmetries in the classes of interactions, the systematic investigation of social dilemmas for potential games, the relevance of high-entropy associations, the spontaneous formation of strategy associations in the presence of cyclic games, the extension of relaxation process for quenched random interactions, the co-evolutionary processes including the evolution of connectivity networks, payoffs, dynamical rules and emergence of new strategies, etc. The study of these intriguing questions offers further promising challenges.

\section*{Acknowledgements}
\addcontentsline{toc}{section}{Acknowledgements}

Discussions with Ben Allen, Kinga Bod{\'o}, Bal{\'a}zs Kir{\'a}ly, Martin Nowak, Attila Szolnoki, and Jeromos Vukov are gratefully acknowledged. This work was supported by the John Templeton Foundation (FQEB Grant \#RFP-12-22) and the Hungarian National Research Fund (OTKA TK-101490).

\phantomsection
\section*{References}
\addcontentsline{toc}{section}{References}


\end{document}